\documentclass[a4paper,fleqn,usenatbib]{mnras}

\usepackage[T1]{fontenc}
\usepackage{ae,aecompl}

\usepackage{graphicx}	
\usepackage{amsmath}	
\usepackage{amssymb}	
\usepackage{pdflscape}
\usepackage{siunitx}
\usepackage[normalem]{ulem}

\newcommand\gaia{\textit{Gaia~}}
\newcommand\egdr[1]{\gaia~EDR#1}
\newcommand{\mv}{$M_{V}-{\rm[Fe/H]}$}
\newcommand{\mg}{$M_{G}-{\rm[Fe/H]}$}

\title[RR Lyrae LZ and PW(Z) based on \textit{Gaia} EDR3 parallaxes]{New LZ and PW(Z) relations of 
RR Lyrae stars calibrated  with  \textit{Gaia} EDR3 parallaxes}

\author[A. Garofalo et al.]{
A. Garofalo $^{1}$,
H.E. Delgado $^{2}$,
L.M. Sarro $^{2}$,
G. Clementini $^{1}$,
T. Muraveva $^{1}$,
\newauthor
M. Marconi $^{3}$,
V. Ripepi $^{3}$\\
$^{1}$INAF,
Osservatorio di Astrofisica e Scienza dello Spazio di Bologna, via Piero Gobetti 93/3,
40129 Bologna, Italy. \\
$^{2}$Dpto. de Inteligencia Artificial, UNED, c/
Juan del Rosal, 16, 28040 Madrid, Spain.\\
$^{3}$INAF-Osservatorio Astronomico di Capodimonte, Via Moiariello 16, I-80131 Naples, Italy}

\date{Accepted XXX. Received YYY; in original form ZZZ}

\pubyear{2021}

\begin{document}
\label{firstpage}
\pagerange{\pageref{firstpage}--\pageref{lastpage}}
\maketitle

\begin{abstract}
We present new luminosity-metallicity ($LZ$; $M_{V}-\rm[Fe/H]$ and $M_{G}-\rm[Fe/H]$) relations and, for the first time,  empirical,  \textit{Gaia} 
 three-band ($G,G_{\rm BP},G_{\rm RP}$) period-Wesenheit-metallicity ($PWZ$) relations  of RR Lyrae stars (RRLs) derived using  a hierarchical Bayesian approach and new accurate parallaxes published for these variables in the {\textit Gaia} Early Data Release 3 (EDR3). In a previous study  we obtained Bayesian hierarchically-derived  
$LZ$ relations from a sample of about four hundred  Milky Way field RRLs  with $G$-band light curves and trigonometric parallaxes 
published in the {\textit Gaia} Data Release 2 (DR2), using $V$ mean magnitudes, metallicities, absorptions and pulsation periods available in the literature. We now extend that study in two directions. Firstly, we update our previous results using trigonometric parallaxes from \textit{Gaia} 
EDR3 and incorporate the Bayesian analysis of a first empirical 
$PWZ$ relation derived 
using those field RRLs with $G$, $G_{\rm BP}$ and $G_{\rm RP}$ time-series photometry available in {\textit Gaia} DR2. Secondly, we 
use  Bayesian inference to derive 
$LZ$ relations and  
empirical PW {\textit Gaia} three-band relations
 from 
385 RRLs belonging to 15 Milky Way globular clusters (GC) with literature-compiled spectroscopic metallicities ranging from $-$0.36 to $-2.39$ dex and prior distances extending from $2.2$ to $41.2$ kpc. 
From the samples of RRLs analysed in this paper we infer a mean {\textit Gaia} EDR3 zero-point offset of 
 $-$0.028 mas with median values ranging from $-$0.033 ($LZ$ and $PWZ$ models for field stars) to $-$0.024  mas ($LZ$ model in the $V$-band  for GC RRLs).

\end{abstract}

\begin{keywords}
stars: variables: RR Lyrae --
(Galaxy:) globular clusters: general -- 
parallaxes -- 
methods: statistical -- methods: data analysis

\end{keywords}






\section{Introduction}
\label{sec:intro}

On December 2020,  the early installment of the \textit{Gaia} third data release (Early Data Release 3, EDR3; \citealt{GaiaC-Brown-2020}) 
published 
updated astrometry and photometry for nearly 2 billion sources 
down to a limiting magnitude $G \sim$ 21 mag. 
Among them, are tens of thousands of RR Lyrae stars (RRLs). 
The number and accuracy of RRL distances directly estimated from  parallax measurements have been steadily increasing with subsequent  \textit{Gaia} Data Releases, 
leading to a renewed role of RRLs as standard candles of the cosmic distance ladder. 
 At the same time, 
 the intriguing discord between the value of the Hubble constant, $H_0$,  from measurements of the cosmic microwave background anisotropy and from distance indicators in the local Universe
 (\citealt[][and references therein,  but see the 
 paper 
 by \citet{Freedman-2021}, for a different view)] {Verde-et-al-2019,DiValentino-2021,Riess-Adam-al-2021,Riess-et-al-2021} has shown the need for  both reducing uncertainties and systematics in the  stellar standard candles traditionally used in the distance ladder measurement of $H_0$ (the Population I Cepheids) and for   exploring alternative  paths to $H_0$ independent of Cepheids, such as that represented by  Population II RRLs  \citep[and references therein]{Beaton-et-al-2016} 
 with luminosity  calibrated on \textit{Gaia}  parallaxes.
\\
A number of different characteristic relations from the optical to the mid-infrared bands make RRLs standard candles widely used to measure distances within the Milky Way (MW) and its nearest neighbours up to about 1 Mpc in distance  \citep[see e.g.][]{Ordonez-Sarajedini-2016}. In optical passbands the absolute visual magnitude of RRLs is linked to the metallicity by a relation, traditionally  known as Luminosity - Metallicity ($LZ$) relation, $M_V - {\rm[Fe/H]}$ \citep[see e.g.][and references therein]{Sandage-1993, Caputo-et-al-2000, Cacciari-Clementini-2003,Clementini-et-al-2003,Catelan-et-al-2004,Marconi-et-al-2015, Muraveva-et-al-2018a} while, from red ($R$ and $I$ bands) to near-mid infrared passbands, RRLs follow Period-Luminosity ($PL$), Period-Luminosity-Metallicity relations \citep[$PLZ$, see e.g.][and references therein]{Longmore_et-al-1986,Bono-et-al-2003,Catelan-et-al-2004,Sollima-et-al-2006, Coppola-et-al-2011,Madore-et-al-2013,Braga-et-al-2015,Marconi-et-al-2015,Muraveva2015, Neeley-et-al-2019}.
Intrinsic errors and systematics affecting distances based on the $LZ$ relation are  widely discussed in the literature \citep[see e.g.][and references therein]{Bono-2003}. They include uncertainty of the metallicity measurements, lack of information about the $\alpha$-element enhancement, dispersion of the RRL mean luminosity caused by poor knowledge reagarding the off  zero-age horizontal branch (ZAHB) evolution of these variables and, also, a possible non-linearity  of the relation itself over the whole metallicity range spanned by the MW field RRLs.
Although the  $PL(Z)$ relations for red and infrared passbands have well known advantages in comparison with the optical $LZ$ relation (primarily a smaller dependence of the luminosity on metallicity and evolutionary effects and a much  weaker impact of reddening at infrared passbands),
uncertainties on these quantities still affect the accuracy with which the 
$PLZ$ coefficients are determined and  the dispersion of the  relations themselves \citep[see e.g.][for a discussion]{Marconi-et-al-2015, Neeley-et-al-2017}. 
The introduction of the Wesenheit function  by \citet{van-den-Bergh-1975} and \citet{Madore-1982} has allowed to minimize the reddening dependence  of  the $PL$ relation, by defining a Period-Wesenheit ($PW$) relation which is reddening-free by construction, 
assuming that the extinction law is known.
However, all  RRL relations need a better determination of their coefficients to reduce systematics and an accurate calibration of their zero points to firmly anchor the distance ladder. 

The advent of \textit{Gaia} has  triggered  new 
empirical and theoretical investigations of the RRL  $PL(Z)$, $PW(Z)$ and $LZ$ relations 
and improved geometrical calibrations  
using \textit{Gaia} astrometry and photometry.
In the recent years there has also been an increasing 
application of Bayesian methods to improve the cosmic distance ladder. In 
particular, Bayesian hierarchical methods have been used to  increase the potential of  type Ia supernovae   \citep[SN Ia,][]{Mandel-et-al-2009} 
and red clump (RC) stars \citep{Hawkins-at-al-2017, Hall-et-al-2019}  as distance indicators, to obtain a first calibration of the $PL$, $PLZ$ and $PW$ relations of Cepheids
and the $LZ$ relation of RRLs based on parallaxes from the Tycho-\textit{Gaia}  Astrometric Solution (TGAS; \citealt{GaiaC-Clementini-2017}), to constrain the 
$PLZ$  relations of RRLs 
\citep{Sesar-et-al-2017,Muraveva-et-al-2018a,Delgado-et-al-2019} and to clarify 
the Hubble constant tension \citep{Feeney-et-al-2018}.  The main 
advantage of the Bayesian hierarchical method, a model-based 
machine learning approach \citep{bishop-2013,ghahramani-2015}, 
is the use of a network model 
that includes problem parameters and measurements, dependency relationships 
between them and prior information gathered from previous studies. In this way, 
when the inference is carried out, it is guaranteed that the observational 
uncertainties are simultaneously propagated through the network and the 
uncertainties on the parameters of interest are estimated in a way
that is consistent with the measurement uncertainties.\\
Applying a Bayesian modelling approach along with \textit{Gaia} DR2 parallaxes and $G$-band time-series photometry and, $V$-band time-series photometry from literature, in  \cite{Muraveva-et-al-2018a} we  derived new \mv ~and infrared $PLZ$ relations using a sample of about 400 RRLs in the field of the MW. In that work, we also presented for the first time, a \mg   ~relation for RRLs and found a dependence of the luminosity on metallicity higher than usually found in the literature in all the considered passbands.
 Still on  the  empirical  side, \citet{Neeley-et-al-2019}, using the DR2 parallaxes of 55 Galactic RRLs, calibrated multi-band $PW(Z)$ and infrared $PL(Z)$ relations. Also these authors found  a metallicity dependence at optical wavelengths larger than expected and predicted by current theoretical models, although consistent with them within 1$\sigma$.
Most recently, on the theoretical side,  \citet{Marconi-et-al-2021} derived the first theoretical $PW$ relations in the $G$, $G_{\rm BP}$ and $G_{\rm RP}$ bands varying both metal and helium contents. They noticed that these $PW$ relations show a dependence of the zero point on metal abundance. In particular, they found that a metallicity variation can change the zero point of the relation by a few tenths of magnitude, unlike the $PW$ relation based on optical bands  which is independent of metallicity \citep{Marconi-et-al-2015}. 
Lastly, \citet{Bhardwaj-et-al-2020} inferred  for the first time a new empirical $PL_{Ks}Z$ relation anchored to EDR3 distances and theoretical $PLZ$ relations,  using 5 globular clusters (M3,M4, M5, $\omega$Cen, and M53). They found a rather weak metallicity dependence ($\sim$0.15-0.17 mag dex$^{-1}$) that is in good agreement with theoretical predictions for the $PL_{Ks}Z$ relation \citep[$\sim$0.18 mag dex$^{-1}$,][]{Marconi-et-al-2015}. 

To further investigate the metallicity dependence of the $LZ$ and $PW(Z)$ relations we have exploited the EDR3 parallaxes available  for RRLs in two different environments: the MW field and a sample of Galactic globular clusters (GCs). In this paper we thus  present new $LZ$ ($M_{V}-\rm[Fe/H]$ and $M_{G}-\rm[Fe/H]$) and $PWZ$ ($PW_{\left(G,G_{\rm{BP}},G_{\rm{RP}}\right)}Z$) relations obtained for field RRLs and new $LZ$ ($M_{V}-\rm[Fe/H]$ and $M_{G}-\rm[Fe/H]$) and $PW$ relations obtained for cluster RRLs, using a hierarchical Bayesian inference and the new and accurate \textit{Gaia} EDR3 parallaxes of these variable stars.

 The structure of the paper is as follows: in Section~\ref{sec:Data} we describe the  samples of MW  RRLs (field and globular clusters) we have analysed,  while in Section~\ref{sec:method} we introduce the method used to infer the  RRL relations. In Section~\ref{sec:results} we present the new  RRL relations, the \textit{Gaia} EDR3 zero-point offset we have derived and discuss our  results.
 In Section~\ref{sec:validation} we discuss our distance estimates to GCs and test 
 our new relations by comparing the distances to the Large Magellanic Cloud (LMC) they predict with  
 accurate distance determinations available for this galaxy in the literature.  
 Finally,  we summarise our results in Section~\ref{sec:summary}.

\section{Data}
\label{sec:Data}
 \begin{figure}
	\begin{center}
		\includegraphics[scale=0.4]{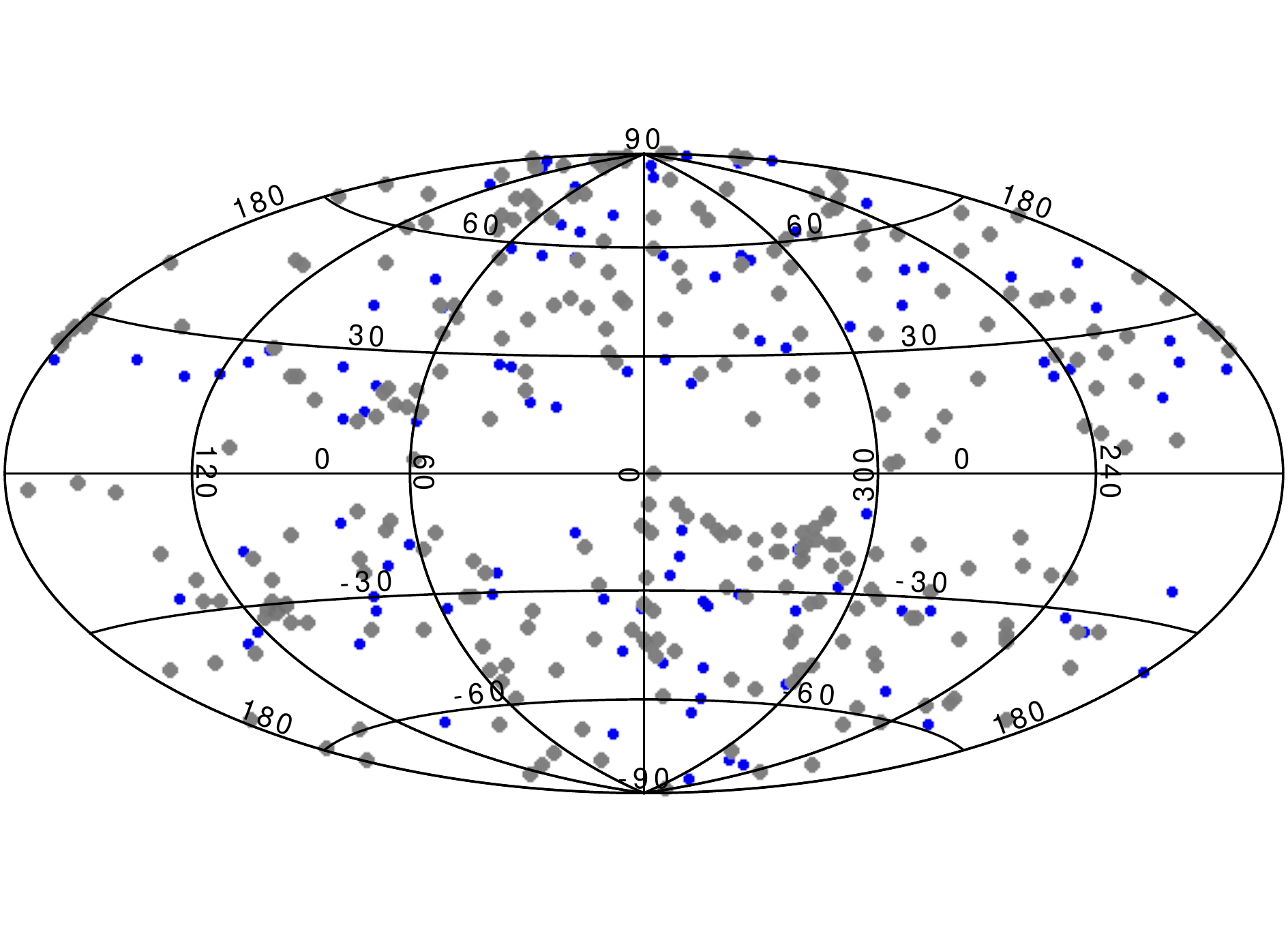}
		\includegraphics[scale=0.4]{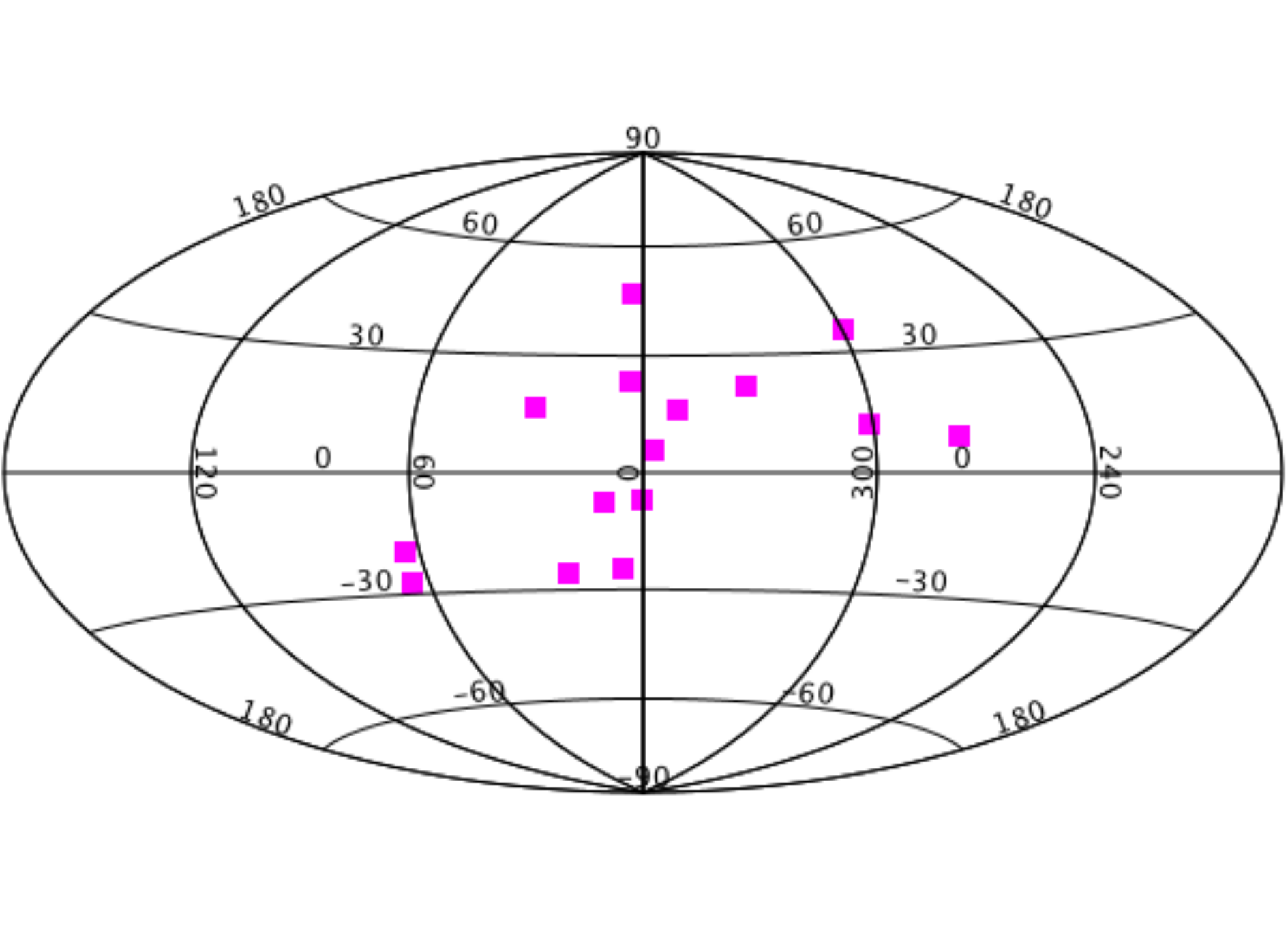}
		\caption{\textit{Top panel:} Hammer-Aitoff projection in galactic coordinates of the sky distribution for the sample of 397 Galactic field RRLs  and the subsample of 291 RRLs with {\tt ruwe} $<$ 1.4   ( blue and grey symbols in the upper plot, respectively). \textit{Bottom panel:} The 15 globular clusters (magenta symbols, lower plot) hosting the RRLs considered in the present study.}
		\label{fig:MAP}
	\end{center}
\end{figure}

\begin{table*}
 \caption{Properties of 15 Galactic GCs used in this study.}
 \label{tab:GC-prop}
 \begin{tabular}{lccccccc}
  \hline
    Cluster ID & $E(B-V)$&A$_{V}$ & $\rm[Fe/H]$ & $\mu_{V}$ &distance&RRLs$^{*}$\\
  & (mag) & (mag)&(dex)& (mag) & (kpc) &\\
  \hline
NGC~3201 & 0.220 $\pm$ 0.015& 0.682&$-$1.51 $\pm$ 0.02&14.20&4.9& 86\\
NGC~4590 (M68) &0.053 $\pm$ 0.001 &0.162&$-$2.27 $\pm$ 0.04&15.21&10.3& 42\\
NGC~5824 & 0.148 $\pm$ 0.003 &0.454& $-$1.94 $\pm$ 0.14&17.94&32.1& 48\\
NGC~5904 (M5)& 0.032 $\pm$ 0.001 & 0.097&$-$1.33 $\pm$ 0.02&14.46&7.5& 129\\
NGC~6121 (M4)& 0.428 $\pm$ 0.150 & 1.549& $-$1.18 $\pm$ 0.02&12.82&2.2& 47\\
NGC~6171 (M107)& 0.395 $\pm$ 0.006&1.264 &$-$1.03 $\pm$ 0.02&15.01&6.4& 23\\
NGC~6316&  0.653 $\pm$ 0.031& 2.089& $-$0.36$\pm$ 0.14& 16.77&10.4 &2\\
NGC~6426 & 0.298 $\pm$ 0.012 &0.934& $-$2.39 $\pm$ 0.30&17.68&20.6& 16\\
NGC~6569& 0.369$\pm$0.008& 1.181& $-$0.72$\pm$ 0.14& 16.83&10.9 &21\\
NGC~6656 (M22)& 0.280 $\pm$ 0.100 & 0.896 &$-$1.70 $\pm$ 0.08&13.60&3.2& 31\\
NGC~6864 (M75)& 0.130 $\pm$ 0.004 &0.403& $-$1.29 $\pm$ 0.14&17.09&20.9& 41\\
NGC~7006 & 0.071 $\pm$ 0.004 &0.214& $-$1.46 $\pm$ 0.06&18.23&41.2& 64\\
NGC~7078 (M15)&0.093 $\pm$ 0.003&0.291 & $-$2.33 $\pm$ 0.02&15.39&10.4& 143\\
Rup~106 & 0.150 $\pm$ 0.003 &0.458& $-$1.78 $\pm$ 0.08&17.25&21.2& 13\\
Ter~8 & 0.127 $\pm$ 0.003 &0.392& $-$2.06 $\pm$ 0.30&17.47&26.3& 3\\
  \hline
  \multicolumn{7}{l}{NOTES. $E(B-V)$ values are from \citet{Schlafly-and-Finkbeiner-2011}. See text for details. }\\   \multicolumn{7}{l}{Spectroscopic metallicities from \citet{Carretta-et-al-2009} and  \citet{Dias-et-al-2016} on \citet{Carretta-et-al-2009} scale.}\\
  \multicolumn{7}{l}{The $\mu_{V}$  and distance values are from \citet[version 2010]{Harris-1996}.}\\
  \multicolumn{7}{l}{
 $^{*}$ Number of RRLs known. References:}\\ \multicolumn{7}{l}{NGC~3201 \citep{Layden-et-al-2003,Arellano-et-al-2014}, NGC~4590 \citep{Kains-et-al-2015},}\\
 \multicolumn{7}{l}{ NGC~5824 \citep{Walker-et-al-2017}, NGC~5904 \citep{Oosterhoff-1941,Kaluzny-et-al-2000,Arellano-et-al-2016},}\\
  \multicolumn{7}{l}{NGC~6121 \citep{Yao-et-al-1988,Kaluzny-et-al-2013,Stetson-et-al-2014},}\\
  \multicolumn{7}{l}{NGC~6171 \citep{Dickens-1970,Clement-et-al-1997,Samus-et-al-2009,Drake-et-al-2013},  NGC~6316 \citep{Layden-et-al-2003-bis},}\\
  \multicolumn{7}{l}{NGC~6426 \citep{Grubissich-1958,Papadakis-et-al-2000,Samus-et-al-2009},  NGC~6569 \citep{Hazen-Liller-1985,Soszynski-et-al-2014}},\\  
  \multicolumn{7}{l}{NGC~6656 \citep{Kunder-et-al-2013,Soszynski-et-al-2014},}\\
    \multicolumn{7}{l}{NGC~6864 \citep{Corwin-et-al-2003,Scott-et-al-2006}, NGC~7006 \citep{Wehlau-1999,Samus-et-al-2009},}\\
    \multicolumn{7}{l}{NGC~7078 \citep{Corwin-et-al-2008}, Rup~106 \citep{Kaluzny-et-al-1995}, Ter~8 \citep{Salinas-et-2005,Torrealba-et-al-2015}}\\ 
 \end{tabular}

\end{table*}

 In this section we present the two separate data samples: RRLs in the MW field and in Galactic GCs used for the derivation
of the corresponding $LZ$ relations in the  $V$ and $G$ passbands and $PWZ$ relations in the  {\it Gaia} $G$, $G_{\rm BP}$ and $G_{\rm RP}$ bands.

\subsection{Field RR Lyrae stars}
\label{sec:FS-data}
\begin{figure}
	\begin{center}
		\includegraphics[ width=8.3cm]{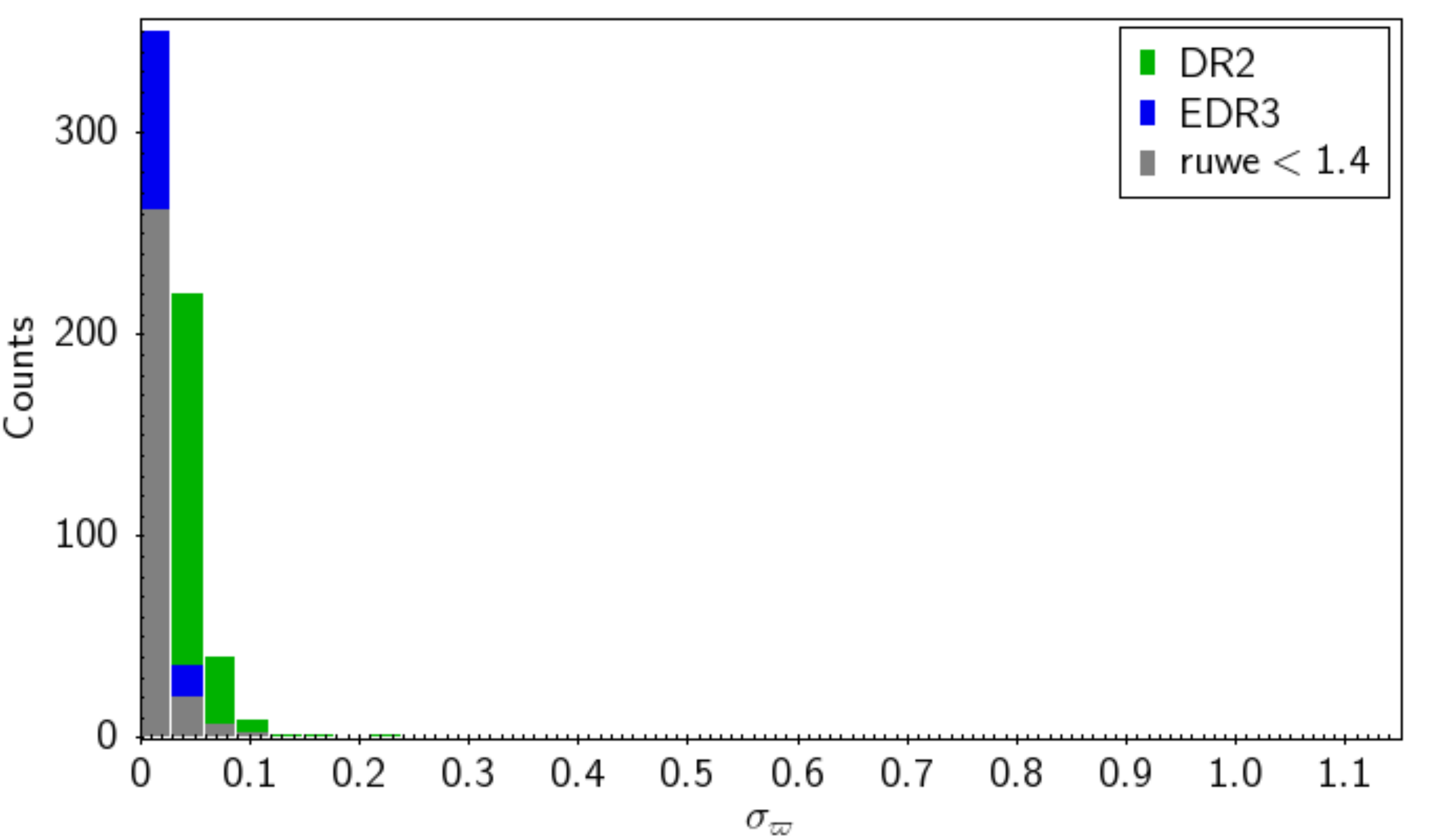}
		\includegraphics[width=8cm]{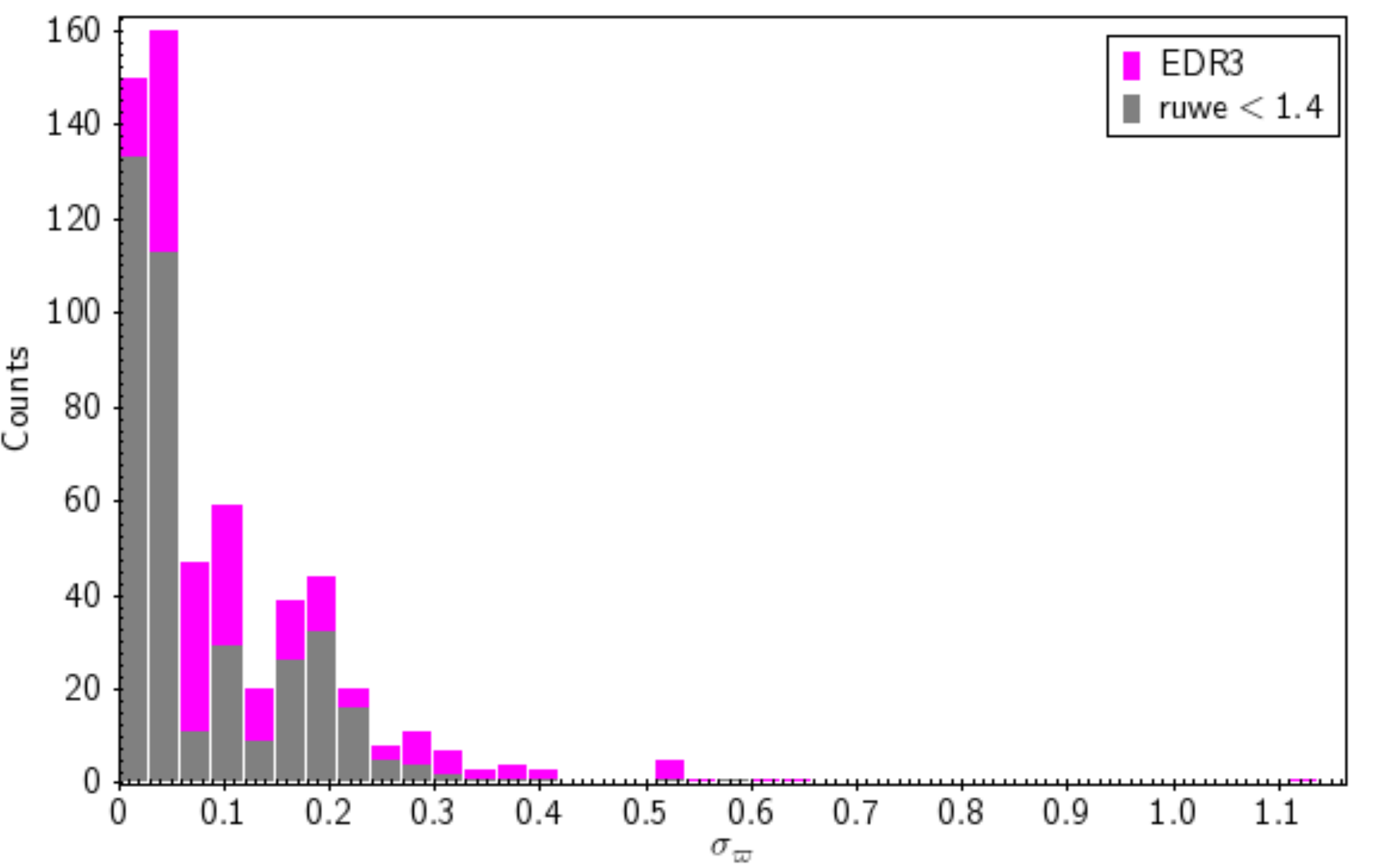}
		\caption{\textit{Top panel:} The blue histogram shows the distribution of the EDR3 parallax uncertainties ($\sigma_{\varpi_{EDR3}}$) for our initial sample of 397 field RRLs. For comparison the DR2 parallax uncertainties of the stars are represented by a  green histogram. The grey histogram shows the $\sigma_{\varpi_{EDR3}}$ distribution for the subsample of 291 RRLs  
		with \texttt{ruwe}<1.4. \textit{Bottom panel:} The magenta histogram shows the distribution of the EDR3 parallax uncertainties for our initial  sample of 582 RRLs in GCs. 
		The grey histogram shows the  $\sigma_{\varpi_{EDR3}}$ distribution for the subsample of 385 GC  RRLs  
		with \texttt{ruwe}<1.4. In both panels the bin size is 0.03 mas.}
		\label{fig:Parallax-error-RUWE}
	\end{center}
 \end{figure}

We have built our field RRL sample starting from the 401 MW RRLs analysed in  \citet{Muraveva-et-al-2018a} which were based on the \citet{Dambis-et-al-2013} sample. 
We first replaced their  {\it Gaia} DR2 parallaxes with the new EDR3 values  ($\varpi_{EDR3}$), which are available for the whole sample.  We  then discarded stars DI~Leo and VX~Ind because  
they were re-classified by \citet{Ripepi-at-al-2019}  as a classical Cepheid and a Population~II Cepheid of BL Herculis type, respectively.  
We also excluded CK~UMa, because  the star lacks a spectroscopic  metallicity estimate, and V363~Cas that \citet{Dambis-et-al-2013}    classifies as fundamental mode RRL (RRab) but it is in fact a  Galactic classical Cepheid according to the literature (\citealt{Ripepi-at-al-2019} and references therein). The RRLs in our sample have parallaxes ranging from $-$0.05 mas to 3.98 mas, with  only three of them having a  negative parallax. 
We note that our MW field sample  
includes RR~Lyr itself, the bright variable star after which the whole class is named. Unlike in {\it Gaia} DR2, the EDR3  parallax of  RR~Lyr has a positive and reliable value of  $3.98\pm0.03$ mas, which is 
in good agreement  with  the {\it Gaia} DR1 and the  Hubble Space Telescope (HST) estimates \citep[see][table~1]{GaiaC-Clementini-2017}. Our sample thus includes 397 MW RRLs  homogeneously distributed all over the sky (filled circles in the upper panel of  Fig.~\ref{fig:MAP}). Uncertainties of the {\it Gaia} EDR3 parallaxes for the 397 RRLs in our sample are shown by the blue histogram in Fig~\ref{fig:Parallax-error-RUWE}. They range from 0.01 to 0.20 mas. For comparison, the green histogram shows 
the error distribution of the DR2 parallaxes of these RRLs.\\
To restrict the sample of field RRLs  used to calibrate the $LZ$ and $PLZ$ relations only  
 to sources with best quality parallaxes we took into account the   
re-normalised unit weight error (\texttt{ruwe}) 
\citep{Lindegren-TN-DPAC-2018} and retained only   sources satisfying the 
 condition 
 \texttt{ruwe} < 1.4. With this selection the sample is  reduced to 291 stars (73$\%$;  grey filled circles in the upper plot of Fig.~\ref{fig:MAP}), 260 of which are fundamental mode (RRab) and 31 are first-overtone (RRc) pulsators. The distribution of the parallax uncertainties  for these 291 RRLs is shown by the grey histogram in the top panel of Fig~\ref{fig:Parallax-error-RUWE}. We provide information for the complete dataset of  291 
 MW field RRLs in Table~\ref{tab:data-F}. For each RRL we list the literature name (column 1), 
   pulsation mode and period (columns 7 and 8), mean apparent  magnitude in the $V$-band  $\langle V \rangle$ (column  9), absorption in $V$-band (column  11) and  metallicity estimation  in the \citet{Zinn-West-84} metallicity scale (column  19), taken from 
  table~1 of \citet{Muraveva-et-al-2018a}. In addition we include the {\it Gaia} EDR3 source identifier (column 2), the EDR3  coordinates (columns 3 and 4),
 the EDR3 parallax and its error (columns 5 and 6), and the  intensity-averaged  mean  magnitudes in the $\langle G \rangle$, $\langle G_{\rm{BP}} \rangle$ and $\langle G_{\rm{RP}} \rangle$ bands and their respective  errors taken from the Gaia DR2 \texttt{vari\_rrlyrae} table 
\citep{Clementini-et-al-2019}. To derive the $M_{V}-\rm[Fe/H]$  and $M_{G}-\rm[Fe/H]$  relations, among the 291 field RRLs, we have used 274 sources having $V$ magnitude (see Section~\ref{sec:mv}) and 223 sources  with $G$ magnitude (see Section~\ref{sec:mg}),  respectively. A subsample of 190 field RRLs having $\langle G \rangle$, $\langle G_{\rm{BP}} \rangle$ and $\langle G_{\rm{RP}} \rangle$ magnitudes were instead used to infer the $PW_{\left(G,G_{\rm{BP}},G_{\rm{RP}}\right)}Z$ relation (see Section~\ref{sec:pwz}). 

\subsection{RR Lyrae stars in Globular Clusters}
\label{sec:GC-data}

\noindent 
We selected a sample of 15 Galactic GCs with distances ranging  
from a few to $\sim$ 41 kpc around the Galactic centre and spatially distributed as shown in the bottom panel of Fig.~\ref{fig:MAP}.
The 15 GCs cover a large range in 
metal abundance,  from $-$2.39 to $-$0.36 dex in [Fe/H].
Their $V$ luminosity  at  the horizontal branch (HB) level spans about 5 magnitudes, going from 13.30 mag for NGC~6121 to 18.80 mag for NGC~7006 \citep{Harris-1996}. 


According to the literature, the selected GCs host a total number of  
709 RRLs. 
For each of the 15 GCs  
we list in Table ~\ref{tab:GC-prop} reddening value  
$E(B-V)$ (taken from \citealt{Schlafly-and-Finkbeiner-2011} using as reference the GC centers), absorption in the $V$-band, spectroscopic metallicity (in the \citealt{Carretta-et-al-2009} scale) from \citet{Dias-et-al-2016}, visual distance modulus and distance (in kpc) from \citet{Harris-1996} catalogue
\footnote{Revision of December 2010}. 
\\In the last column of  Table~\ref{tab:GC-prop} we also give  the number of RRLs in each cluster. References to the papers used to collect the samples of GC RRLs are provided  in the table notes.
We cross-matched the  709 cluster  RRLs  
against the {\it Gaia} EDR3 
catalogue using a cross-match  radius of 2 arcseconds and 
retained only  sources unambiguously  classified by the  literature as fundamental and first-overtone RRLs, including also those  exhibiting the 
Blazhko effect (\citealt{Blazhko-1907}; labels ``RRAB'', ``RRAB Bl'', ``RRC" and ``RRC Bl''). With this 
filtering the GC RRL sample reduced to 643 sources. 
We then dropped 3 sources with unknown period and further 58 RRLs for which a 
parallax is not available in the {\it Gaia} EDR3 catalogue, thus remaining with a total sample of 582 GC RRLs  (372 RRab and 210 RRc stars). 
Finally, as done for the field RRLs we  cleaned the GC RRLs discarding sources with  \texttt{ruwe} $\geq 1.4$.
With this latter filtering the sample was reduced to 385
stars (66$\%$, 249 RRab and 136 RRc stars). 
The  distributions of  the  EDR3  parallax  uncertainties  for  the  582 and 385  GC RRL samples are shown 
by the magenta and grey histograms 
in Fig. \ref{fig:Parallax-error-RUWE}, respectively.  
The  complete  dataset  with the relevant information (sourceids, coordinates, parallax and parallax errors from EDR3, pulsation  mode, period, $\langle V \rangle$, and DR2 $\langle G \rangle$ ,$\langle G_{\rm{BP}} \rangle$, $\langle G_{\rm{RP}} \rangle$ mean  magnitudes) is summarised in Table~\ref{tab:data-C}.

To compute the $LZ$ and $PW$ relations 
our Bayesian method (see Section \ref{sec:method} below) corrects the mean apparent magnitudes of the GC RRLs using the  $V$-absorption values provided for each GC in columns 3 of 
 Table~\ref{tab:GC-prop},  
which are taken from \citet{Schlafly-and-Finkbeiner-2011}  assuming a standard total-to-selective absorption ($R_{V}=$3.1). Five GCs of our sample, which are located closer to the center of the Galaxy, are Bulge systems. They are heavily affected by extinction and have a higher metal content. Different studies (\citealt{Barbury-et-al-1998,Bica-et-al-2016} and references therein) have measured a total-to-selective absorption for these GCs higher than 3.1.
For NGC~6171, NGC~6316, NGC~6656 and NGC~6569 we converted $E(B-V)$ into $A_{V}$ using $R_{V}=$3.2 as proposed by \citet{Bica-et-al-2016} while for NGC~6121 we adopted $R_{V}=$3.62 as found by \citet{Hendricks-2012}. We assumed that in each GC the absorption is constant with the exclusion of NGC~6121 and NGC~6569 which are known to be severely affected by differential reddening. For them, we estimated the $E(B-V)$ reddening and its uncertainty individually for each RRL. In particular, we adopted as reddening zero-points the values of \citet{Schlafly-and-Finkbeiner-2011} measured at the NGC~6121 and NGC~6569 centres and derived individual $E(B-V)$ values for each RRL in these clusters using their differential reddening maps provided by \citet{Garcia-et-al-2012}\footnote{In Table~\ref{tab:GC-prop} we report, as reddening uncertainty the reddening dispersion measured by  \citet{Garcia-et-al-2012} for NGC~6121 and NGC~6569. }.
\\Contrary to the field RRLs,
the RRLs  in our GC sample lack  
individual and homogeneous metal abundance estimations. Hence, on the assumption that the clusters considered in our study are mono-metallic, we assigned to their RRLs 
 the spectroscopic metal abundances (on the  \citealt{Carretta-et-al-2009} scale)  derived for the host clusters by \citet{Dias-et-al-2016} and listed  
 in column 4 of 
Table~\ref{tab:GC-prop}.  

\begin{figure*}
	\begin{center}
		\includegraphics[trim= 0 0 0 0 clip, width=7.2cm]{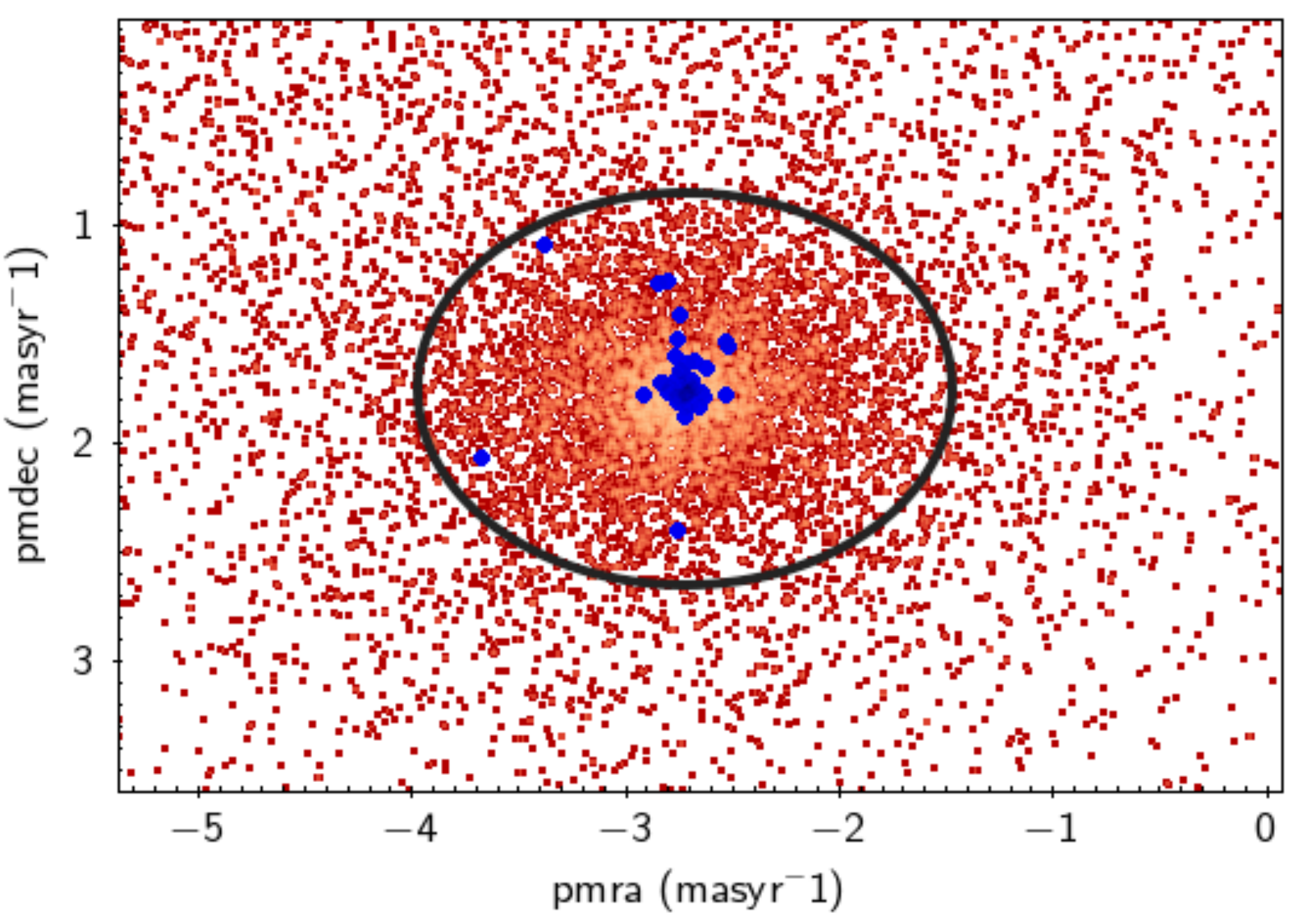}~\includegraphics[trim= 0 0 0 0 clip, width=7.5cm]{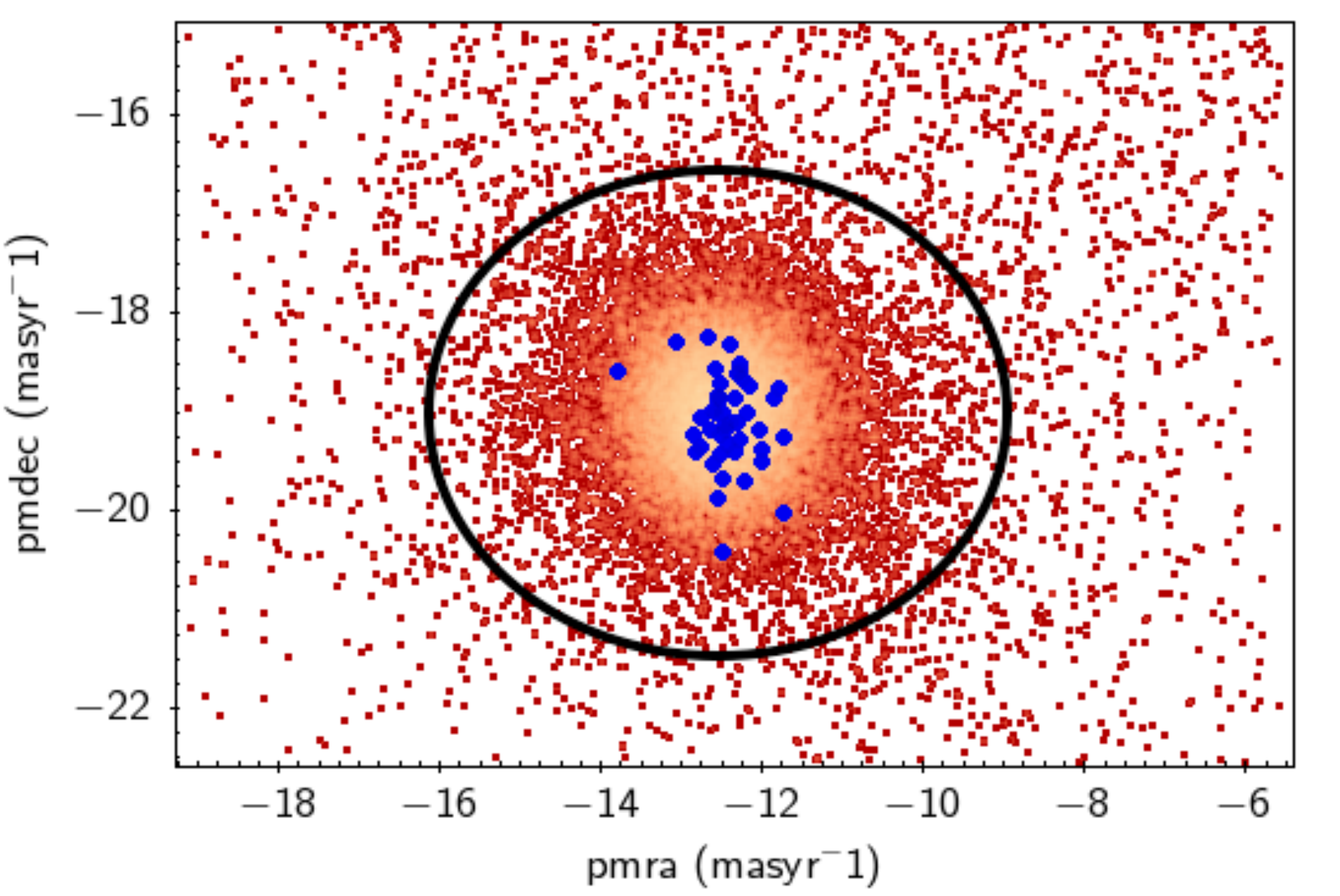}
		\includegraphics[trim= 0 0 0 0 clip, width=7.4cm]{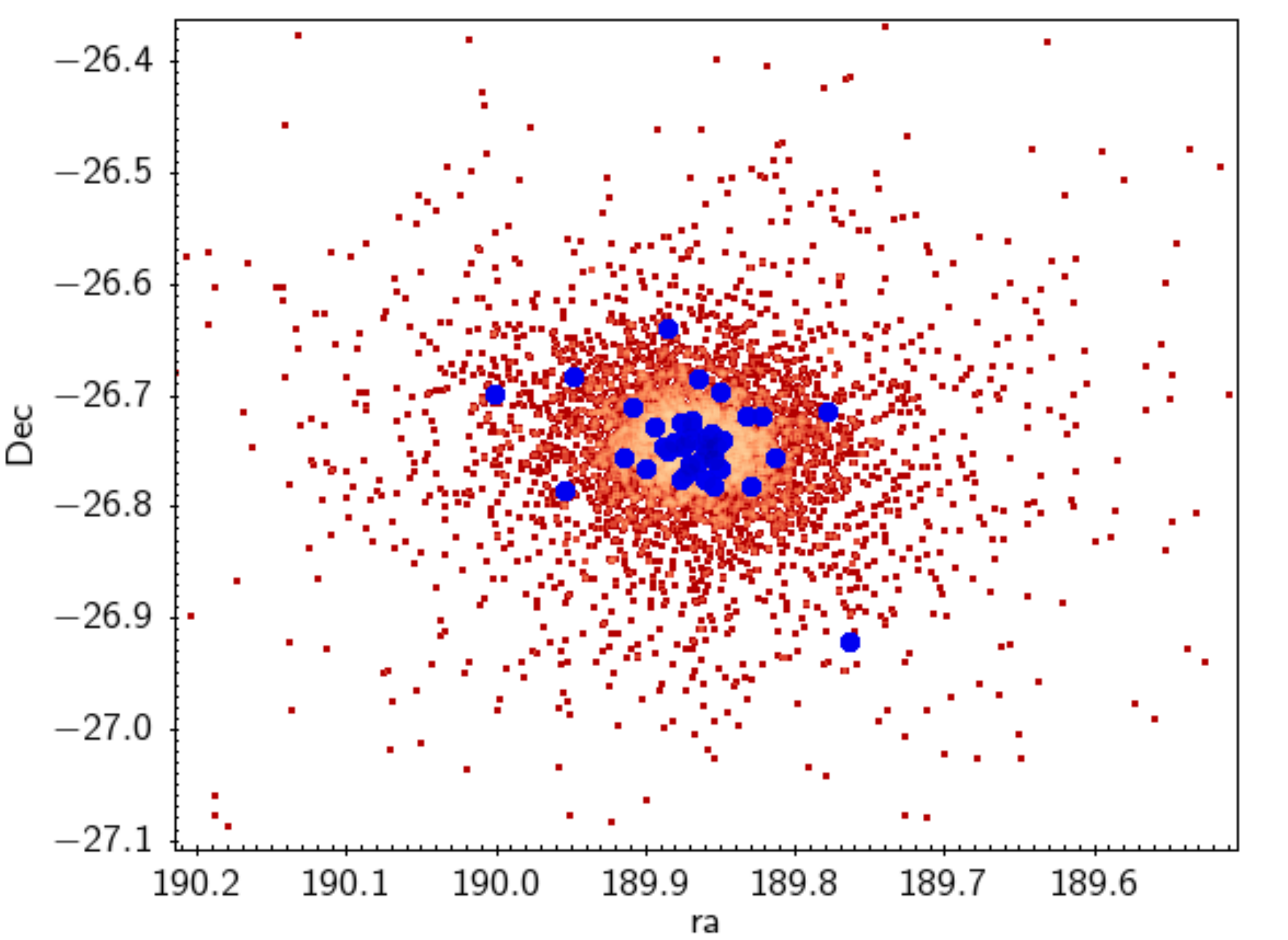}~\includegraphics[trim= 10 0 0 0 clip, width=7.9cm]{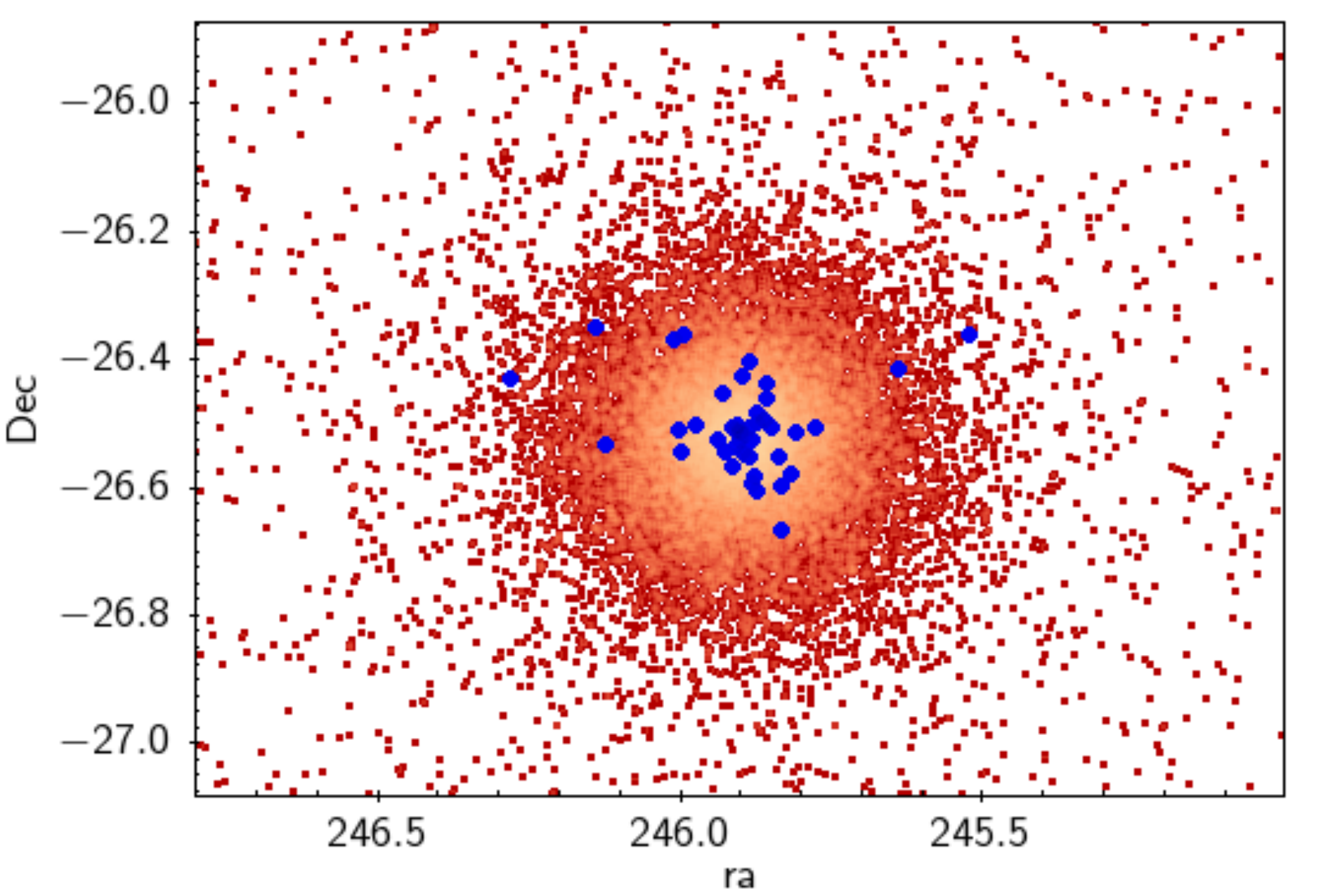}
		\includegraphics[trim= 30 0 0 0 clip, width=7.cm]{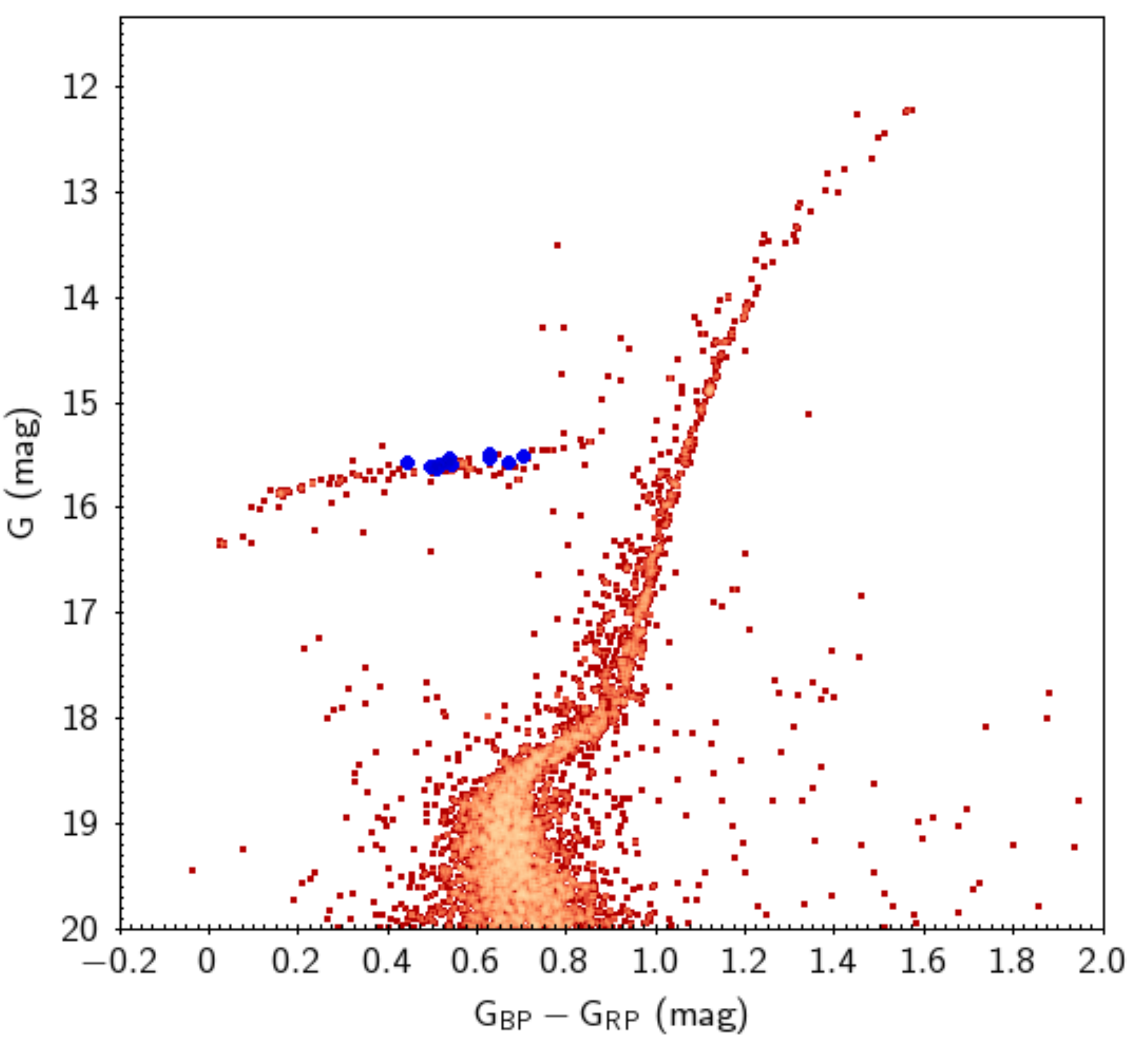}~\includegraphics[trim= 0 0 0 0 clip, width=6.9cm]{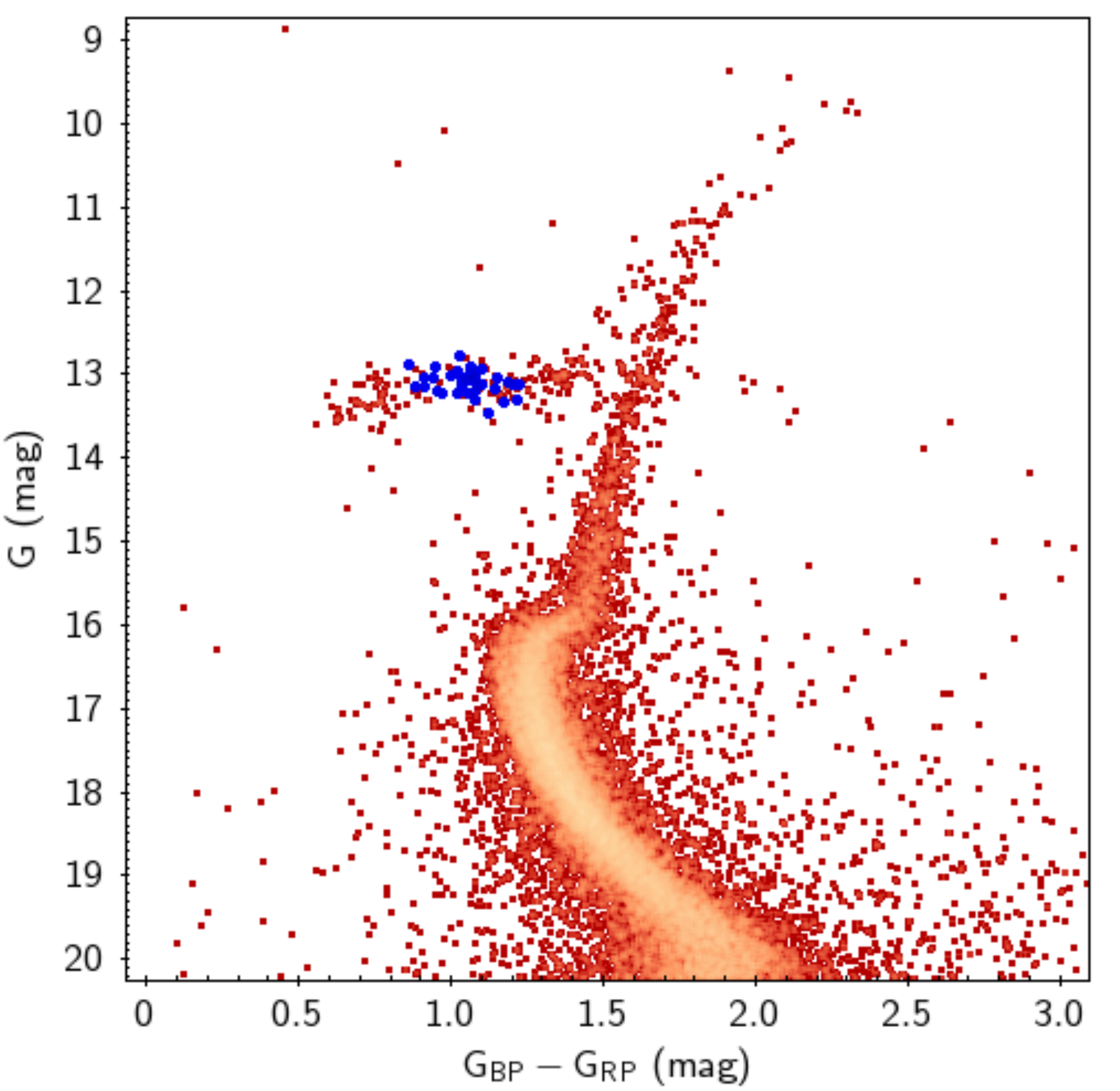}
		\caption{Proper motions (top), spatial distribution (middle) and $G$, $G_{\rm BP}-G_{\rm RP}$ colour-magnitude diagram (CMD; bottom) of two GCs in our sample:
		NGC~4590 (${\rm[Fe/H]}=-$2.27; left) and  NGC~6121 (${\rm[Fe/H]}=-$1.18; right).  The CMDs are drawn using for constant stars magnitudes from the EDR3 catalogue.  Black ellipses in the top panels encircle stars plotted in the CMDs in the bottom panels of the figure.   Blue dots show the cluster RRLs (42 RRLs in the top and middle  panels and 13 RRLs with 
		mean  $G$, $G_{\rm BP}$ and $G_{\rm RP}$ magnitudes from the {\it Gaia} DR2 \texttt{vari\_rrlyrae} table,  in the CMD  of NGC4590; 47 RRLs in the top and middle  panels and 37 RRLs in the CMD of NGC~6121).
			}
		\label{fig:PM-CMD}
	\end{center}
 \end{figure*}

Figure~\ref{fig:PM-CMD} shows 
the EDR3 proper motions (top  panels), the spatial distributions (middle panels) and the $G, G_{\rm BP}-G_{\rm RP}$ colour-magnitude diagrams (bottom panels) for two GCs in our sample, namely: NGC~4590 (left panels) and NGC~6121 (right panels). We have marked with blue dots the GC RRLs observed by {\it Gaia} that we used in our analysis.

We 
adopted the literature period for all  cluster RRLs except for V27 in NGC~5824. 
\citet{Walker-et-al-2017} reports this RRL to have a noisy light curve perhaps due to an incorrect period. We 
thus adopted for the star the period provided by the {\it Gaia} DR2 \texttt{vary-rrlyrae} table.
Periods of all RRc stars (31 sources among the field RRLs and 136 sources among the cluster RRLs) 
were ``fundamentalised'' by adding 0.127 to the decadic 
logarithm of the period and uncertainties on $\log\left(P\right)$ were estimated as 
$\sigma_{\log\left(P\right)}=0.01$, which
is equivalent to an uncertainty of 2.3\% in the period.  This choice, rather conservative, to apply a large error to the periods is due both to the frequent absence in the literature of period uncertainties and to the heterogeneity of the values measured in different works. 

V75 in NGC~6121 ({\it Gaia} \texttt{sourceid} 6045480963805292416)  
was removed from our sample of 385 cluster RRLs because the mean $V$ magnitude reported for the star by \citet{Yao-et-al-1988} is almost 1 mag  fainter than for the other RRLs in the cluster  while in a more recent work of \citet{Stetson-et-al-2014} its luminosity is considerably brighter than the others likely due to limitations of its time sampling.
After removing also 18 sources 
for which a $V$ 
magnitude is not available in the literature, or  the V-mean magnitude coming from literature is not so accurate [as V34 in NGC~6121, for this star \citet{Stetson-et-al-2014} had insufficient data to fit the light curve], the sample of cluster RRLs reduced  
to 366 sources that we used to infer the $M_{V}-{\rm[Fe/H]}$ relation presented in Section~\ref{sec:mv}. 
As for the $LZ$ relation in the $G$-band,  
 after dropping sources
without a $\langle G \rangle$ magnitude in the DR2 \texttt{vary-rrlyrae} table, the  final sample used to define the $M_{G}-{\rm[Fe/H]}$ relation consisted of 227 cluster RRLs 
(Section~\ref{sec:mg}). 
$\langle G_{\rm{BP}} \rangle$ and $\langle G_{\rm{RP}} \rangle$ mean  magnitudes are available for a subsample of 168 cluster RRLs. They were 
used 
to compute the $PW_{\left(G,G_{\rm{BP}},G_{\rm{RP}}\right)}$ relations described in Section~\ref{sec:pwz}. The $G$-band absorption was estimated from the absorption in the $V$-band using the relation: ${A}_{G}=0.840{A}_{V}$ \citep{Bono-at-al-2019}.



\section{Method}
\label{sec:method}
In this section we summarise the fundamentals of the hierarchical Bayesian method and describe the model we developed for inferring $PL(Z)$ relations from our sample of GC RRLs. 
The hierarchical model for inferring $PL(Z)$ relations for field RRLs used in this paper is essentially the same 
as described in section 4 of \cite{Muraveva-et-al-2018a}. The only difference is the treatment of the parallax offset. While in the model of \cite{Muraveva-et-al-2018a} we included a unique offset, namely a global parallax zero-point offset, we now allow the offset to vary for each source in the sample.
 The RRLs in our samples are distributed all-sky covering a quite wide range in G-magnitude with different extinction values and crowding conditions. Since all these factors affect the zero-point offset we have decided to adopt a varying offset to minimize these effects.
But as reported in Section \ref{sec:plx_offset} below, the model with a unique offset is still used to get an initial guess of the global offset value, which is then set as prior information in the models with varying offset.

The hierarchical Bayesian method starts by partitioning the parameter space associated to the problem into several hierarchical levels of statistical variability. This partition applies, in the most general case, both to the parameters themselves (random variables) and to the measurements (data). In a further stage  the modelling process assigns a set of probabilistic conditional dependency relationships between parameters and data, at the same or different level of the hierarchy. The final result is a model for the problem expressed as the joint probability density function (PDF) of parameters and measurements 
factorized as the product of a set of conditional probability distributions, namely a Bayesian network \citep{Pearl-1988,Lauritzen-1996}.  Assuming that the data are partitioned into $J$ clusters with $N_j$ sets of measurements in each (one for each individual star) from a total of $N=\sum_{j=1}^{J}N_{j}$ stars, and that the random parameters are partitioned into true astrophysical parameters ($\Theta$)
and hyperparameters ($\Phi$), we express the network as
\begin{equation}
	\begin{split}
	p\left(\mathcal{D},\mathbf{\Theta},\Phi\right)=&\prod_{j=1}^{J}p\left(
	\mathbf{d}_{j}\mid\mathbf{\Theta}_{j}\right)\cdot
	\prod_{i=1}^{N_{j}}p\left(
	\mathbf{d}_{ij}\mid\mathbf{\Theta}_{ij}\right)\cdot\\&
	 p\left(\mathbf{\Theta}\mid\Phi\right)\cdot p\left(\Phi\right)
	\,,\\
   \end{split}
	\label{eq:joint-pdf-1}
\end{equation}

\noindent where 
 $\mathbf{d}_{ij}$ are the observational data of star $i$ in cluster $j$ (apparent magnitudes, parallaxes and periods as explained below), ${\mathbf d}_j$ represent the observations pertaining to cluster $j$ (its metallicity assumed to be the same for all stars in the   cluster) and $\mathcal{D}=\left\{ \mathbf{d}_{j},\mathcal{D}_{j}\right\} _{j=1}^{J}$, with
 $\mathcal{D}_{j}=\left\{ \mathbf{d}_{ij}\right\} _{i=1}^{N_{j}}$. 
The astrophysical parameters denoted by $\mathbf{\Theta}_{ij}$ in Eq. \eqref{eq:joint-pdf-1} include the true distance, true period and absolute magnitude associated to each RRL in every GC, while the parameters denoted by $\mathbf{\Theta}_{j}$ include the true metallicity, the coefficients of the $PL$ relationship and the parameters characterizing the GC spatial distribution associated to each GC. The hyperparameters $\Phi$ include the coefficients of the global $PLZ$ relation.  
 The first two products on the right hand side of Eq. \eqref{eq:joint-pdf-1} constitute the conditional distribution of the data given the parameters (the so called likelihood), $p\left(\mathbf{\Theta}\mid\Phi\right)$ is the prior distribution of the true parameters given the hyperparameters and $p\left(\Phi\right)$ is the unconditional hyperprior distribution of the hyperparameters. A model like the one described in  Eq. 
 \eqref{eq:joint-pdf-1} dictates a certain dependency structure between 
 variables which is formally expressed by a directed acyclic
 graph (DAG) whose nodes encode model parameters, measurements,
 or constants, and whose directed links represent conditional
 dependence relationships.  While Eq. \eqref{eq:joint-pdf-1}, and the associated 
 DAG in particular,
 represent  the problem formulation, the solution is based on 
 the application 
 of Bayes' rule:
 \begin{equation}
 	p\left(\mathbf{\Theta},\Phi\mid\mathcal{D}\right)\propto
 	p\left(\mathcal{D}\mid\mathbf{\Theta}\right)\cdot
 	p\left(\mathbf{\Theta}\mid\Phi\right)\cdot p\left(\Phi\right)\,,
 	\label{eq:joint-posterior-pdf}
 \end{equation}
 
 \noindent being the goal to infer the marginal posterior distribution
 $p\left(\Lambda\mid\mathcal{D}\right)$ of some subset 
 [$\Lambda\subseteq\left(\mathbf{\Theta},\Phi\right)$] of 
 the parameters of interest. For multi-level models with a 
 relatively complex dependency structure between their parameters, the posterior 
 distribution usually is not available
 in an analytically tractable closed form  but can be 
 evaluated using Markov chain Monte Carlo (MCMC) simulation techniques.

 
\begin{figure}
	\begin{center}
		\includegraphics[scale=0.7]{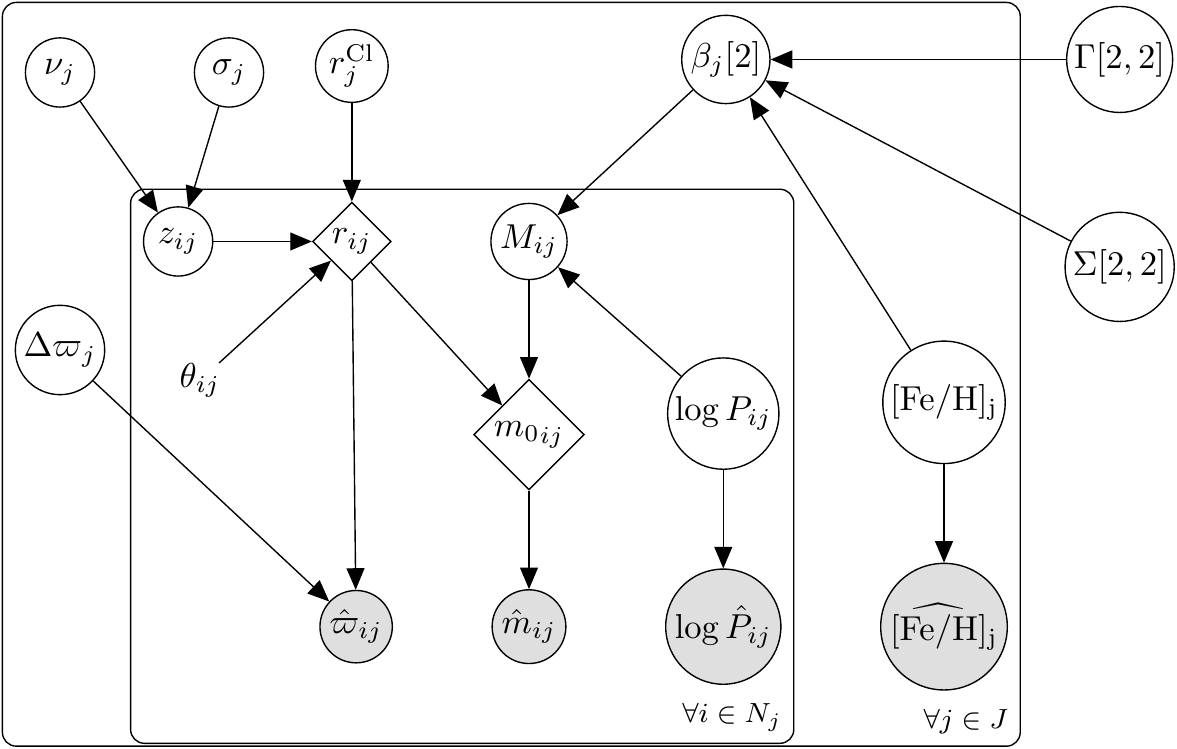}
		\caption{Directed Acyclic Graph (DAG)  that     represents the generic  hierarchical Bayesian model (HBM)  used to infer the coefficients of the $PL$, $LZ$ and $PLZ$ relations. A full description of the model parameters is included in the text. }
		\label{DAG}
	\end{center}
\end{figure}

In what follows we describe the construction of the hierarchical 
probabilistic  graphical model (abbreviated hereafter as HM) applied in this 
paper  to the inference of $PL(Z)$ relations for RRLs in GCs. 
Figure \ref{DAG} shows its associated  DAG representing the 
conditional dependence relationships between the parameters and data. 
Parameters 
and data are represented by  DAG 
nodes enclosed
into  two nested rectangles (plate notation)  to make evident the grouped 
nature of the problem. The nodes located into the outer plate but not into the 
inner plate represent properties at a 
GC level like the 
cluster mean iron-to-hydrogen content and the coefficients of an underlying 
$PL$ relationship for the population of RRLs in the cluster. 
The outer plate replicates in the DAG as many times ($J$) as the total number 
of clusters (groups), with the subindex $j$ identifying
a particular cluster. The nodes placed into the inner plate represent 
properties of individual stars (e.g. the  time averaged absolute magnitude $M$). For the 
$j$th cluster, the inner plate replicates as many times  ($N_j$) as the size 
of the  cluster RRL sample,  with the subindex $i$ identifying a particular   
star. 

Figure \ref{DAG}  shows the measurements as shaded nodes at the 
bottom of the DAG. At the 
individual star level (inner plate) these measurements 
are the decadic logarithm of period $\log\hat{P}_{ij}$, the 
apparent magnitude $\hat{m}_{ij}$ and the parallax $\hat{\varpi}_{ij}$. At the 
cluster level (outer plate) the only measurement is the GC mean 
iron-to-hydrogen 
content $\widehat{\left[\mathrm{Fe/H}\right]}_{j}$. Quantities that are 
considered constant in the model are not represented
in the DAG. These 
quantities include all the measurement uncertainties, the measured interstellar 
absorption $\hat{A}_{m_{ij}}$  and all the 
model hyperprior parameters (for example the parameter values used to define the hyperprior for the distance to the centre of each GC).  The measurements 
are assumed to be realizations from Gaussian distributions
centred at their true (unknown) values (this is only an approximation in the case of the apparent magnitudes; only the uncertainties in the measured fluxes are Gaussian), 
which are represented without the 
circumflex accent (\verb!^!) in the DAG. The adopted standard deviations are 
the measurement uncertainties provided by our catalogue. In the particular case 
of the parallax, we assume that its true value is equal to the reciprocal of 
the true distance and contemplate the possibility that its measured value may be 
affected by a parallax offset which varies with the globular cluster. This is 
formulated as

\begin{equation}
\hat{\varpi}_{ij}\sim\mathcal{N}\left(\frac{1}{r_{ij}}+
{\Delta\varpi}_{j},\sigma_{\varpi_{ij}}\right)\,,
\label{eq:obs-pi}
\end{equation}
 
\noindent where $\sim$ should be read as 'is distributed as',
$\mathcal{N}$ represents the normal (Gaussian) distribution, $r_{ij}$ is the 
true distance to the star and ${\Delta\varpi}_{j}$ is the parallax offset 
associated  to the $j$th GC.  The parallax offsets are given as a  Gaussian 
prior centred at the value (mas) inferred by the field stars HM described in section 4 of \cite{Muraveva-et-al-2018a} applied to the inference of a $PW_{(G,G_{\rm 
BP},G_{\rm RP})}Z$ relation (the details of that inference are given  
in Section \ref{sec:plx_offset} further below). 
The standard deviation that accounts for the offset variability between clusters is fixed at a value of $0.03$ mas.  This value,
which is about an order of magnitude larger than the standard deviation obtained from the preliminary calibration of relations for field RRLs,  
takes into account  the large uncertainties of the parallax measurements of RRLs in the GC sample.
The Heliocentric distances $r_{ij}$ are 
modeled as a function of the distance to the centre of each cluster assuming 
that the cluster geometry is spherical as  explained at the end of the 
section.  The measured apparent magnitude $\hat{m}_{ij}$ of the $i$th star 
belonging to the $j$th  globular cluster is modelled by 
a Gaussian distribution  centred at its true value ${m_0}_{ij}$, while the true 
value is generated deterministically as   
\begin{equation}
{m_0}_{ij}  =  M_{ij} + \hat{A}_{m_{ij}} + 5\log(r_{ij})+10\,,
\label{eq:M-to-m0}
\end{equation}

\noindent where $M_{ij}$, $\hat{A}_{m_{ij}}$, and $r_{ij}$ denote  
the true absolute magnitude, the measured interstellar absorption  and the true 
distance (kpc),   respectively. For each globular cluster the model 
contemplates a different $PL$ relationship which is parameterized as

\begin{equation}
M_{ij}\sim\mathcal{N}\left(\left(\begin{array}{cc}
\beta_{0j} & \beta_{1j}\end{array}\right)\cdot\left(\begin{array}{c}
1\\
\log P_{ij}
\end{array}\right),\sigma_{j}\right)\,.
\label{eq:M-distr}
\end{equation}

\noindent The coefficients of the $PL$ relations are  assigned  a 
bivariate Normal  prior in turn, whose mean vector depends on the 
true metallicity of the globular cluster  in the most general case:  

\begin{equation}
\left(\begin{array}{c}
\beta_{0j}\\
\beta_{1j}
\end{array}\right)\sim\mathcal{N}\left(\left(\begin{array}{cc}
\gamma_{00} & \gamma_{01}\\
\gamma_{10} & \gamma_{11}
\end{array}\right)\cdot\left(\begin{array}{c}
1\\
\left[\mathrm{Fe/H}\right]_{j}
\end{array}\right),\Sigma\right)\,,
\label{eq:Beta-distr-PLZ}
\end{equation}

\noindent where 
$\Sigma=\mathrm{diag}\left({\tau}_{\beta_0}^{2},{\tau}_{\beta_1}^{2}\right)$ 
 and the standard deviations $\tau_{\beta_\cdot}$ are drawn from an 
exponential prior  with inverse-scale parameter  equal to 1. For  the 
slopes  $\gamma_{\cdot1}$ and 
	zero-points  $\gamma_{\cdot0}$ population-level parameters we specify standard Cauchy
	prior distributions centred at 0 with scale parameter equal to 1 and 10, 
	respectively.

We emphasize that the metallicities enter the model in an upper hierarchical 
level than periods, something made explicit in the DAG depicted in Fig. 
\ref{DAG} where the $N_j$ RRL stars of the $j$th cluster are all inside the 
inner plate but $\left[\mathrm{Fe/H}\right]$ is placed into the 
outer plate.  The model assumes that all the RRLs of a given globular 
cluster have the same metallicity and that the slope 
($\beta_{1j}$) and intercept  ($\beta_{0j}$) of the PL relation  
depend on the cluster metallicity $\left[\mathrm{Fe/H}\right]_{j}$.  The 
method 
contemplates the possibility of modelling the influence of the cluster 
metallicity only on the intercept of the $PL$ relations  which   would be 
implemented by setting $\gamma_{11}=0$. 
For a model that does not take into account the 
	influence of the pulsation period, namely a $LZ$ relationship,  
	Eqs. \ref{eq:M-distr} and \ref{eq:Beta-distr-PLZ}  particularise  to
\begin{equation}
M_{ij}\sim\mathcal{N}\left(\beta_{0j},\sigma_{j}\right)
\label{eq:M-distr-LZ}
\end{equation}
and
\begin{equation}
\beta_{0j}\sim\mathcal{N}\left(\gamma_{00}+\gamma_{01}
\left[\mathrm{Fe/H}\right]_{j},\tau\right)\,,
\label{eq:beta-distr-LZ}
\end{equation}

\noindent respectively.

 The Bayesian method proposed in this paper constrains the $PL(Z)$ relations by 
 modelling explicitly the spatial distribution of the RR~Lyrae star population of 
 each globular cluster in our sample. The GC spatial distribution  is assumed 
 to be spherical. We define two reference frames: one centred at the solar system barycentre. In this reference system the distance to the $j$th cluster centre is defined as $r_j$ and the distance to the $i$th star in this $j$th cluster as $r_{ij}$. The second reference system has its origin at the cluster centre 
 and the $z$-axis pointing towards the observer. We assume that the 3D probability 
 distribution of the coordinates of RR~Lyrae stars in this reference system is given by the product of three independent 
 non-standardized Student's $t$-distributions, one for each Cartesian 
 component,  with identical degrees  of freedom $\nu_{j}$ and scale parameter 
 $\sigma_{j}$ (see below for a justification of this choice). In particular, the  non-standardized Student's t probability density function (PDF)  corresponding  to the coordinate $z_{ij}$  of a 
 source (which in the adopted reference frame represents the projection of the 
 radial distance from the source to the cluster centre onto the line of 
 sight) is given by
   \begin{equation}
   	f\left(z_{ij}\right)=\frac{1}{\sqrt{\nu_{j}\sigma_{j}^{2}}\mathrm{B}
   		\left(\nu_{j},\frac{1}{2}\right)}\left(1+\frac{z_{ij}^{2}}{\nu_{j}s_{j}^{2}}
   	\right)^{-\frac{\nu_{j}+1}{2}}\,,
   	\label{eq:Pearson-type-VII-distr}
   \end{equation}
\noindent where B denotes the Beta function. 
We note that through the reparameterisation  
$s_{j}=\sqrt{\nu_{j}\sigma_{j}}$ and $m_{j}=\frac{\nu_{j}+1}{2}$  the 
 PDF of Eq. \eqref{eq:Pearson-type-VII-distr} is equivalent to the so called 
 generalized Schuster density \citep{Ninkovic-1998} with the parameters
 $s_j$ and $m_j$ representing the cluster core length scale and the 
 steepness of the $j$th GC stellar distribution, respectively, which for 
 $m=\frac{3}{2}$   approximates  the  King stellar surface density profile 
 \citep{Ninkovic-1998}. 
 While the parameters $\sigma_{j}$ and  $\nu_{j}$ are directly inferred by  
 our HM, $s_j$ and $m_j$ are calculated deterministically from the former. 
 Our HM assigns for  $\sigma_{j}$ and  $\nu_{j}$   an 
 exponential prior  with inverse scale $\lambda=50$ $\mathrm{kpc}^{-1}$ and an  
 exponential prior shifted $2$ units  towards positive values with inverse scale 
 $\lambda=1$, respectively.  
 
 \noindent The value of the barycentric  distance $r_{ij}$ to the  
 $i$th RR~Lyrae star of the $j$th cluster is calculated deterministically from its $z_{ij}$ coordinate, the  barycentric distance to the cluster centre 
 $r_{j}^{\mathrm{Cl}}$ and the angular separation $\theta_{ij}$ between the 
 star and the cluster centre by
 \begin{equation}
 r_{ij}=\frac{r_{j}^{\mathrm{Cl}}+z_{ij}}{\cos\left(\theta_{ij}\right)}\,, 
 \label{eq:dist-to-star}
 \end{equation}
 
 \noindent where the angular distances $\theta_{ij}$ are kept constant in the 
 model and calculated from 
 \begin{equation}
 \begin{split}
 \cos\left(\theta_{ij}\right)=&\cos\left(\frac{\pi}{2}-\delta_{ij}\right)
 \cos\left(\frac{\pi}{2}-\delta_{j}^{\mathrm{Cl}}\right)+\\
 &\sin\left(\frac{\pi}{2}-\delta_{ij}\right)\sin\left(\frac{\pi}{2}-
 \delta_{j}^{\mathrm{Cl}}\right)\cos\left(\alpha-\alpha_{j}^{\mathrm{Cl}}\right).\\
 \end{split}
 \label{eq:angular-dist}
 \end{equation}
 
In Eq. \eqref{eq:angular-dist}, $\alpha_{j}^{\mathrm{Cl}}$, $\delta_{j}^{\mathrm{Cl}}$,    	$\alpha_{ij}$ and $\delta_{ij}$  
 	are the right ascension and declination of the cluster centre and the 
 	individual RR~Lyrae stars, respectively.  The distance $r_{j}^{\mathrm{Cl}}$ to 
  each cluster centre 
 is assigned a Gamma prior distribution with shape $k=2$ and scale parameter 
 $L$ equal to the distance to the cluster provided by \citet[2010 edition]{Harris-1996}.
 
 We have encoded our HM in the Stan probabilistic modelling language
 \citep{Carpenter-et-al-17} and sampled the model posterior distribution by 
 means of the adaptive Hamiltonian Monte
 Carlo (HMC) No-U-Turn Sampler (NUTS) of \cite{Hoffman-Gelman-14} through the 
 use of the R \citep{R-language} interface  Rstan  \citep{RStan-package}.

\section{Results}
\label{sec:results}

\subsection{{\egdr{3}} parallax offset}
\label{sec:plx_offset}

To obtain an independent estimate of the \egdr{3} parallax offset 
distribution \citep[see e.g.][among others]{Lindegren-et-al-2020,Bhardwaj-et-al-2020,Zinn-2021,Stassun-Torres-2021}  we 
used a two-stage procedure.  
In the first stage we applied the HM 
described in section 4 of \cite{Muraveva-et-al-2018a} for calibrating a 
preliminary $PW_{(G,G_{\rm BP},G_{\rm RP})}Z$ relation (using our field RRL 
sample) and inferring the associated global parallax offset 
parameter included in the model. In this first stage we assigned the offset a 
weakly informative Gaussian prior centred at $\mu_{\varpi_{0}}=0$ mas with 
standard deviation $\sigma_{\varpi_{0}}=0.1$ mas and inferred the posterior 
credible interval $\Delta_{\varpi_0}=-0.033_{-0.006}^{+0.005}$ mas, 
 defined here, and hereinafter, as the median plus/minus the difference in absolute value between the median and the 84th and 16th percentile of the posterior distribution of the parameter. 
Then,  in  the  second  stage, we adopted  a Gaussian prior
distribution centred at  ${\mu_{\varpi_{0}}}_j=-0.033$ mas with ${\sigma_{\varpi_{0}}}_j=0.03$ mas for the individual
offsets of the GC and field RRL models described in Section \ref{sec:method}, 
based on the offset inferred in our first stage. The results of this second stage
are reported in the following sections.
We use the HM with individual parallax offset to allow the offset to 
vary with the sky position (among other factors). For that we could have 
proceeded directly in a single step by using the weakly informative Gaussian 
prior centred at 0 mas.  However, this results in a largely unconstrained 
model space, poor convergence and unphysical posterior distributions due to the
significant fractional uncertainties of the parallax measurements in our
sample.


\subsection{\mv~relation}
\label{sec:mv}
 
The statistical results of the {\mv}  relation parameters inferred by our HM at GC level for our selection of 366 cluster RRLs in the $V$ band are  presented in Table \ref{tab:V-FeH-GC-CL}.
Each parameter is summarised by 
the median plus minus the difference in absolute value between the median and the 84th and 16th percentile of its posterior distribution, as defined in Section \ref{sec:plx_offset}.
The table includes:
the parameter $\beta_{0j}$  of Eq. \ref{eq:M-distr-LZ} representing 
the mean absolute magnitude (zero-point) of each GC RRL star population  
 ($M_j$), the intrinsic dispersions $\sigma_j$ (Eq. \ref{eq:M-distr-LZ}) of the individual RRL absolute magnitudes around the GC mean absolute magnitude (${\sigma_M}_j$), the true distance modulus of each GC ($\mu_j$) calculated deterministically from the inferred barycentric distance to the GC centre ($r_{j}^{\rm Cl}$), the median
projections of the radial distances from each source to the GC centre onto the line of sight (${z_{\cdot j}}$) and the 
cluster-level  parallax offsets inferred by our HM (${\Delta\varpi}_j$).

\begin{figure}
 	\begin{center}
 	 	\includegraphics[trim=20 50 40 10,  
 	 	width=\linewidth]{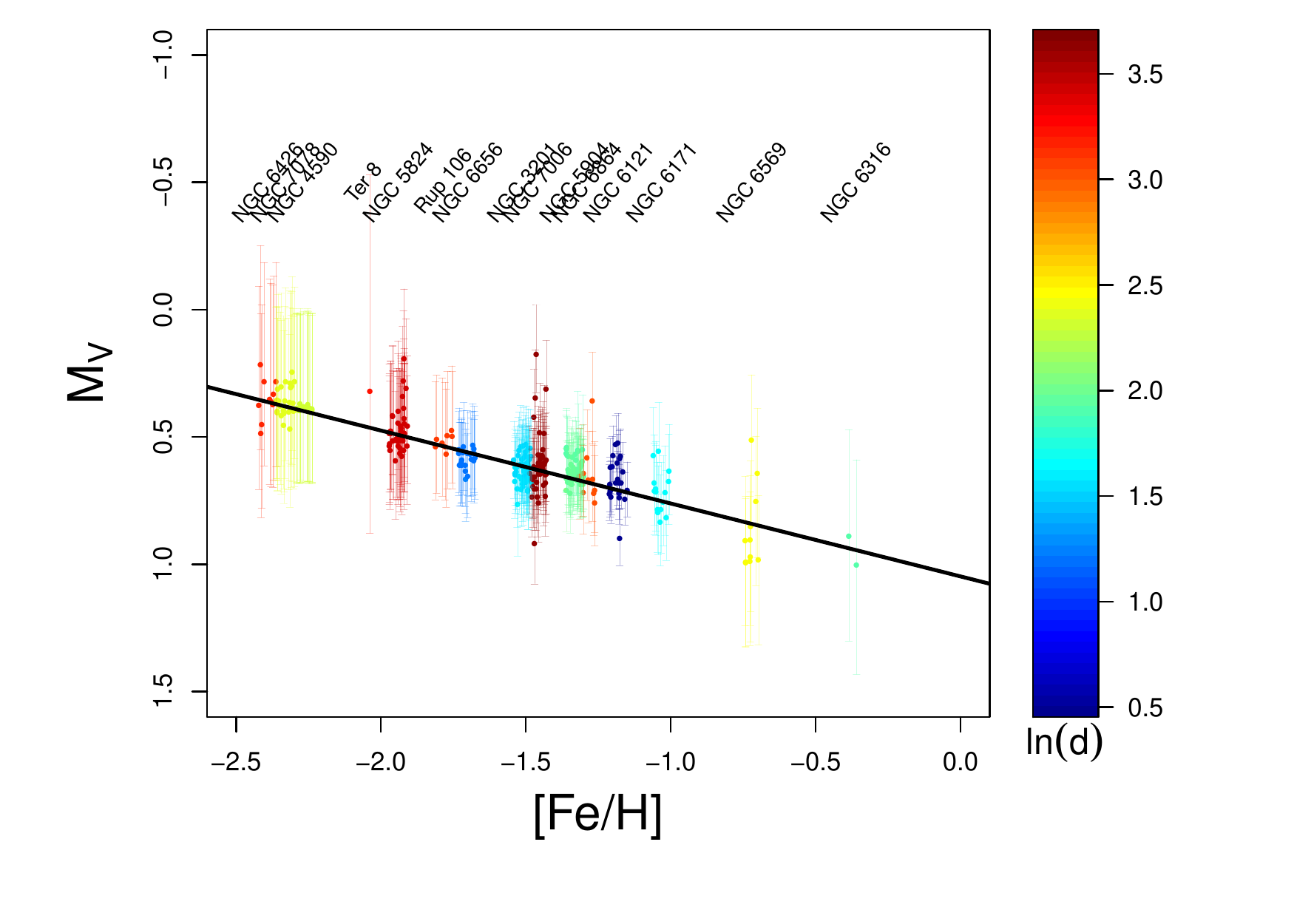}
 		\caption{ 
 		$M_V - {\rm [Fe/H]}$ relation defined  by globular cluster RRLs as summarized in Table \ref{tab:V-FeH-GC-CL}. The solid line 
 			represents the linear fit corresponding to 
 			the parameters of Table
 			\ref{tab:V-band-all}. The colour scale 
 			encodes the (natural) logarithm of the
 			inferred (true) barycentric distance to individual RRL stars 
 			measured in units of kpc.  	}
 		\label{fig:LZ-V-band-M-vs-FeH-GC}
 	\end{center}
 \end{figure}

 The upper portion of Table~\ref{tab:V-band-all}
summarises the results of the inference at the top (population-level) of our HM. It reports the slope, zero-point and intrinsic dispersion obtained for the {\mv} relation, and the inferred population-level parallax zero-point offset ${{\Delta\varpi}_0=-0.024_{-0.008}^{+0.008}}$ mas calculated as the  median of the distribution of the medians of each sample of the joint posterior distribution of ${\Delta\varpi}_j$. 
Figure \ref{fig:LZ-V-band-M-vs-FeH-GC} shows the corresponding fit. The credible intervals obtained for the
metallicity slope and the intercept are   ${\gamma_{1}^{\rm UVES} = 
0.29_{-0.29}^{+0.35}}$ mag/dex and ${\gamma_{0}^{\rm UVES} = 
1.05_{-0.46}^{+0.43}}$ mag, respectively, with an associated intrinsic 
dispersion of ${\sigma_M = 0.11_{-0.08}^{+0.15}}$ mag.  Our inferred GC {\mv} 
relation for metal abundances on the \cite{Carretta-et-al-2009} metallicity 
scale is then given by

\begin{equation}
	M_{V}=\left(0.29_{-0.29}^{+0.35}\right)\left[\mathrm{Fe/H}\right]+
	 	 \left(1.05_{-0.46}^{+0.43}\right)\,. 
\label{eq:V-FeH-GC}
\end{equation}

\noindent We recall that these results correspond to an inferred global 
	parallax offset   ${{\Delta\varpi}_0=-0.024_{-0.008}^{+0.008}}$ mas
	 calculated as explained above. 

\begin{figure}
  	\begin{center}
  		\includegraphics[trim=20 50 40 10,  
  		width=\linewidth]{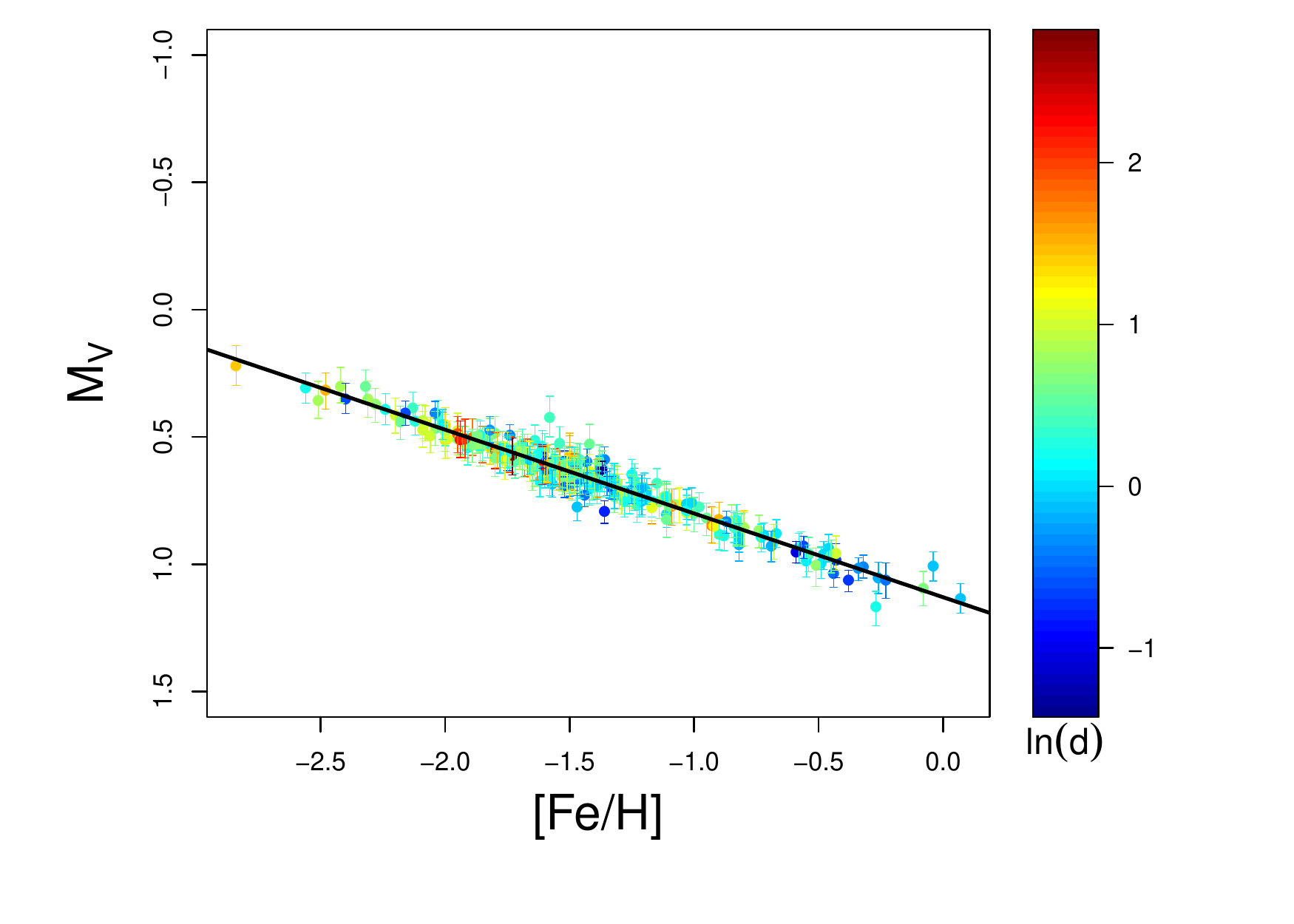}
  		\caption{ $M_V - {\rm [Fe/H]}$ relation defined by our field RRL sample. The 
  			solid line represents the linear fit corresponding to the parameters of Table \ref{tab:V-band-all}.
  			The colour scale 
  			encodes the (natural) logarithm of the
  			inferred (true) barycentric distance to individual RRL stars 
  			measured in units of kpc.  	}
  		\label{fig:LZ-V-band-M-vs-FeH-FS}
  	\end{center}
  \end{figure}

In Table \ref{tab:V-band-all} (lower portion) 
 and Fig. 
\ref{fig:LZ-V-band-M-vs-FeH-FS} we present the results of our updated Bayesian 
analysis using \egdr{3} parallaxes applied  to the sample of the 274 field RRLs that satisfy the criterion \texttt{ruwe} < 1.4 for which the $V$ magnitude was available.
The credible intervals obtained for the
metallicity slope and the intercept are   $\gamma_{1}^{\rm ZW} = 
0.33_{-0.02}^{+0.02}$ mag/dex and $\gamma_{0}^{\rm ZW} = 
1.13_{-0.03}^{+0.02}$ mag, respectively, with an associated intrinsic 
dispersion of $\sigma_M = 0.02_{-0.01}^{+0.01}$ mag.  Our inferred  {\mv} 
relation for field stars with metal abundances on the \cite{Zinn-West-84} 
metallicity scale is then given by
\begin{equation}
M_{V}=\left(0.33_{-0.02}^{+0.02}\right)\left[\mathrm{Fe/H}\right]+
\left(1.13_{-0.03}^{+0.02}\right)\,.
\label{eq:V-FeH-FS}
\end{equation} 

\noindent For the {\mv} relation for field stars we report the parallax zero-point offset 
${\Delta\varpi}_0=-0.033_{-0.002}^{+0.002}$ mas (Table \ref{tab:V-band-all}, last column  of the lower portion) 
calculated as the median of the distribution of the medians of each sample of the joint posterior distribution of the parallax offsets inferred by our HM for the field RRLs in the sample.

To compare the relation for cluster RRLs,  Eq. \eqref{eq:V-FeH-GC},  with the field  RRL {\mv} relation of Eq. \eqref{eq:V-FeH-FS}  we have to account  for the difference between the metallicity scales of the two samples. From the linear relation  	
\begin{equation}
\left[\mathrm{Fe/H}\right]_{\mathrm{UVES}}=1.105\left[\mathrm{Fe/H}\right]_{\mathrm{ZW}}+0.160 
\label{eq:FeH-ZW-to-UVES}
\end{equation}

\noindent of \cite{Carretta-et-al-2009} we obtain that the relationship between
the slopes and intercepts in the \cite{Zinn-West-84} and 
\cite{Carretta-et-al-2009} 
metallicity scales is given by
 \begin{equation}
 \gamma_{1}^{\mathrm{ZW}}=1.105\gamma_{1}^{\mathrm{UVES}}
 \label{eq:slope-UVES-2-ZW}
 \end{equation}
\noindent and
  \begin{equation} 
   \gamma_{0}^{\mathrm{ZW}}=0.160\gamma_{1}^{\mathrm{UVES}}+
   \gamma_{0}^{\mathrm{UVES}}\,,
  \label{eq:intercept-UVES-2-ZW}
  \end{equation}
 
\noindent  respectively, where we use the superscript UVES to refer to the mean GC  iron abundance ([Fe/H])  derived from high  resolution spectra of  red giant branch (RGB) stars  obtained with the Ultraviolet and Visual Echelle Spectrograph (UVES) using the procedure described in \citet[and references therein]{Carretta-et-al-2009}.  Applying the former conversion to the inferred slope of Table~\ref{tab:V-band-all} 
we obtain  
${\gamma_{1}^{\rm ZW(UVES)} = 0.32_{-0.32}^{+0.39}}$ mag/dex and ${\gamma_{0}^{\rm ZW(UVES)} = 
1.10_{-0.46}^{+0.43}}$ mag, where the superindex $\rm ZW(UVES)$ denotes a transformed coefficient. We next attempt to compare these coefficients with the coefficients of the {\mv} relation for field stars of Eq. \eqref{eq:V-FeH-FS}, denoted  jointly by vector notation here and hereafter  by $\boldsymbol{\gamma}^{\mathrm{ZW}}$. 
We first analysed the overlap between the two joint (bivariate)  posterior distributions. Our results were inconclusive due to: i) the much larger spread of the  posterior distribution associated to the transformed coefficients $\boldsymbol{\gamma}^{\mathrm{ZW(UVES)}}$, and ii) the support of the distribution of $\boldsymbol{\gamma}^{\mathrm{ZW(UVES)}}$ subsumes the effective support of the posterior distribution of $\boldsymbol{\gamma}^{\mathrm{ZW}}$. Therefore, we opted to directly  evaluate the difference  between the population medians of both pairs of coefficients, which will be denoted, using vector notation, here and hereafter by $\boldsymbol{\gamma}^{\mathrm{diff}}=\boldsymbol{\gamma}^{\mathrm{ZW}}-\boldsymbol{\gamma}^{\mathrm{ZW(UVES)}}$,   where the first and second component of each vector are the $\rm[Fe/H]$ slope and the intercept in Eq.~\eqref{eq:V-FeH-FS} and the slope and the intercept of Eq.~\eqref{eq:V-FeH-GC} transformed to the ZW metallcity scale, respectively.

Since we trained both models only once, we only have a single estimate for the median of each population. To assess the consistency  between $\boldsymbol{\gamma}^{\mathrm{ZW}}$  and $\boldsymbol{\gamma}^{\mathrm{ZW(UVES)}}$  throughout
the difference $\boldsymbol{\gamma}^{\mathrm{diff}}$ we have used the bootstrapping method and re-sampled 1000 times our 10000 MCMC posterior realisations associated to each pair of coefficients, calculating the sample median estimate $\hat{\boldsymbol{\gamma}}^{\rm{diff}}$ on each iteration. Then,  using the Mahalanobis distance \citep{Mahalanobis-1936}, hereafter abbreviated as MD, 
we have constructed the 99 per cent level confidence ellipse for the difference between the population medians $\boldsymbol{\gamma}^{\mathrm{diff}}$ given by 
\begin{equation}
\begin{split}
\left(\begin{array}{c}
\gamma_{1}^{\mathrm{diff}}\\
\gamma_{0}^{\mathrm{diff}}
\end{array}\right)=&\left(\begin{array}{cc}
0.497 & -0.867\\
0.867 & 0.497
\end{array}\right)\cdot\left(\begin{array}{c}
0.020\cdot\sin\left(\phi\right)\\
0.005\cdot\cos\left(\phi\right)
\end{array}\right)+\\&\left(\begin{array}{c}
0.008 \\ 
0.034
\end{array}\right)\,,  
 \end{split}
\label{eq:VZ-confidence-region}
\end{equation}

\noindent where $\phi$ is the eccentric angle associated to an arbitrary point on the ellipse.
The confidence ellipse of Eq. \eqref{eq:VZ-confidence-region} is centred at (0.008 mag/dex, 0.034 mag), the MD from its centre is equal to 3.20 units, its position angle with respect to the horizontal axis is equal to 60.18 deg and the lengths of its semi-major and minor axis are equal to  0.023 and  0.006 units, respectively. The coordinate origin (0,0) is away from the centre of the ellipse by a MD equal to 7.94 units clearly outside the confidence ellipse,  allowing us to conclude that the two population medians are significantly different.

\begin{figure}
 	\begin{center}
 	 	\includegraphics[trim=20 50 40 10,  
 	 	width=\linewidth]{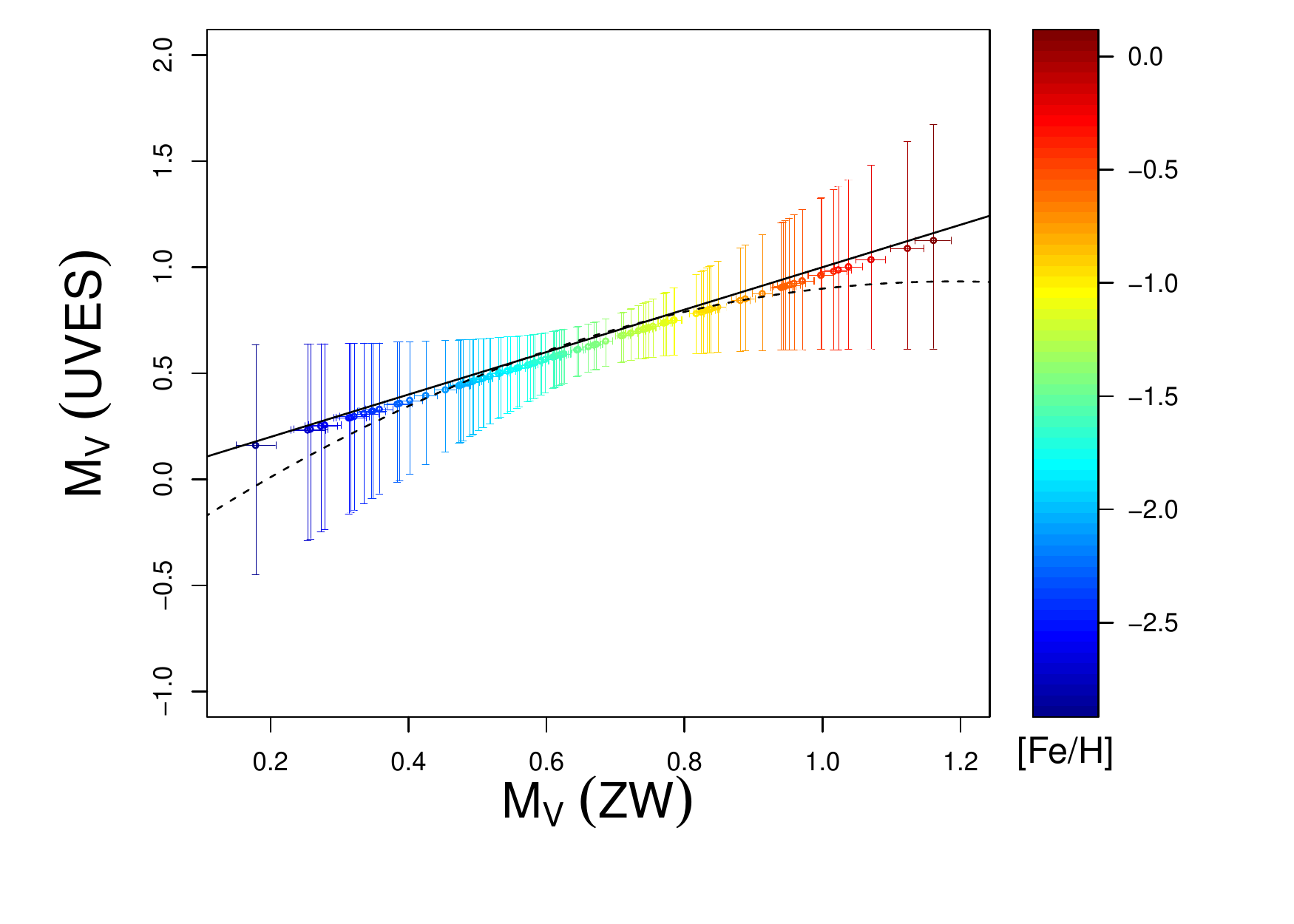}
 		\caption{
 		Comparison between $V$-band absolute magnitudes predicted by the linear relationships of Eqs. \eqref{eq:V-FeH-FS} and \eqref{eq:V-FeH-GC} when a linear conversion between metallicity scales is applied.  The solid and dashed lines represent the bisector and the relation between median magnitudes when a quadratic relation is used to transform between metallicity scales. The colour encodes the metallcity in the ZW scale.	}
 		\label{fig:V-mag-UVES-vs-mag-ZW}
 	\end{center}
 \end{figure}

We have also tackled the comparison problem by assessing the ability of the  {\mv} relation for field RRL stars of Eq. \eqref{eq:V-FeH-FS}
 (coefficients  $\boldsymbol{\gamma}^{\mathrm{ZW}}$) 
and the relation obtained using the transformed coefficients $\boldsymbol{\gamma}^{\mathrm{ZW(UVES)}}$ to predict mutually consistent absolute magnitudes. 
For that we have generated a sample of one hundred synthetic metallicities in the ZW scale (range -3 to 0.2 dex)  and predicted two sets of absolute magnitudes using directly the relation of Eq. \eqref{eq:V-FeH-FS}  and then applying the relation of Eq. \eqref{eq:V-FeH-GC} after the conversion between metallicity scales by the linear relation of Eq. \eqref{eq:FeH-ZW-to-UVES}.  
Figure \ref{fig:V-mag-UVES-vs-mag-ZW}  shows the result of our comparison. The medians of the two sets of predicted absolute magnitudes are in very good agreement presenting only a slight deviation with respect to the bisector line at their most extreme values.  We note that the larger uncertainties at the tails are consequence of the correlation between the slopes and intercepts for both models. Nevertheless we face a different scenario when using the quadratic relation of \citet[][cf. Eq. 3]{Carretta-et-al-2009} to transform between metallicity scales. 
In this latter case the median absolute magnitudes predicted by our linear GC $LZ$ relation in $V$ band ($VZ$) are significantly brighter than the values predicted by the field RRL relation for the lower and higher metallicity ranges, as  shown by 
the dashed line in the figure.  To achieve consistency between both magnitude estimates in these metallicity ranges, the metallicity slope for the $VZ$ relation calibrated in the \cite{Carretta-et-al-2009} should be flatter (steeper) for the lower (higher) metallicity ranges, which would be consistent with a non-linear $VZ$ relation as suggested by some theoretical models (see e.g. \citealt{Bono-et-al-2003}; \citealt{Catelan-et-al-2004}). Given the high accuracy of our linear {\mv} relation of Eq. \eqref{eq:V-FeH-FS} inferred from field RRLs with metallicities in the \cite{Zinn-West-84} scale, we discard the existence of higher order metallicity terms  for this latter relation.

\begin{table*}
\centering
\caption{
Summary statistics associated to the $M_V-[\mathrm{Fe/H}]$ 
             relationship by  globular cluster RRLs.
             $M$, $\sigma_{M}$, $\mu_{0}$, ${r}^{\rm{Cl}}$, ${z}^{\rm{RRL}}$ and $\Delta\varpi$ 
             represent the absolute magnitude zero-point,
             its dispersion, the true distance modulus, the distance to the centre,
             the RRL median radial projections and the parallax offset
             inferred by the model for each cluster of our sample, respectively. 
             } 
\label{tab:V-FeH-GC-CL}
\begin{tabular}{lrlllrrr}
  \hline

 \multicolumn{1}{c}{GC} & \multicolumn{1}{c}{RRLs} & 
     \multicolumn{1}{c}{$M$} & \multicolumn{1}{c}{$\sigma_{M}$}
      & \multicolumn{1}{c}{$\mu_{0}$} & \multicolumn{1}{c}{${r}^{\rm{Cl}}$} & \multicolumn{1}{c}{$z^{\rm{RRL}}$} & \multicolumn{1}{c}{$\Delta\varpi$} \\
   & & \multicolumn{1}{c}{(mag)}   & \multicolumn{1}{c}{(mag)}  & \multicolumn{1}{c}{(mag)}    & \multicolumn{1}{c} {(kpc)} & \multicolumn{1}{c} {(pc)} & \multicolumn{1}{c} {(mas)} \\
  \hline
  NGC 3201 &   59 & $0.63_{-0.15}^{+0.13}$ & $0.067_{-0.018}^{+0.017}$ & $13.45_{-0.13}^{+0.15}$ & $4.90_{-0.29}^{+0.36}$ & $0.09_{-10.210}^{+9.832}$ & $-0.014_{-0.013}^{+0.014}$ \\ 
  NGC 4590 &   24 & $0.39_{-0.38}^{+0.29}$ & $0.016_{-0.010}^{+0.015}$ & $15.13_{-0.30}^{+0.38}$ & $10.63_{-1.35}^{+2.02}$ & $0.12_{-0.710}^{+0.472}$ & $-0.019_{-0.015}^{+0.016}$ \\ 
  NGC 5824 &   42 & $0.48_{-0.28}^{+0.23}$ & $0.083_{-0.009}^{+0.010}$ & $17.45_{-0.23}^{+0.28}$ & $30.93_{-3.06}^{+4.21}$ & $-0.00_{-0.267}^{+0.274}$ & $-0.027_{-0.020}^{+0.020}$ \\ 
  NGC 5904 &   68 & $0.64_{-0.15}^{+0.13}$ & $0.056_{-0.032}^{+0.030}$ & $14.35_{-0.13}^{+0.15}$ & $7.41_{-0.42}^{+0.53}$ & $2.71_{-70.047}^{+57.861}$ & $-0.035_{-0.009}^{+0.009}$ \\ 
   NGC 6121 &  24  & $0.67_{-0.11}^{+0.09}$ & $0.097_{-0.017}^{+0.022}$ & $11.13_{-0.09}^{+0.11}$ & $1.68_{-0.07}^{+0.08}$ & $-1.53_{-7.533}^{+12.410}$ & $-0.038_{-0.026}^{+0.028}$ \\ 
  NGC 6171 &   15 & $0.72_{-0.19}^{+0.16}$ & $0.094_{-0.017}^{+0.024}$ & $13.67_{-0.16}^{+0.19}$ & $5.43_{-0.40}^{+0.49}$ & $-0.06_{-0.450}^{+0.975}$ & $-0.041_{-0.016}^{+0.016}$ \\ 
   NGC 6316 & 2 & $0.95_{-0.35}^{+0.36}$ & $0.230_{-0.168}^{+0.452}$ & $14.24_{-0.42}^{+0.42}$ & $7.05_{-1.25}^{+1.50}$ & $0.06_{-0.196}^{+0.196}$ & $-0.021_{-0.027}^{+0.026}$ \\ 
  NGC 6426 &   10 & $0.35_{-0.47}^{+0.32}$ & $0.094_{-0.024}^{+0.032}$ & $16.89_{-0.33}^{+0.47}$ & $23.86_{-3.36}^{+5.77}$ & $-0.03_{-0.080}^{+0.212}$ & $-0.032_{-0.022}^{+0.021}$ \\ 
  NGC 6569 & 11  & $0.86_{-0.25}^{+0.31}$ & $0.179_{-0.036}^{+0.050}$ & $15.42_{-0.33}^{+0.25}$ & $12.12_{-1.70}^{+1.51}$ & $0.05_{-0.242}^{+0.235}$ & $0.011_{-0.022}^{+0.021}$ \\ 
  NGC 6656 &   22 & $0.59_{-0.17}^{+0.16}$ & $0.048_{-0.017}^{+0.017}$ & $12.69_{-0.16}^{+0.17}$ & $3.46_{-0.24}^{+0.27}$ & $-0.22_{-5.474}^{+4.922}$ & $-0.017_{-0.022}^{+0.021}$ \\ 
  NGC 6864 &   10 & $0.65_{-0.19}^{+0.15}$ & $0.126_{-0.026}^{+0.042}$ & $16.68_{-0.16}^{+0.19}$ & $21.70_{-1.55}^{+2.00}$ & $0.09_{-0.221}^{+0.059}$ & $-0.033_{-0.023}^{+0.023}$ \\ 
  NGC 7006 &   43 & $0.61_{-0.19}^{+0.15}$ & $0.128_{-0.013}^{+0.017}$ & $18.00_{-0.15}^{+0.19}$ & $39.81_{-2.74}^{+3.65}$ & $0.01_{-0.066}^{+0.075}$ & $0.001_{-0.021}^{+0.021}$ \\ 
  NGC 7078 &   27 & $0.36_{-0.37}^{+0.30}$ & $0.061_{-0.009}^{+0.011}$ & $15.17_{-0.30}^{+0.37}$ & $10.80_{-1.41}^{+2.02}$ & $0.05_{-0.546}^{+0.473}$ & $-0.024_{-0.015}^{+0.015}$ \\ 
  Rup 106 &    8 & $0.52_{-0.26}^{+0.20}$ & $0.040_{-0.014}^{+0.019}$ & $16.83_{-0.21}^{+0.25}$ & $23.20_{-2.11}^{+2.89}$ & $0.02_{-0.235}^{+0.257}$ & $-0.036_{-0.022}^{+0.022}$ \\ 
  Ter 8 &    1 & $0.45_{-0.36}^{+0.26}$ & $0.633_{-0.450}^{+0.894}$ & $17.13_{-0.62}^{+0.89}$ & $26.66_{-6.62}^{+13.51}$ & $0.07_{-0.000}^{+0.000}$ & $-0.028_{-0.029}^{+0.027}$ \\ 
   \hline
\end{tabular}
\end{table*}
 
 \begin{table*}
\centering
\caption{Upper portion: Summary statistics of the population-level RRL parameters inferred by the
                             $M_{V}-[\mathrm{Fe/H}]$ HM. $\gamma_{1}$, $\gamma_{0}$ and $\tau$  represent the slope, the intercept and the dispersion of the the linear relationship  between the mean globular cluster RRL absolute magnitude $\beta_{0}$ and $[\rm{Fe/H}]$ according to Eq. \eqref{eq:beta-distr-LZ}, respectively. $\Delta\varpi_{0}$ denotes the parallax zero-point offset calculated as described in the text. Lower Portion: Population-level parameters inferred by the $M_{V}-[\mathrm{Fe/H}]$ HM 
                             from our field RRL sample.  $\gamma_{1}$, $\gamma_{0}$ and $\tau$ 
                             represent the $[\rm{Fe/H}]$ slope, the intercept and the intrinsic dispersion 
                             in Eq. (6) of \citet{Muraveva-et-al-2018a}, respectively.
                             $\Delta\varpi_{0}$ denotes the parallax zero-point offset calculated as described in the text.}. 
\label{tab:V-band-all}
\begin{tabular}{lllll}
  \hline
  \multicolumn{1}{c}{RRLs sample} &
   \multicolumn{1}{c}{$\gamma_{1}$} & \multicolumn{1}{c}{$\gamma_{0}$} & 
     \multicolumn{1}{c}{$\tau$} & \multicolumn{1}{c}{${\Delta\varpi}_{0}$} \\
     \multicolumn{1}{c}{} &
     \multicolumn{1}{c}{(mag/dex)}   & \multicolumn{1}{c}{(mag)}  & \multicolumn{1}{c}{(mag)}    & \multicolumn{1}{c} {(mas)}   \\

  \hline
 GC& $0.29_{-0.29}^{+0.35}$ & $1.05_{-0.46}^{+0.43}$ & $0.108_{-0.080}^{+0.154}$ & $-0.024_{-0.008}^{+0.008}$   \\ 
   \hline

  \hline
Field&$0.33_{-0.02}^{+0.02}$ & $1.13_{-0.03}^{+0.02}$ & $0.023_{-0.010}^{+0.012}$ & $-0.033_{-0.002}^{+0.002}$  \\ 
   \hline
\end{tabular}
\end{table*}

\subsection{\mg ~relation}
\label{sec:mg}  
The GC-level parameters associated to the {\mg} relation inferred by our HM 
 from the selection of 227 cluster RRLs in the $G$-band 
 are summarised in Table  \ref{tab:G-FeH-GC-CL}, which includes the same 
 parameters 
 as Table \ref{tab:V-FeH-GC-CL}. The GC mean absolute magnitudes ${M_G}_j$ (parameter ${\beta_{0j}}$ of Eq. \ref{eq:M-distr-LZ}) are systematically brighter than the ${M_V}_j$ magnitudes 
 of Table \ref{tab:V-FeH-GC-CL}, as expected given the larger passband of the $G$ compared to the $V$ filter. From the 
 cluster-level parallax offsets reported in the rightmost column of the table 
 we infer for the $G$-band data a global parallax offset of 
 ${-0.028_{-0.008}^{+0.009}}$ mas (last column of upper portion of Table~\ref{tab:G-band-all}). 
 The inferred metallicity slope, zero-point and 
 intrinsic 
dispersion of our GC RRL {\mg} relation for metal abundances on the 
\cite{Carretta-et-al-2009}  
scale  are summarised in the upper portion of Table ~\ref{tab:G-band-all} 
which translates into
the following relationship
 \begin{equation}
 {
 M_{G}=\left(0.28_{-0.36}^{+0.36}\right)\left[\mathrm{Fe/H}\right]+
 \left(0.97_{-0.52}^{+0.49}\right)\,, }
 \label{eq:G-FeH-GC}
 \end{equation} 

\noindent with an associated intrinsic dispersion ${\sigma_M  = 
0.13_{-0.10}^{+0.18}}$ mag. 
 
\begin{figure}
  	\begin{center}
         \includegraphics[trim=20 50 40 10,  
  		width=\linewidth]{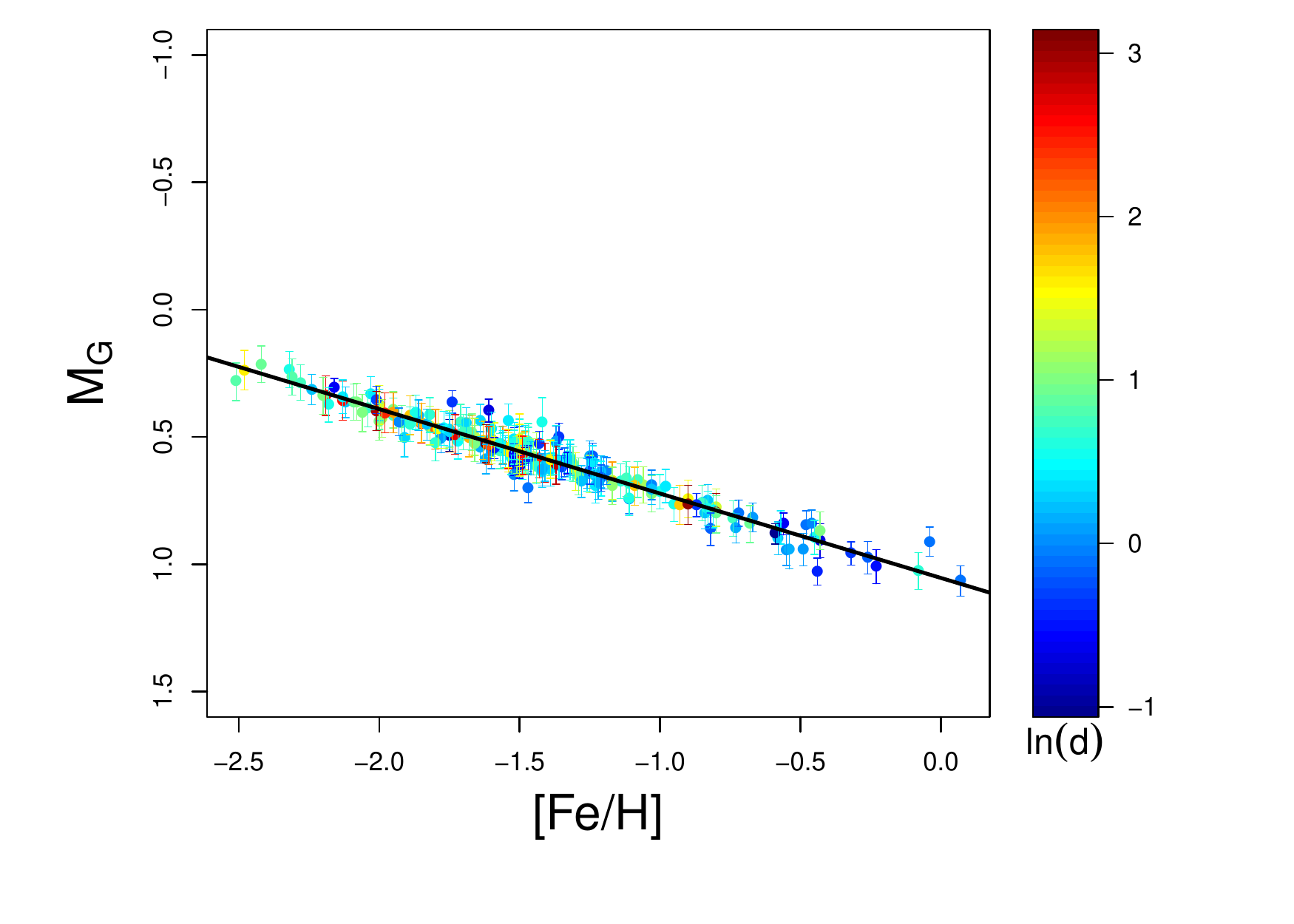} 		
  		\caption{ $M_G - {\rm [Fe/H]}$ relation defined by our field RRL sample. The solid 
  			line represents the linear fit corresponding to 
  			the parameters of Table~\ref{tab:G-band-all} (lower portion). 
  			The colour scale 
  			encodes the (natural) logarithm of the
  			inferred (true) barycentric distance to individual RRL stars 
  			measured in units of kpc. 	}
  		\label{fig:LZ-G-band-M-vs-FeH-FS}
  	\end{center}
  \end{figure}

We have also updated our previous study of 
the {\mg} relation for field RRLs in \cite{Muraveva-et-al-2018a} using the \egdr{3} 
parallaxes. We present the new results, corresponding to our selection of 223 field RRLs, in the  lower portion of Table~\ref{tab:G-band-all} 
and Fig. \ref{fig:LZ-G-band-M-vs-FeH-FS}. The credible intervals obtained for the
metallicity slope and the intercept are  $\gamma_{1}^{\rm ZW} = 
0.33_{-0.02}^{+0.02}$ mag/dex and $\gamma_{0}^{\rm ZW} = 
1.05_{-0.03}^{+0.03}$ mag, respectively, with an associated intrinsic 
dispersion of $\sigma_M = 0.03_{-0.01}^{+0.01}$ mag.  Our new inferred  {\mg} 
relation for field stars with metal abundances on the \cite{Zinn-West-84} 
metallicity scale is then given by
\begin{equation}
M_{G}=\left(0.33_{-0.02}^{+0.02}\right)\left[\mathrm{Fe/H}\right]+
\left(1.05_{-0.03}^{+0.03}\right)\,.
\label{eq:G-FeH-FS}
\end{equation} 

\noindent  For the {\mg} relation for field stars we obtained a parallax zero-point offset ${\Delta\varpi}_0=-0.033_{-0.002}^{+0.002}$ mas (Table ~\ref{tab:G-band-all}, 
lower portion, last column) calculated using the same procedure used in the previous section.

\noindent To compare the relation of Eq. \eqref{eq:G-FeH-FS}  with the GC RRL relation of Eq. \eqref{eq:G-FeH-GC} 
we applied the
metallicity conversions in Eqs. 
\eqref{eq:slope-UVES-2-ZW} and \eqref{eq:intercept-UVES-2-ZW} to its slope and intercept, and obtained ${\gamma_{1}^{\rm{ZW(UVES)}} = 0.31_{-0.40}^{+0.40}}$ mag/dex and 
${\gamma_{0}^{\rm{ZW(UVES)}} = 1.02_{-0.52}^{+0.49}}$ mag. 
Using the bootstrapping method we obtained this time a 99 per cent level confidence ellipse for $\boldsymbol{\gamma}^{\mathrm{diff}}$ centred at (0.027 mag/dex, 0.040 mag) with associated MD from its centre equal to 3.03 units. The distance from the coordinate origin (0,0) to the centre of the confidence ellipse was equal to 6.084 units in this case. This latter MD is  shorter than the value  of 7.94 units reported in Section \ref{sec:mv} but enough to conclude that the medians $\boldsymbol{\gamma}^{\mathrm{ZW}}$ and $\boldsymbol{\gamma}^{\mathrm{ZW(UVES)}}$ of both pairs of coefficients are  significantly different. With regard to the ability of both models to predict consistent absolute magnitudes our results are similar to those reported in the previous section.

  \begin{figure}
  	\begin{center}
  		\includegraphics[trim=20 50 40 10,  
  		width=\linewidth]{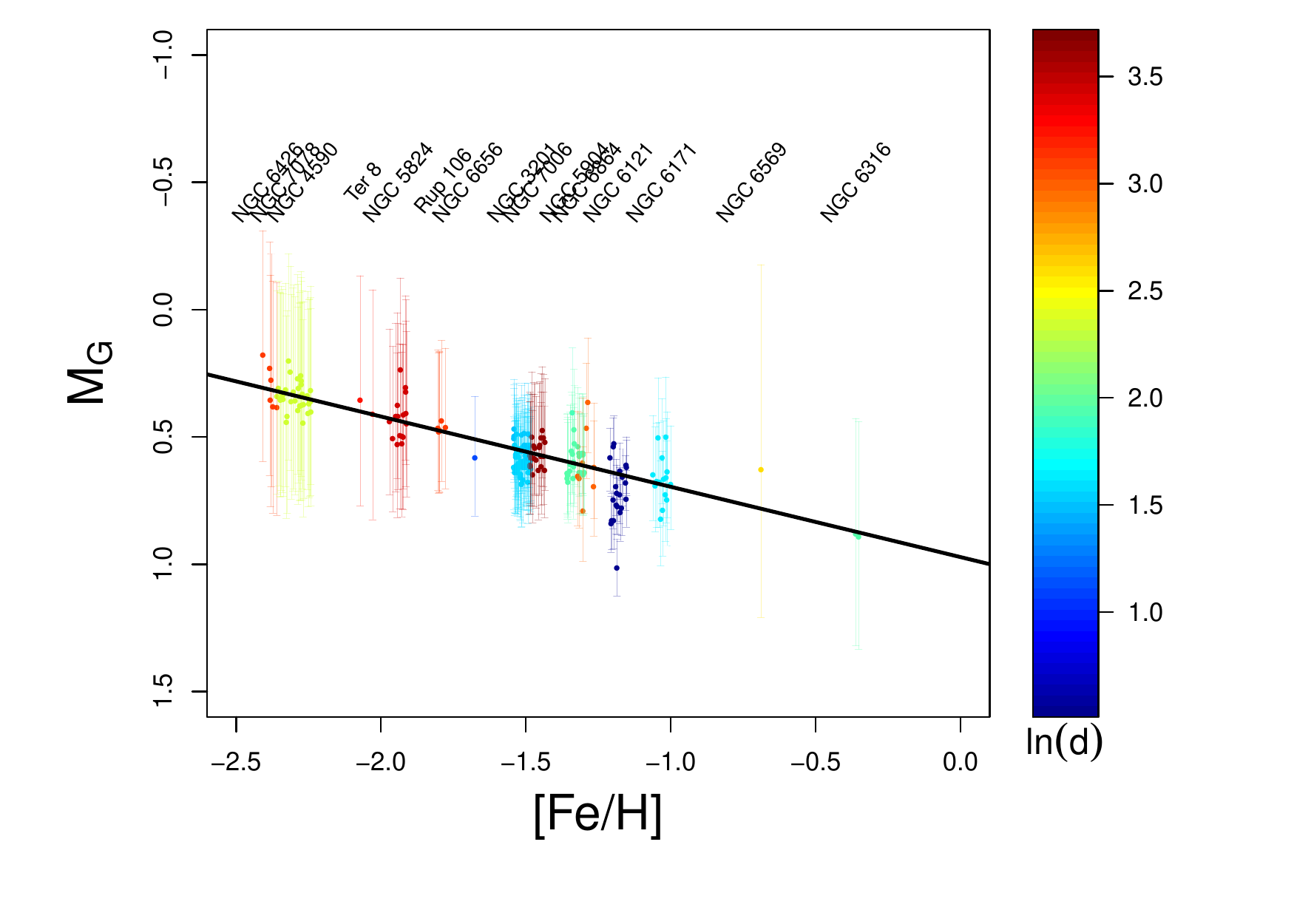}
  		\caption{ 
  		$M_G - {\rm [Fe/H]}$ relation defined by globular cluster RRLs 
  		as summarized in Table \ref{tab:G-FeH-GC-CL}. The solid 
  		line 
  		represents the linear fit corresponding to 
  			the parameters of Table~\ref{tab:G-band-all} (upper portion). 
  			The colour scale 
  			encodes the (natural) logarithm of the
  			inferred (true) barycentric distance to individual RRL stars 
  			measured in units of kpc. 	}
  		\label{fig:LZ-G-band-M-vs-FeH-GC}
  	\end{center}
  \end{figure}

\begin{table*}
\centering
\caption{ 
Summary statistics associated to the 
             the $M_G-[\mathrm{Fe/H}]$ relationship by globular cluster RRLs. The inferred parameters of each cluster are named as in Table \ref{tab:V-FeH-GC-CL}.} 
\label{tab:G-FeH-GC-CL}
\begin{tabular}{lrlllrrr}
  \hline

\multicolumn{1}{c}{GC} & \multicolumn{1}{c}{RRLs} & 
     \multicolumn{1}{c}{$M$} & \multicolumn{1}{c}{$\sigma_{M}$}
      & \multicolumn{1}{c}{$\mu_{0}$} & \multicolumn{1}{c}{${r}^{\rm{Cl}}$} & \multicolumn{1}{c}{$z^{\rm{RRL}}$} & \multicolumn{1}{c}{$\Delta\varpi$} \\
   & & \multicolumn{1}{c}{(mag)}   & \multicolumn{1}{c}{(mag)}  & \multicolumn{1}{c}{(mag)}    & \multicolumn{1}{c} {(kpc)} & \multicolumn{1}{c} {(pc)} & \multicolumn{1}{c} {(mas)} \\

  \hline
  NGC 3201 &   64 & $0.59_{-0.20}^{+0.16}$ & $0.063_{-0.008}^{+0.009}$ & $13.40_{-0.16}^{+0.20}$ & $4.80_{-0.34}^{+0.46}$ & $0.17_{-1.370}^{+0.695}$ & $-0.021_{-0.016}^{+0.018}$ \\ 
  NGC 4590 &   21 & $0.35_{-0.41}^{+0.37}$ & $0.051_{-0.008}^{+0.010}$ & $15.10_{-0.37}^{+0.41}$ & $10.46_{-1.63}^{+2.17}$ & $0.15_{-1.178}^{+0.517}$ & $-0.018_{-0.018}^{+0.017}$ \\ 
  NGC 5824 &   16 & $0.43_{-0.36}^{+0.29}$ & $0.093_{-0.016}^{+0.023}$ & $17.50_{-0.29}^{+0.36}$ & $31.58_{-3.90}^{+5.74}$ & $0.02_{-0.156}^{+0.218}$ & $-0.034_{-0.024}^{+0.025}$ \\ 
  NGC 5904 &   27 & $0.60_{-0.23}^{+0.16}$ & $0.072_{-0.022}^{+0.016}$ & $14.28_{-0.16}^{+0.23}$ & $7.16_{-0.50}^{+0.81}$ & $0.40_{-2.735}^{+1.668}$ & $-0.041_{-0.012}^{+0.015}$ \\ 
  NGC 6121 &   21 & $0.72_{-0.11}^{+0.11}$ & $0.125_{-0.019}^{+0.026}$ & $11.20_{-0.10}^{+0.11}$ & $1.74_{-0.08}^{+0.09}$ & $-0.06_{-1.241}^{+1.454}$ & $-0.019_{-0.028}^{+0.028}$ \\ 
  NGC 6171 &   15 & $0.67_{-0.23}^{+0.18}$ & $0.099_{-0.018}^{+0.024}$ & $13.65_{-0.18}^{+0.23}$ & $5.36_{-0.42}^{+0.60}$ & $0.19_{-0.939}^{+0.599}$ & $-0.043_{-0.017}^{+0.019}$ \\ 
  NGC 6316 &  2  & $0.88_{-0.41}^{+0.38}$ & $0.166_{-0.126}^{+0.396}$ & $14.26_{-0.43}^{+0.47}$ & $7.13_{-1.29}^{+1.71}$ & $-0.03_{-0.027}^{+0.027}$ & $-0.022_{-0.027}^{+0.027}$ \\ 
  NGC 6426 &    6 & $0.30_{-0.49}^{+0.41}$ & $0.108_{-0.034}^{+0.058}$ & $16.77_{-0.42}^{+0.49}$ & $22.63_{-3.95}^{+5.78}$ & $0.07_{-0.245}^{+0.091}$ & $-0.031_{-0.024}^{+0.024}$ \\ 
   NGC 6569 &  1 & $0.77_{-0.33}^{+0.28}$ & $0.596_{-0.437}^{+0.919}$ & $15.64_{-0.58}^{+0.81}$ & $13.43_{-3.13}^{+6.10}$ & $0.01_{-0.000}^{+0.000}$ & $-0.033_{-0.029}^{+0.028}$ \\ 
  NGC 6656 &  1 & $0.53_{-0.25}^{+0.20}$ & $0.358_{-0.256}^{+0.640}$ & $12.52_{-0.23}^{+0.25}$ & $3.18_{-0.33}^{+0.38}$ & $-0.00_{-0.000}^{+0.000}$ & $-0.025_{-0.028}^{+0.028}$ \\ 
  NGC 6864 &    8 & $0.61_{-0.25}^{+0.18}$ & $0.152_{-0.036}^{+0.062}$ & $16.56_{-0.19}^{+0.26}$ & $20.53_{-1.74}^{+2.60}$ & $-0.02_{-0.069}^{+0.089}$ & $-0.036_{-0.025}^{+0.024}$ \\ 
  NGC 7006 &   22 & $0.57_{-0.26}^{+0.18}$ & $0.055_{-0.009}^{+0.011}$ & $18.02_{-0.18}^{+0.25}$ & $40.19_{-3.28}^{+4.96}$ & $-0.02_{-0.076}^{+0.120}$ & $-0.009_{-0.024}^{+0.025}$ \\ 
  NGC 7078 &   15 & $0.34_{-0.42}^{+0.38}$ & $0.065_{-0.012}^{+0.016}$ & $15.18_{-0.38}^{+0.42}$ & $10.88_{-1.74}^{+2.33}$ & $0.12_{-0.449}^{+0.308}$ & $-0.018_{-0.018}^{+0.017}$ \\ 
  Rup 106 &    6 & $0.47_{-0.31}^{+0.24}$ & $0.023_{-0.009}^{+0.014}$ & $16.78_{-0.24}^{+0.31}$ & $22.67_{-2.34}^{+3.51}$ & $0.13_{-0.467}^{+0.402}$ & $-0.045_{-0.023}^{+0.024}$ \\ 
  Ter 8 &    2 & $0.42_{-0.41}^{+0.32}$ & $0.153_{-0.107}^{+0.444}$ & $17.09_{-0.41}^{+0.49}$ & $26.12_{-4.52}^{+6.65}$ & $0.06_{-0.076}^{+0.076}$ & $-0.018_{-0.028}^{+0.029}$ \\ 
   \hline
\end{tabular}
\end{table*}
 
\begin{table*}
\centering

\caption{Upper portion: Summary statistics of the population-level RRL parameters inferred by the $M_{G}-[\mathrm{Fe/H}]$ HM. Lower portion: Population-level parameters inferred by the by the $M_{G}-[\mathrm{Fe/H}]$ HM 
                             from our field RRL sample. Each parameter is named as in Table \ref{tab:V-band-all}. } 
\label{tab:G-band-all}
\begin{tabular}{lllll}
  \hline
    \multicolumn{1}{c}{RRLs sample} &

  \multicolumn{1}{c}{$\gamma_{1}$} & \multicolumn{1}{c}{$\gamma_{0}$} & 
     \multicolumn{1}{c}{$\tau$} & \multicolumn{1}{c}{${\Delta\varpi}_{0}$}  \\
     \multicolumn{1}{c}{} &
     \multicolumn{1}{c}{(mag/dex)}   & \multicolumn{1}{c}{(mag)}  & \multicolumn{1}{c}{(mag)}    & \multicolumn{1}{c} {(mas)}  \\

  \hline
GC&$0.28_{-0.36}^{+0.36}$ & $0.97_{-0.52}^{+0.49}$ & $0.131_{-0.095}^{+0.183}$ & $-0.028_{-0.008}^{+0.009}$ \\ 
   \hline
Field&$0.33_{-0.02}^{+0.02}$ & $1.05_{-0.03}^{+0.03}$ & $0.029_{-0.014}^{+0.015}$ & $-0.033_{-0.002}^{+0.002}$   \\ 
   \hline
\end{tabular}
\end{table*}


\subsection{PWZ relations}
\label{sec:pwz}

The literature lacks  empirical Period-Metallicity-Wesenheit relations for RR~Lyrae stars in the \gaia bands. In this section we use the samples of field and GC RRLs with {\it Gaia} EDR3 parallaxes described in Sections \ref{sec:FS-data} and \ref{sec:GC-data}, 
to calculate a Wesenheit magnitude when 
intensity-averaged  mean  magnitudes in the $G$, $G_{\rm BP}$ and $G_{\rm RP}$ bands are 
simultaneously available for these RRL selections in the {\it Gaia} DR2 \texttt{vari\_rrlyrae} table. This was possible for 191 field and  168 GC RRLs. To calculate the Wesenheit magnitude we followed the 
formulation 
by \cite{Ripepi-at-al-2019}
\begin{equation}
	W\left(G,G_{\rm{BP}},G_{\rm{RP}}\right)=G-\lambda\left(G_{\rm{BP}}-G_{\rm{RP}}\right)\,,
	\label{eq:Wesenheit-mag-def}
\end{equation}

\noindent where 
$\lambda=\frac{A\left(G\right)}{E\left(G_{\rm{BP}}-G_{\rm{RP}}\right)}$. 
The $\lambda$ value is known to be $\simeq$2 over a wide range of effective temperatures \citep{Andrae-et-al-2018}. Recently, \cite{Ripepi-at-al-2019} and \cite{Wang-et-al-2019}  with independent methods and different stellar samples (Cepheids and Red Clump stars, respectively) obtained $\lambda\simeq 1.90$. To infer a $\lambda$ value appropriate for RRLs we considered table~13 from  \citet{Jordi-et-al-2010} where $\lambda$ estimates are obtained  as a combination of effective temperature ($T_{eff}$), gravity (logg) and metallicity. We assumed typical values for RRLs ($T_{eff}$ from 5500 to 7000K; logg from 2.5 to 3)  obtaining an average $\lambda$ value of $1.922\pm 0.045$.  Since $\lambda$ estimates from  \citet{Jordi-et-al-2010} are based on the average galactic value of $R_{V} =$ 3.1 we adopt the value of $1.922\pm 0.045$ to infer the $PW$ relation for field RRLs sample and for GC RRLs with the exception for the bulge GCs. Indeed we used $\lambda=1.960$ to determine the Wesenheit magnitudes of RRLs members of NGC~6171, NGC~6316, NGC~6656 and NGC~6569 while we adopted $\lambda=2.104$ for RRLs in NGC~6121\footnote{ These values as been obtained by interpolating from table~3 of \citet{Cardelli1989} $A_{G}/A_{V}$, $A_{G_{\rm BP}}/A_{V}$ and $A_{G_{\rm RP}}/A_{V}$  with $G$, $G_{\rm BP}$ and $G_{\rm RP}$ central wavelenghts reported by \citet{Bono-at-al-2019}, for $R_{V} =$ 3.1, $R_{V} =$ 3.2 and $R_{V} =$ 3.62. Then we derived the following ratios $(A_{G}/E(G_{\rm BP}-G_{\rm RP}))_{R_V=3.2} $/$(A_{G}/E(G_{\rm BP}-G_{\rm RP}))_{R_V=3.1}$=1.020 and $(A_{G}$/$E(G_{\rm BP}-G_{\rm RP}))_{R_V=3.62}$ /$(A_{G}$/$E(G_{\rm BP}-G_{\rm RP}))_{R_V=3.1}$=1.095 that we used to obtain $\lambda=1.960$ and $\lambda=2.104$ for $R_{V} =$ 3.2 and $R_{V} =$ 3.62 respectively.}. 
Finally, the uncertainties on the  Wesenheit magnitudes were computed using the  
linear error propagation method.

Our selection of 168  cluster RRLs includes only one source  belonging to Ter~8, 
making it impossible to apply our HM to 
infer a $PW$ relation for this GC.  We also removed 15 sources with $G_{\rm BP}$ and $G_{\rm RP}$ 
photometry available but not reliable. For the remaining 153 RRLs with  $W_{(G,G_{\rm BP},G_{\rm RP})}$ magnitudes available, Table 
\ref{tab:PLZ-Wesenheit-GC-CL} summarises the  parameters inferred by our HM at GC level.  Similarly to 
the information presented in 
Tables \ref{tab:V-FeH-GC-CL} and \ref{tab:G-FeH-GC-CL} 
for the $V $ and $G$ bands, now the third to the fifth columns   
provide the period slope ${\beta_1}_j $, the zero-point ${\beta_0}_j$ and the 
intrinsic dispersion ${\sigma_M}_j$ of the  
$PW_{(G,G_{\rm BP},G_{\rm RP})}$ relation inferred for each GC according to Eq. 
\eqref{eq:M-distr}, and column six gives the  average of the GC
 individual-star mean absolute magnitudes. As in Table \ref{tab:V-FeH-GC-CL} 
 we  
 also report the distance modulus of each GC (column 7) and the 
 cluster-level parallax offsets (rightmost column) from which a    
 population-level median parallax offset 
 	${\Delta_{\varpi_0}=-0.026_{-0.010}^{+0.010}}$ mas is inferred. 
 Table \ref{tab:PLZ-Wesenheit-GC-CL} also summarises, for each 
 GC, the inferred barycentric distance to its centre ($r_{j}^{\rm Cl}$) and  the median
of the radial distances from the centre to each RRL projected onto the line of sight  (${z_{\cdot j}}$).

  We do not report the coefficients inferred  by our HM  at population-level, because they are not reliable values for the slopes and intercept of a $PWZ$ relation that would be incorrectly derived from a sample of RRLs belonging to different absorption environment. Nevertheless, the use of the appropriate
ratio of total-to-selective absorption $\lambda$ for each GC in our sample guarantees that the Wesenheit magnitudes in Eq. (\ref{eq:Wesenheit-mag-def}) are correctly defined in the sense that they can be expressed, alternatively, both as a function of the observed or the intrinsic colour (reddening-free property). Hence, the  $PW$ relations in Table \ref{tab:PLZ-Wesenheit-GC-CL} are correctly derived, and we still can use a subset of them for validation purposes, provided that the $\lambda$ associated with the $PW$ relations in the subset is the same than the $\lambda$ used to derive the Wesenheit magnitudes in the validation dataset (see Section~\ref{sec:validation})\footnote{
 In a previous version of this paper we had used, for all GCs in our sample, the same reddening law. 
In that way, the reddening-free property is no longer preserved when we use an incorrect ratio of total-to-selective absorption (that we call $\lambda_2$) in a dust environment characterised by $\lambda_1$, being the Wesenheit magnitudes biased to fainter values by a quantity equal to $\left(\lambda_{1}-\lambda_{2}\right)\left[\left(G_{\rm{BP}}-G_{\rm{RP}}\right)_{0}+E\left(G_{\rm{BP}}-G_{\rm{RP}}\right)\right]$, where  $E\left(G_{\rm{BP}}-G_{\rm{RP}}\right)$ is the colour excess of the source.
Hence, the approach to derive the Wesenheit magnitudes of RRL stars belonging to different absorption environments (i.e. NGC~6121) inevitably bias, not only the coefficients of the global $PWZ$ relation, but also the coefficients of any $PW$ relation calibrated using an incorrect value of $\lambda$.
The publication already with the next \textit{Gaia} Data release, DR3, 
 of individual $G$-band absorption values for a wider sample of RRLs and more accurate intensity-averaged mean magnitudes in the $G$, $G_{\rm{BP}}$ and $G_{\rm{RP}}$ bands, as well a larger sample of GC RRLs that we can assume in the same absorption environment 
  will help to reduce the dispersion/bias of the $PW_{(G,G_{\rm BP},G_{\rm RP})}$ relation of the cluster RRLs.}.



\begin{figure}
	\begin{center}
		\includegraphics[trim=20 50 40 10,   
		width=\linewidth]{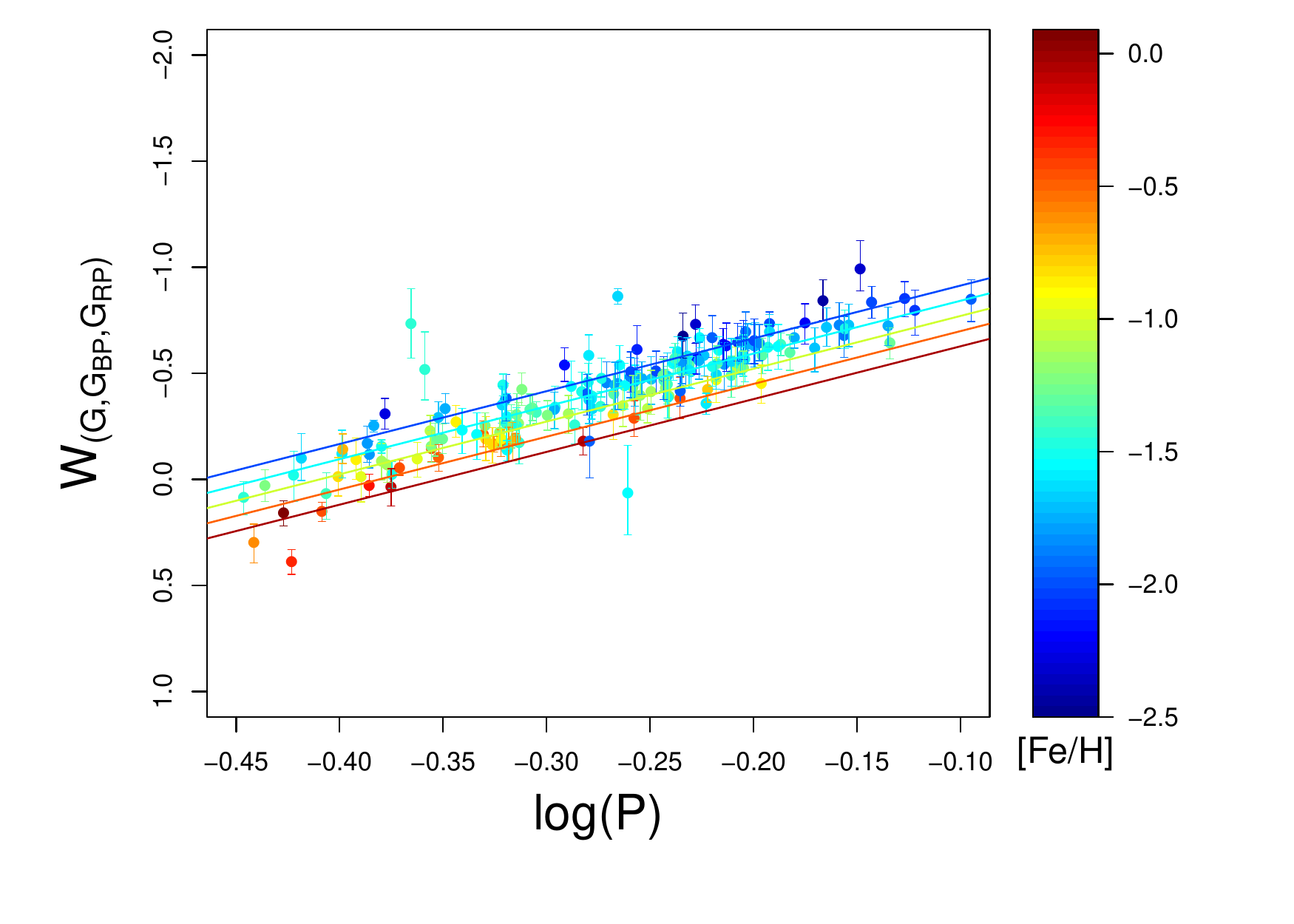}
		\caption{  $PW_{\left(G,G_{\rm{BP}},G_{\rm{RP}}\right)}$  distribution of 169 RRLs from our sample of field RRLs,
				for which absolute $W_{\left(G,G_{\rm{BP}},G_{\rm{RP}}\right)}$ 
				magnitudes were inferred from 
				the model described in Sect. 4 of \citet{Muraveva-et-al-2018a}. 
				The lines represent projections of the 
				fit shown by Eq. \eqref{eq:Period-Wesenheit-Field-RRL}
				onto the Magnitude-Period
				plane. The colour scale encodes values of measured metallicity 
				on the \citet{Zinn-West-84} metallicity scale. }
		\label{fig:PLZ-Wesenheit-M-vs-logP-FS}
	\end{center}
\end{figure}

We now focus 
on the sample of 191 field RRL 
stars for which 
$W_{(G,G_{\rm BP},G_{\rm RP})}$  apparent magnitudes are  available. 
After dropping 
one source (star SW Her)  from the analysis for being largely discrepant with the inferred Period-Wesenheit diagram,  
we applied the varying parallax offset HM for field stars described in Section \ref{sec:method} to the remaining sample of 190 RRL, thus  obtaining the results summarised in Table \ref{tab:PLZ-Wesenheit-FS}. The inferred $PW_{(G,G_{\rm BP},G_{\rm RP})}Z$ relation is given by
\begin{equation}
\begin{split}
W_{\left(G,G_{\rm{BP}},G_{\rm{RP}}\right)}=&\left(-2.49_{-0.20}^{+0.21}\right)\log\left(P\right)+\\
&\left(0.14_{-0.03}^{+0.03}\right)\left[\mathrm{Fe/H}\right]+\left(-0.88_{-0.09}^{+0.08}\right)\,,\\
\end{split}
\label{eq:Period-Wesenheit-Field-RRL}
\end{equation}

\noindent with an intrinsic dispersion of $0.09 \pm{0.01}$ mag and an inferred 
global parallax offset $\Delta\varpi_0 = -0.033 \pm{0.028}$ mas. The 
$PW_{(G,G_{\rm BP},G_{\rm RP})}Z$ relation for the 190 field RRLs is shown in 
Fig. \ref{fig:PLZ-Wesenheit-M-vs-logP-FS}. Our inferred parallax zero-point offset associated to this relation is ${\Delta\varpi}_0=-0.033_{-0.003}^{+0.003}$ mas (Table \ref{tab:PLZ-Wesenheit-FS}, lower row).

 Eq.~\eqref{eq:Period-Wesenheit-Field-RRL} is, as far as we know, the first empirical RRL PWZ relation in the Gaia bands calibrated on EDR3 parallaxes. 

\begin{figure}
	\begin{center}
		\includegraphics[trim=25  30 20 10,    
	 	width=\linewidth]{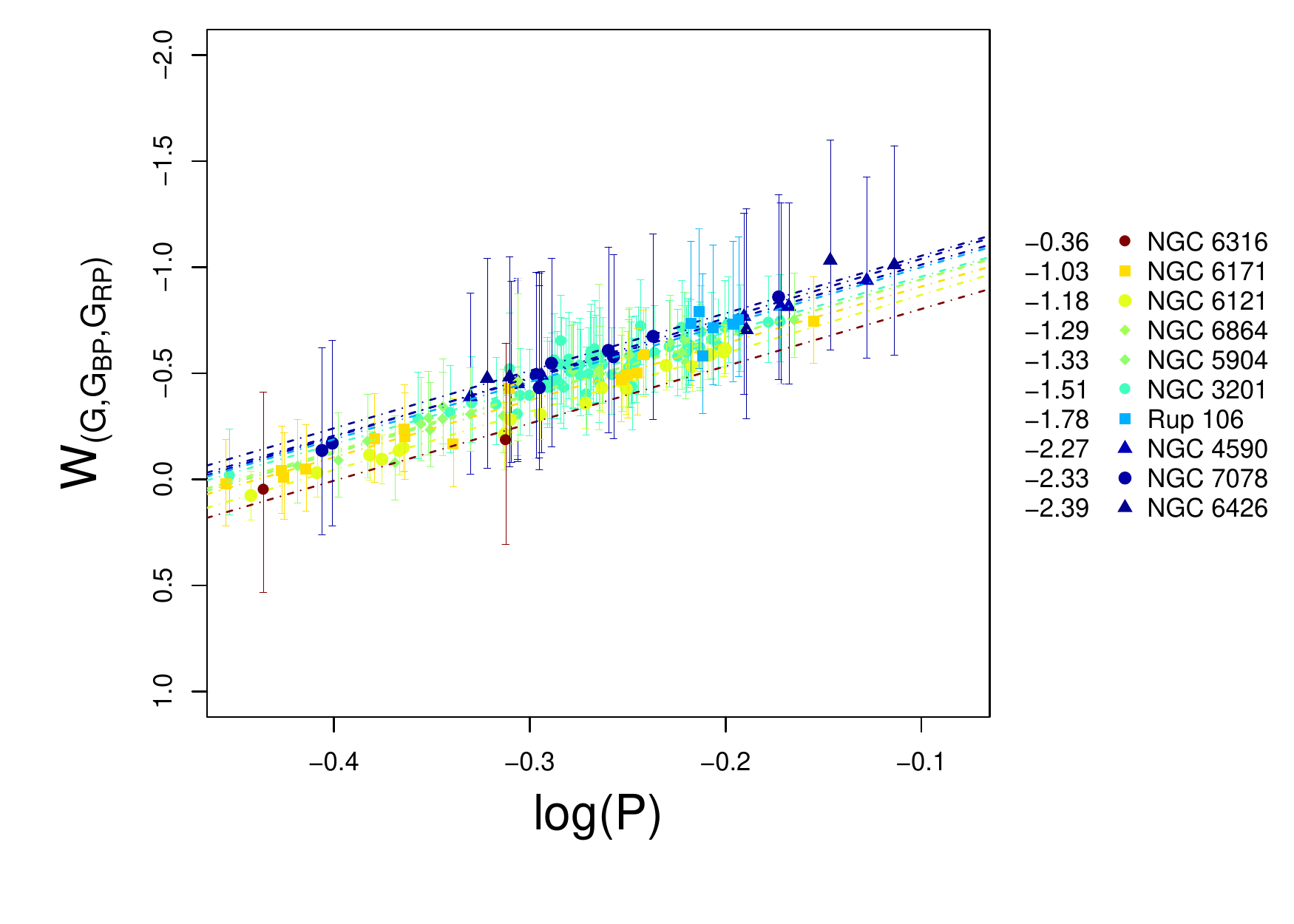}
		\caption{Wesenheit magnitudes inferred by the HM   as a 
			function of the decadic logarithm of the true period. The dashed lines represent the  $PW_{\left(G,G_{\rm{BP}},G_{\rm{RP}}\right)}$ relations of Eq.  \eqref{eq:M-distr} particularised  for the parameters of Table 
			\ref{tab:PLZ-Wesenheit-GC-CL}. 
			Colours encode the true metallicity of each globular cluster on the \citet{Carretta-et-al-2009} metallicity scale according to the scale on the right.}
		\label{fig:PLZ-Wesenheit-M-vs-logP-GC}
	\end{center}
\end{figure}

\begin{table*}
\centering
\caption{ 
Summary statistics of the cluster-level RRL parameters inferred by the  
                         $PW_{\left(G,G_{\rm{BP}},G_{\rm{RP}}\right)}Z$ HM.
                              $\beta_1$, $\beta_0$ and $\sigma_{M}$ represent the 
                              $\log(P)$ slope, the zero-point and the intrinsic dispersion of the $j$th cluster
                              $PW_{\left(G,G_{\rm{BP}},G_{\rm{RP}}\right)}$ relationship, and
                              $M$, $\mu_{0}$, ${r}^{\rm{Cl}}$, ${z}^{\rm{RRL}}$ and $\Delta\varpi$ denote the median magnitude of its individual RRL stars, 
                              the GC true distance modulus, the distance to the centre,
                              the RRL median radial projections and the parallax zero-point offset. } 
\label{tab:PLZ-Wesenheit-GC-CL}
\begin{tabular}{lrlllllrrr}
  \hline

\multicolumn{1}{c}{GC} & \multicolumn{1}{c}{RRLs} & 
\multicolumn{1}{c}{$\beta_1$} &
\multicolumn{1}{c}{$\beta_0$} &
 \multicolumn{1}{c}{$\sigma_{M}$}
      & 
      \multicolumn{1}{c}{$M$}
      & 
      \multicolumn{1}{c}{$\mu_{0}$} & \multicolumn{1}{c}{${ r}^{\rm{Cl}}$} & \multicolumn{1}{c}{$z^{\rm{RRL}}$} & \multicolumn{1}{c}{$\Delta\varpi$} \\
   & & \multicolumn{1}{c}{(mag/dex)}   & \multicolumn{1}{c}{(mag)}  & \multicolumn{1}{c}{(mag)}    &
   \multicolumn{1}{c}{(mag)}    &
   \multicolumn{1}{c}{(mag)}    &
   \multicolumn{1}{c} {(kpc)} & \multicolumn{1}{c} {(pc)} & \multicolumn{1}{c} {(mas)} \\

  \hline
  NGC 3201 &   63 & $-2.63_{-0.13}^{+0.18}$ & $-1.22_{-0.22}^{+0.18}$ & $0.070_{-0.015}^{+0.015}$ & $-0.54_{-0.11}^{+0.11}$ & $13.39_{-0.18}^{+0.22}$ & $4.76_{-0.39}^{+0.51}$ & $1.70_{-14.589}^{+7.825}$ & $-0.022_{-0.018}^{+0.020}$ \\ 
  NGC 4590 &    9 & $-2.69_{-0.11}^{+0.11}$ & $-1.28_{-0.48}^{+0.37}$ & $0.013_{-0.009}^{+0.015}$ & $-0.49_{-0.33}^{+0.04}$ & $15.10_{-0.36}^{+0.49}$ & $10.47_{-1.61}^{+2.65}$ & $-0.10_{-0.488}^{+0.277}$ & $-0.005_{-0.020}^{+0.019}$ \\ 
  NGC 5904 &   22 & $-2.73_{-0.14}^{+0.12}$ & $-1.22_{-0.22}^{+0.17}$ & $0.059_{-0.012}^{+0.014}$ & $-0.42_{-0.18}^{+0.22}$ & $14.34_{-0.17}^{+0.22}$ & $7.38_{-0.57}^{+0.79}$ & $-0.03_{-0.587}^{+0.435}$ & $-0.035_{-0.013}^{+0.014}$ \\ 
  NGC 6121 &   20 & $-2.75_{-0.12}^{+0.11}$ & $-1.14_{-0.12}^{+0.12}$ & $0.035_{-0.017}^{+0.028}$ & $-0.39_{-0.20}^{+0.28}$ & $11.21_{-0.11}^{+0.11}$ & $1.75_{-0.09}^{+0.09}$ & $-0.32_{-1.292}^{+2.368}$ & $-0.013_{-0.031}^{+0.029}$ \\ 
  NGC 6171 &   14 & $-2.68_{-0.12}^{+0.13}$ & $-1.18_{-0.21}^{+0.20}$ & $0.059_{-0.020}^{+0.021}$ & $-0.22_{-0.28}^{+0.18}$ & $13.66_{-0.20}^{+0.21}$ & $5.39_{-0.47}^{+0.55}$ & $0.35_{-1.549}^{+0.615}$ & $-0.036_{-0.018}^{+0.019}$ \\ 
 NGC 6316  &  2  & $-2.70_{-0.16}^{+0.18}$ & $-1.07_{-0.41}^{+0.42}$ & $0.201_{-0.134}^{+0.424}$ & $-0.07_{-0.08}^{+0.08}$ & $14.31_{-0.48}^{+0.46}$ & $7.27_{-1.44}^{+1.73}$ & $0.10_{-0.190}^{+0.190}$ & $-0.019_{-0.028}^{+0.029}$ \\ 
  NGC 6426 &  5 & $-2.71_{-0.16}^{+0.16}$ & $-1.32_{-0.56}^{+0.41}$ & $0.105_{-0.038}^{+0.073}$ & $-0.71_{-0.31}^{+0.23}$ & $16.76_{-0.42}^{+0.56}$ & $22.46_{-3.96}^{+6.66}$ & $-0.02_{-0.125}^{+0.113}$ & $-0.020_{-0.026}^{+0.027}$ \\ 
  NGC 6864 &  2  & $-2.70_{-0.16}^{+0.18}$ & $-1.21_{-0.26}^{+0.19}$ & $0.159_{-0.119}^{+0.473}$ & $-0.49_{-0.02}^{+0.02}$ & $16.95_{-0.28}^{+0.41}$ & $24.55_{-2.97}^{+5.09}$ & $-0.05_{-0.066}^{+0.066}$ & $-0.045_{-0.030}^{+0.029}$ \\ 
  NGC 7078 &   10 & $-2.80_{-0.23}^{+0.15}$ & $-1.32_{-0.50}^{+0.39}$ & $0.046_{-0.022}^{+0.029}$ & $-0.52_{-0.12}^{+0.23}$ & $15.14_{-0.39}^{+0.48}$ & $10.68_{-1.74}^{+2.66}$ & $-0.24_{-0.427}^{+0.830}$ & $-0.018_{-0.020}^{+0.020}$ \\ 
  Rup 106 &    6 & $-2.70_{-0.16}^{+0.17}$ & $-1.27_{-0.38}^{+0.26}$ & $0.092_{-0.029}^{+0.051}$ & $-0.73_{-0.03}^{+0.04}$ & $16.72_{-0.27}^{+0.39}$ & $22.09_{-2.57}^{+4.36}$ & $-0.05_{-0.058}^{+0.160}$ & $-0.047_{-0.023}^{+0.024}$ \\ 
   \hline
\end{tabular}
\end{table*}


\begin{table*}
\centering
\caption{Population-level parameters inferred by the $PW_{\left(G,G_{\rm{BP}},G_{\rm{RP}}\right)}Z$ HM
             from our field RRL sample.  
             $\gamma_{2}$, $\gamma_{1}$ and $\gamma_{0}$  represent the $\log(P)$ and $[\rm{Fe/H}]$ slopes and the intercept
             in Eq. (6) of \citet{Muraveva-et-al-2018a} and $\tau$ denotes the intrinsic dispersion.
             $\Delta\varpi_{0}$ denotes the parallax zero-point offset.} 
\label{tab:PLZ-Wesenheit-FS}
\begin{tabular}{lllll}
  \hline

 \multicolumn{1}{c}{$\gamma_{2}$} & \multicolumn{1}{c}{$\gamma_{1}$} & \multicolumn{1}{c}{$\gamma_{0}$} &
     \multicolumn{1}{c}{$\tau$} & \multicolumn{1}{c}{${\Delta\varpi}_{0}$} \\
      \multicolumn{1}{c}{(mag/dex)}   & \multicolumn{1}{c}{(mag/dex)}  &  \multicolumn{1}{c}{(mag)}    & \multicolumn{1}{c}{(mag)}    & \multicolumn{1}{c} {(mas)}   \\
    
  \hline
$-2.49_{-0.20}^{+0.21}$ & $0.14_{-0.03}^{+0.03}$ & $-0.88_{-0.09}^{+0.08}$ & $0.09_{-0.01}^{+0.01}$ & $-0.033_{-0.003}^{+0.003}$  \\ 
   \hline
\end{tabular}
\end{table*}

 
%

\subsection{Spatial distribution of our samples}
\label{sec:3D-Distrib}

One of the advantages of our Bayesian approach is the ability to infer 
the posterior probability distribution of any parameter of interest in the HM. 
In this section we present the results of validating the 3D spatial distributions 
of our field RRLs and the RRLs of some GCs studied in the paper.  

For each RRL in the field sample we derived the individual barycentric distance
and located the star position in the Galaxy.  The measured coordinates $(\alpha,\delta)$ of our RRL stars were first transformed to Galactic coordinates
 $(l,b)$.  For each RR~Lyrae we then derived  
 the posterior distribution of its rectangular coordinates $(x_i,y_i,z_i)$ from  its posterior parallax ($\varpi_i$) in the Cartesian Galactocentric coordinate system of \cite{Juric-2008} by
 \begin{equation}
 	\begin{split}
 		x_{i}^{n}&=R_\odot-d_{i}^{n}\cos\left(l_{i}\right)\cos\left(b_{i}\right)\\
 		y_{i}^{n}&=-d_{i}^{n}\sin\left(l_{i}\right)\cos\left(b_{i}\right)\\
 		z_{i}^{n}&=d_{i}^{n}\sin\left(b_{i}\right) \,,\\
 	\end{split}
 	\label{eq:GC.coord}
 \end{equation}
 
 \noindent where $d_{i}^{n}=1/\varpi_{i}^{n}$ denotes  the $n$th posterior distance realisation calculated as the reciprocal of the $n$th MCMC posterior parallax realisation, and $R_\odot=8.178$ kpc is the adopted distance to the Galactic centre   \citep{Abuter-et-al-2019}. The posterior distribution of the radial distance of each star to the Galactic centre were calculated as
 \begin{equation}
 	{r_{\mathrm{Gal.Cen.}}}_{i}^{n}=\sqrt{\left(x_{i}^{n}\right)^{2}+
 		\left(y_{i}^{n}\right)^{2}}\,.
 	\label{eq:r.GC}
 \end{equation}

\begin{figure}
 	\begin{center}
 	 	\includegraphics[trim=30 50 40 10,  
 	 	width=\linewidth]{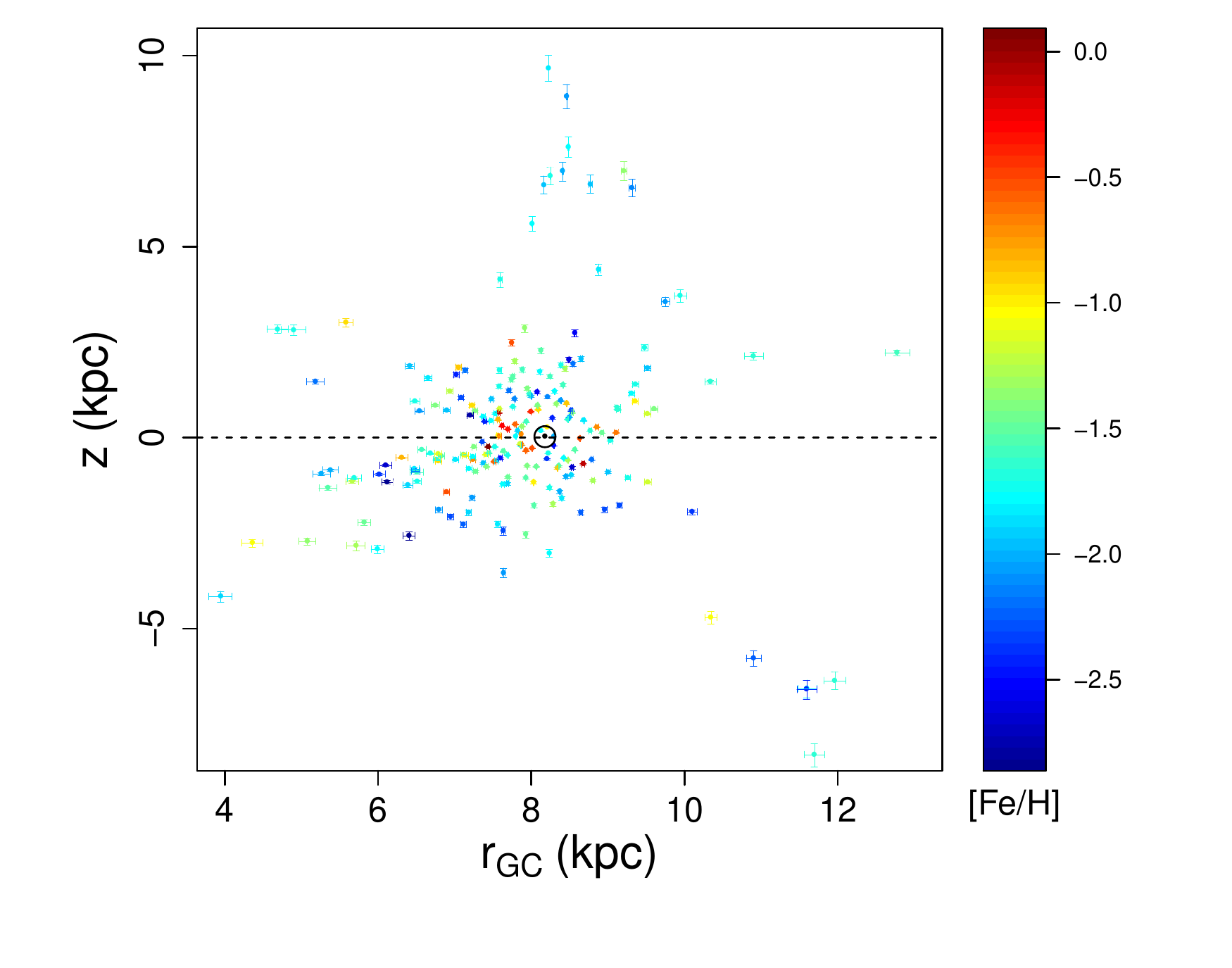}
 		\caption{
 		Spatial distribution and metallicity of the field  RR~Lyrae sample represented in a Galactocentric reference
 			frame.  $r_{\rm Gal.Cen.}$ and $z$ denote the radial
 			distance to the Galactic  centre and the vertical distance
 			with regard to the Galactic midplane $x-y$, respectively. The 
 			Cartesian Galactocentric coordinates and the radial distances to the Galactic centre have been summarised by the median plus/minus the difference in absolute value between the median and the 84th and 16th percentile of the posterior samples of Eqs. \eqref{eq:GC.coord} and \eqref{eq:r.GC}. }
 		\label{fig:field-rrl-galact}
 	\end{center}   
 \end{figure}

 Figure \ref{fig:field-rrl-galact} represents the spatial distribution and metal abundance of our field RR~Lyrae sample in the plane $z-r_{\mathrm{Gal.Cen.}}$ associated with the Galactocentric reference frame.

\begin{figure*}
	\begin{center}
	 	\includegraphics[trim=0 30 30 0,   
	 	width=\linewidth]{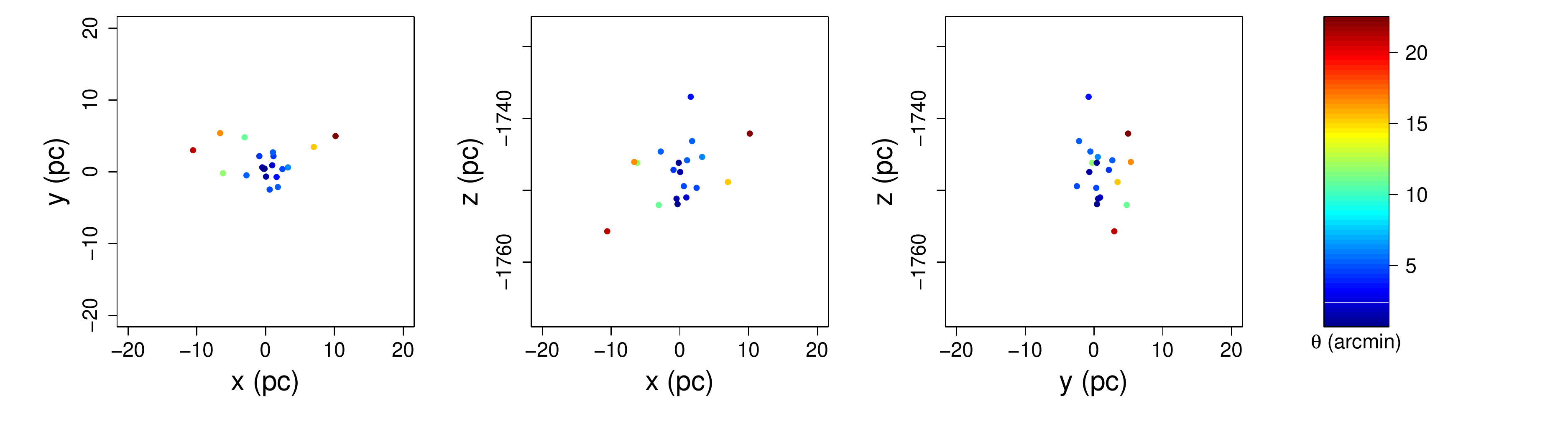}
		\caption{
		Spatial distribution inferred by our 
			$PW_{\left(G,G_{\rm BP},G_{\rm RP}\right)}Z$  HM  for 20 RRLs in  NGC 6121  
			projected onto the coordinate planes 
		associated to the reference frame described in 
		the text. 
        The colour scale encodes the angular separation between the 
		RRLs and the cluster centre. Posterior credible intervals are omitted for clarity,  but along the z-axis (line of sight) are on the order of $100$ pc.
		   }
		\label{fig:NGC6121-3D-distr}
	\end{center}
\end{figure*}

Aiming to conduct the validation analysis of the 3D spatial distribution for some of the GCs in our sample, we found  
that, for a fixed GC, the median distances inferred to its centre by our three hierarchical models, which are all consistent within 1$\sigma$, differ by quantities ranging from the order of several tens of pc (NGC 6121 and NGC 5904) to a few kpc (NGC 5824) (see Table \ref{tab:dist-comp-V-G-Wesenheit-Harris} in Section  \ref{sec:GC-distances} below). Our most consistent results correspond to NGC 6121, for which the median barycentric distances to its centre inferred by our {\mg} and 
 $PW_{\left(G,G_{\rm BP},G_{\rm RP}\right)}Z$ 
hierarchical models are in excellent agreement differing   by  $\approx$ 10 pc. 
In what follows we present the results corresponding to the 
application of our HM for the inference of  the $PW_{\left(G,G_{\rm BP},G_{\rm RP}\right)}Z$  relation to the analysis of NGC 
6121 spatial distribution. 
From the measured coordinates $(\alpha_i,\delta_i)$ of the $i$th RR~Lyrae star
belonging to NGC 6121 and the barycentric
distance MCMC posterior samples $r_{i}^{n}$  we first derived samples 
$(x_i^n,y_i^n,z_i^n)$  of the posterior 
distribution of its rectangular coordinates. Then, based on 
\cite{Weinberg-Nikolaev-2001} we defined, from the spherical 
coordinates  $(\alpha^{\rm Cl},\delta^{\rm Cl})$ of the NGC  6121  centre, a   barycentric  Cartesian reference frame with 
the x-axis antiparallel to the $\alpha^{\rm Cl}$ axis, the y-axis 
parallel to the $\delta^{\rm Cl}$ axis and the z-axis pointing toward the 
observer, and transformed the samples $(x_i^n,y_i^n,z_i^n)$ to the new frame.
We note that the z coordinates in this system are given, from Eq. \eqref{eq:dist-to-star}, by $r_{j}^{\mathrm{Cl}}+z_{ij}$ where $z_{ij}$ is the projection of the radial distance from the $i$th source to the cluster centre onto the line of sight.
Figure \ref{fig:NGC6121-3D-distr} represents the NGC  6121 3D spatial 
distribution inferred by our $PW_{\left(G,G_{\rm{BP}},G_{\rm{RP}}\right)}Z$ HM projected onto the coordinate planes of the reference frame described above.  The three panels of the figure are in 
the same scale which covers an area of 1600 ${\rm pc}^2$. While the marginal 
RRL distribution associated with the plane of the sky XY is consistent with  
the $(\alpha, \delta)$  distribution in the middle-right  panel of Fig.   \ref{fig:Parallax-error-RUWE},  
for the other two 
planes XZ and YZ we notice that the distribution is elongated along the line of 
sight (z axis).  This non-physical artifact is more 
prominent for the RRLs closer to the cluster centre as apparent 
from the angular distances encoded by the colour of the figure.

\section{Validation}
\label{sec:validation}
\subsection{GC distances}
 \label{sec:GC-distances}

In column 6 of Tables~\ref{tab:V-FeH-GC-CL}, \ref{tab:G-FeH-GC-CL} and column 8 of Table~\ref{tab:PLZ-Wesenheit-GC-CL} we reported the credible interval for the distance to the centre of each GC in our sample, inferred by our HM  applied to the inference of the {\mv}, {\mg} and $PW_{\left(G,G_{\rm{BP}},G_{\rm{RP}}\right)}Z$ relations. In this section we compare our three distance estimates with the prior values taken from  \cite{Harris-1996} (2010 edition, hereafter Harris10) and with GC heliocentric distances reported by the recent literature. 
 Distances in \citet{Harris-1996} collection are not the most updated results, but, we think they still form valid priors for our model, since the errors associated with these distances make them all compatible with the most recent literature value. Moreover, adopting the distances contained in the Harris catalogue we guarantee that our results are independent from the most recent literature values which are mainly derived from Gaia EDR3 parallaxes.

An exhaustive study of this topic is beyond  
the scope of the paper and we only include the minimal information for validation purposes of our own results. We have selected the GC catalogues of \cite{Baumgardt-et-al-2019} (B19; cf. Table 3), \cite{Hernitschek-et-al-2019} (H19; cf. Table 10) and \cite{Valcin-et-al-2020} (V20; cf. Appendix E) which report distances to 53, 59 and 68 Galactic globular clusters, respectively.  
 
Table \ref{tab:dist-comp-V-G-Wesenheit-Harris} presents the results of our study. Columns 2, 3 and 4 
of the table list the credible intervals for the distance to the centre to each GC of our sample inferred by our HM using the $V$, $G$ and $W\left(G,G_{\mathrm{BP}},G_{\mathrm{RP}}\right)$ data and {\textit Gaia} EDR3 parallaxes, respectively. The GC distances of 
Harris10 used as priors in our study are given in column 5, while the literature distances of B19, H19 and V20 
are presented in the three rightmost columns of the table. Our distances for the 4 clusters closest to us 
(NGC 6121, 6656, 3201 and 6171)  differ from the literature,  at most,  by 
 1 kpc. 
 For those four closest clusters our models report the smallest distances in the table. In particular, for NGC 6121 our models infer distances ranging from 1.68 to 1.75 kpc which are in better agreement with the values by \cite{Bhardwaj-et-al-2020} which finds 
true distance moduli for this cluster ranging from 11.241 to 11.323 mag (1.77 to 1.84 kpc).  We remark, as pointed out by \cite{Hendricks-2012} in their section  5.5, the strong dependency of the true distance to NGC 6121 on the reddening law. As we adopted $R_V = 3.62$ for this cluster, our distance estimates are also more consistent with the distances reported in table 6 by \cite{Hendricks-2012} for  larger values of  $R_V$.  
The distances we inferred for NGC 5904 (from 7.16 to 7.41 kpc) are also in better agreement with the values  
found for this cluster by \cite{Bhardwaj-et-al-2020} 
(true distance moduli ranging from 14.257 to 14.342 mag corresponding to distances between 7.10 and 7.39 kpc). As in the present work, 
\cite{Bhardwaj-et-al-2020} distance moduli are calibrated using \textit{Gaia} EDR3 parallaxes.

\begin{table*}
\centering
\caption{
Comparison between inferred, prior (Harris10) and literature  (B19, H19 and V20)  distances to  the 15 GCs considered in our study.
} 
\label{tab:dist-comp-V-G-Wesenheit-Harris}
\begin{tabular}{lrrrrrrr}
  \hline

\multicolumn{1}{c}{GC} & \multicolumn{1}{c}{${V}$ band} & 
     \multicolumn{1}{c}{${G}$  band} & \multicolumn{1}{c}{Wesenheit}
      & \multicolumn{1}{c}{Harris10} & \multicolumn{1}{c}{B19} & \multicolumn{1}{c}{H19} & \multicolumn{1}{c}{V20} \\
    & \multicolumn{1}{c}{(kpc)}   & \multicolumn{1}{c}{(kpc)}  & \multicolumn{1}{c}{(kpc)}    & \multicolumn{1}{c}{(kpc)} & \multicolumn{1}{c}{(kpc)} & \multicolumn{1}{c}{(kpc)} & \multicolumn{1}{c}{(kpc)} \\



  \hline
NGC 3201 &    $4.90_{-0.29}^{+0.36}$ & $4.80_{-0.34}^{+0.46}$ & 
$4.76_{-0.39}^{+0.51}$ & $4.9$ & $4.47_{-0.18}^{+0.18}$ & $-$ & $4.91_{-0.08}^{+0.15}$ \\ 
  NGC 4590 &   $10.63_{-1.35}^{+2.02}$ & $10.46_{-1.63}^{+2.17}$ & 
  $10.47_{-1.61}^{+2.65}$ & $10.3$ & $-$ & $10.48_{-0.28}^{+0.26}$ & $11.22_{-0.25}^{+0.16}$ \\ 
  NGC 5824 &   $30.93_{-3.06}^{+4.21}$ & $31.58_{-3.90}^{+5.74}$ & 
  $-$ & $32.1$ & $-$ & $-$ & $-$ \\ 
  NGC 5904 &    $7.41_{-0.42}^{+0.53}$ & $7.16_{-0.50}^{+0.81}$ & 
  $7.38_{-0.57}^{+0.79}$ & $7.5$ & $7.58_{-0.14}^{+0.14}$ & $7.87_{-0.19}^{+0.19}$ & $7.53_{-0.17}^{+0.11}$ \\ 
  NGC 6121 &   $1.68_{-0.07}^{+0.08}$ & $1.74_{-0.08}^{+0.09}$ & 
  $1.75_{-0.09}^{+0.09}$ & $2.2$ & $1.96_{-0.04}^{+0.04}$ & $-$ & $2.05_{-0.05}^{+0.03}$ \\ 
  NGC 6171 &   $5.43_{-0.40}^{+0.49}$ & $5.36_{-0.42}^{+0.60}$ & 
  $5.39_{-0.47}^{+0.55}$ & $6.4$ & $5.92_{-0.38}^{+0.38}$ & $6.01$ & $-$ \\ 
  NGC 6316 &   $7.05_{-1.25}^{+1.50}$ & $7.13_{-1.29}^{+1.71}$ & 
  $7.27_{-1.44}^{+1.73}$ & $10.4$ & $-$ & $-$ & $-$ \\ 
  NGC 6426 &   $23.86_{-3.36}^{+5.77}$ & $22.63_{-3.95}^{+5.78}$ & 
  $22.46_{-3.96}^{+6.66}$ & $20.6$ & $-$ & $19.83$ & $21.99_{-1.02}^{+0.85}$ \\ 
  NGC 6569 &   $12.12_{-1.70}^{+1.51}$ & $13.43_{-3.13}^{+6.10}$ & 
  $-$ & $10.9$ & $10.24_{-1.16}^{+1.16}$ & $-$ & $-$ \\ 
  NGC 6656 &   $3.46_{-0.24}^{+0.27}$ & $3.18_{-0.33}^{+0.38}$ & 
  $-$ & $3.2$ & $3.24_{-0.08}^{+0.08}$ & $-$ & $3.62_{-0.09}^{+0.09}$ \\ 
  NGC 6864 &   $21.70_{-1.55}^{+2.00}$ & $20.53_{-1.74}^{+2.60}$ & 
  $24.55_{-2.97}^{+5.09}$ & $20.9$ & $-$ & $20.79_{-0.35}^{+0.32}$ & $-$ \\ 
  NGC 7006 &  $39.81_{-2.74}^{+3.65}$ &  $40.19_{-3.28}^{+4.96}$ &  
  $-$ & $41.2$ & $-$ & $40.12_{-0.15}^{+0.16}$ & $39.78_{-1.41}^{+2.11}$ \\ 
  NGC 7078 &   $10.80_{-1.41}^{+2.02}$ & $10.88_{-1.74}^{+2.33}$ & 
  $10.68_{-1.74}^{+2.66}$ & $10.4$ & $10.21_{-0.13}^{+0.13}$ & $11.07_{-0.35}^{+0.32}$ & $11.25_{-0.33}^{+0.22}$ \\ 
  Rup 106 &    $23.20_{-2.11}^{+2.89}$ & $22.67_{-2.34}^{+3.51}$ & 
  $22.09_{-2.57}^{+4.36}$ & $21.2$ & $-$ & $-$ & $21.95_{-0.73}^{+0.73}$ \\ 
  Ter 8 &    $26.66_{-6.62}^{+13.51}$ & $26.12_{-4.52}^{+6.65}$ & 
  $-$ & $26.3$ & $-$ & $-$ & $28.74_{-0.79}^{+0.99}$ \\ 
   \hline
\end{tabular}
\end{table*}


  
 \subsection{Distance to the LMC}
 \label{subsec:LMC}
 
 \begin{table*}
 \caption{ 
 Distance moduli of the LMC obtained  from the relations for globular cluster and field RRLs derived in this work.}
 \label{tab:lmc-results}
 \begin{tabular}{lcc}
  \hline
 
  Relation & Form$^{*}$ & $\mu_{\rm LMC}$\\
\hline  
 &Globular Cluster RRLs sample\\
    \hline
    \\
  \mv&$M_{V}=\left(0.29_{-0.29}^{+0.35}\right)\left[\mathrm{Fe/H}\right]+
	 	 \left(1.05_{-0.46}^{+0.43}\right)$ & $18.507_{-0.16}^{+0.18}$\\

 \mg&   $M_{G}=\left(0.28_{-0.36}^{+0.36}\right)\left[\mathrm{Fe/H}\right]+
 \left(0.97_{-0.52}^{+0.49}\right)$ & $18.483_{-0.20}^{+0.26}$  \\

 $PW_{(G,G_{\rm BP},G_{\rm RP})}^{\rm{NGC~6864} } $& $W=\left(-2.70_{-0.16}^{+0.18}\right)\log\left(P\right)+\left(-1.21_{-0.26}^{+0.19}\right)$ &$18.512_{-0.24}^{+0.29}$\\

$PW_{(G,G_{\rm BP},G_{\rm RP})}^{\rm{NGC~5904} } $& $W=\left(-2.73_{-0.14}^{+0.12}\right)\log\left(P\right)+\left(-1.22_{-0.22}^{+0.17}\right)$ &$18.503_{-0.19}^{+0.24}$\\

$PW_{(G,G_{\rm BP},G_{\rm RP})}^{\rm{NGC~3201} } $& $W=\left(-2.63_{-0.13}^{+0.18}\right)\log\left(P\right)+\left(-1.22_{-0.22}^{+0.18}\right)$ &$18.534_{-0.20}^{+0.23}$\\

\\

\hline
      &Field RRLs sample\\
        \hline
        \\
  \mv & $M_{V}=\left(0.33_{-0.02}^{+0.02}\right)\left[\mathrm{Fe/H}\right]+
	 	 \left(1.13_{-0.03}^{+0.02}\right)$ &$18.486_{-0.032}^{+0.031}$\\
   \mg & $M_{G}=\left(0.33_{-0.43}^{+0.48}\right)\left[\mathrm{Fe/H}\right]+
 \left(1.05_{-0.03}^{+0.03}\right)$ & $18.509_{-0.023}^{+0.022}$\\
   $PW_{(G,G_{\rm BP},G_{\rm RP})}Z$ & $W=\left(-2.49_{-0.20}^{+0.21}\right)\log\left(P\right)+
\left(0.14_{-0.03}^{+0.03}\right)\left[\mathrm{Fe/H}\right]+\left(-0.88_{-0.09}^{+0.08}\right)$ &$18.489_{-0.10}^{+0.11}$\\
\\
  \hline
 \multicolumn{3}{l}{$^{*}$Relations for cluster RRLs are reported in the \citet{Carretta-et-al-2009} metallicity 
  scale, relations for field RRLs are in the }\\
 \multicolumn{3}{l}{ 
 \citet{Zinn-West-84} metallicity scale.}
 \end{tabular}
\end{table*}

 \begin{figure}
 	\begin{center}
 	\includegraphics[scale=0.75]{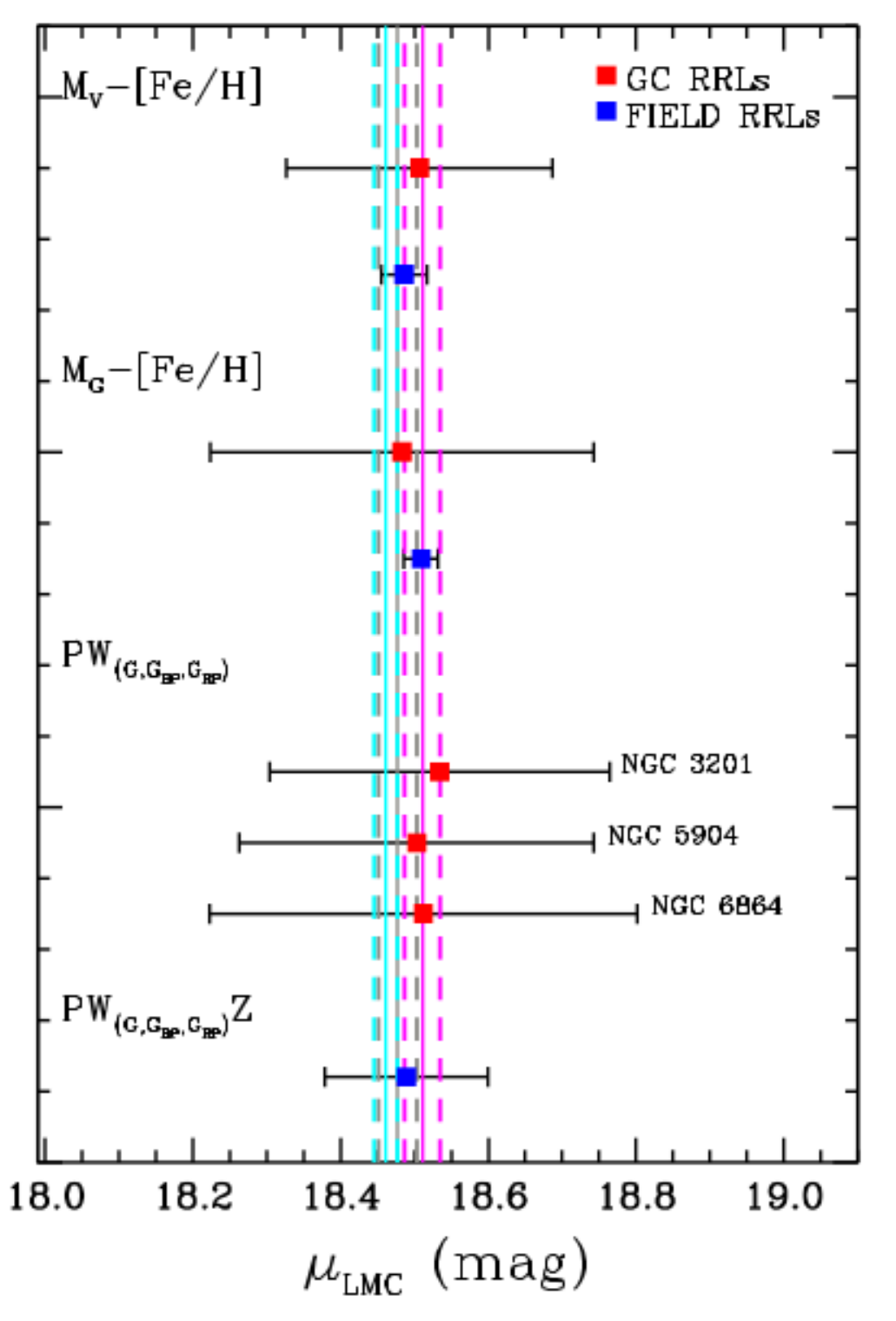} 		\caption{Credible intervals (medians and 16th-84th   percentiles) for the  distance modulus of the LMC derived  according to the procedure described in the text, using the relations for MW field (blue symbols) and GC (red symbols) RRLs inferred in this work.
 		RRL relations and $\mu_{\rm LMC}$ values shown in the figure are listed in columns 2 and 3 of  Table~\ref{tab:lmc-results}.  We show as references the most accurate and recent LMC distance modulus estimations (solid lines)  $\pm 1\sigma$ (dashed lines) by \citet{Pietrzynski-et-al-2019}:  $18.477\pm 0.026$ mag (grey) and by \citet{Riess-et-al-2021}: $18.461\pm 0.016$ (cyan) and $18.511\pm 0.024$ (magenta). 
 		}
 		\label{fig:LMC-mu-comp}
 	\end{center}   
 	
 \end{figure}

%

%
%

In Sections \ref{sec:mv} and \ref{sec:mg} we compared the distributions associated to the median coefficients of two pairs of RRLs relations ($M_{V}-{\rm [Fe/H]}$ and $M_{G}-{\rm [Fe/H]}$) 
derived in this paper and analysed  the consistency between the absolute magnitudes predicted by each pair.  
 In Sect. \ref{sec:pwz}, we presented the results corresponding to the $PW_{\left(G,G_{\rm{BP}},G_{\rm{RP}}\right)}Z$ relation for field stars and a set of  $PW_{\left(G,G_{\rm{BP}},G_{\rm{RP}}\right)}$ relations for GC stars inferred by our HM and explained why we could not derive a reliable $PWZ$ relation from them.
In this section we compare the distances provided by our relations to a universally recognised anchor of the cosmic distance ladder, the Large Magellanic Cloud (LMC). The LMC distance is known with high-accuracy  (1$\%$) from eclipsing binary stars 
(\citealt{Pietrzynski-et-al-2019}), hence provides the best place to test and validate the RRL relations we have obtained in this paper.
 In order to estimate the distance to the LMC we applied the  RRL  relations inferred in this paper to a  sample of 62 stars selected from the catalogue of 70 RRLs located close to the LMC  bar described in  \cite{Muraveva2015,Muraveva-et-al-2018a}. 
 Spectroscopically measured metallicities are available for these 62 LMC RRLs from the work of  \citet{Gratton-et-al-2004}, 
 photometry and absorptions in the $V$-band from  \citet{Clementini-et-al-2003} and pulsation characteristics (periods) from the OGLE-IV catalogue \citep{Soszynski-et-al-2019}.  Since the 
 metallicities of our GC RRL sample are in the \cite{Carretta-et-al-2009} scale 
 we first converted the metallicities of the 62 LMC RRLs to the 
 \cite{Zinn-West-84} scale by subtracting 0.06 dex \citep{Gratton-et-al-2004}  
 and then applied the quadratic relation of Eq. 3 in \cite{Carretta-et-al-2009}  
 to convert them from the 
\cite{Zinn-West-84} to the \cite{Carretta-et-al-2009} metallicity scale.  
 We then crossmatched these 62 RRLs 
 with the 
 {\it Gaia} DR2 \texttt{vari\_rrlyrae} table and recovered 40 RRLs for which  
 intensity-averaged $G$ mean magnitudes are  available and 2 RRLs which 
 also have the  corresponding $G_{\mathrm{BP}}$ and $G_{\mathrm{RP}}$  
 magnitudes. For the 40 RRLs in the $G$ 
 band we estimated the absorption as  ${A}_{G}=0.840{A}_{V}$. For the 2 RRLs 
 for which the $G_{\mathrm{BP}}$ and $G_{\mathrm{RP}}$ magnitudes are  
 available we computed  
 the Wesenheit magnitude $W\left(G,G_{\mathrm{BP}},G_{\mathrm{RP}}\right)$ using Eq. \eqref{eq:Wesenheit-mag-def} and adopting $\lambda=1.922$.  
 
 The pairs 
 $\left(\log P,\left[\mathrm{Fe/H}\right]_{\mathrm{UVES}}\right)$ corresponding to these two RRL for which the Wesenheit magnitude are available are equal to  $\left(-0.228,-1.47\right)$ and $\left(-0.237,-1.35\right)$, respectively. Since  our entire (and representative) sample of 62 RRLs has a median value of the log-period equal to  $-0.249$, in order to alleviate a biased estimate of the distance to the LMC,   we have selected only the source with its $\log P$ closest to that median value, namely the one having $\log P=-0.237$.

 The LMC distance modulus was estimated by Monte Carlo simulations as 
 follows. First, for each (dereddened)  apparent magnitude, 
 logarithm of period and metallicity in the three LMC 
 RRL  samples described above we generated 10000 realizations of Gaussian distributions with standard deviation equal to the measurement uncertainty. 
 Second, we simulated empirical probability distributions for the individual 
 absolute 
 magnitude $M_V$, $M_G$ and $W_{\left(G,G_{\rm{BP}},G_{\rm{RP}}\right)}$ in the 
 LMC samples  
 using the MCMC posterior samples of the coefficients and the intrinsic 
 dispersion associated to each fundamental RRL relation inferred by the 
 HM.  We generated a univariate distribution for each distance modulus of each RRL and combined these univariate distributions into a joint distribution. Finally, we estimated the LMC distance modulus as the median of the distribution of the medians of each sample of that joint distribution. 
 
  The distance moduli of the LMC ($\mu_{\rm LMC}$)  predicted by our  \mv, {\mg}  HMs for GC and field RRLs, the $PW_{(G,G_{\rm BP},G_{\rm RP})}Z$ HM for field RRLs, and a selection of  $PW_{(G,G_{\rm BP},G_{\rm RP})}$ relations for GC RRLs are summarised in the rightmost column  of Table~\ref{tab:lmc-results}. The selection includes the three GCs with the Wesenheit magnitudes calculated using the same ratio of total-to-selective absorption  ($\lambda=1.922$) and having the closest metallicities to the LMC, namely, NGC 6864, NGC 5904 and NGC 3201.\\
 For the $V$-band we report in the first rows of the two portions of the table the credible intervals 
 $\mu_{\rm LMC}^{\rm GC}=18.507_{-0.16}^{+0.18}$ and $\mu_{\rm LMC}^{\rm FS}=18.486_{-0.032}^{+0.031}$, being the difference between their medians of
 $0.021$  mag.  
 Our predicted LMC distance moduli corresponding to the $G$-band  (second rows of the two portions of the table) are $\mu_{\rm LMC}^{\rm GC}=18.483_{-0.20}^{+0.26}$ and $\mu_{\rm LMC}^{\rm FS}=18.509_{-0.14}^{+0.14}$ with a difference between two of  $ -0.026$ mag.  For the $PW_{(G,G_{\rm BP},G_{\rm RP})}$ relations, the inferred 
  LMC distance moduli are $\mu_{\rm LMC}^{\rm NGC~3201}=18.534_{-0.20}^{+0.23}$, $\mu_{\rm LMC}^{\rm NGC~6864}=18.512_{-0.24}^{+0.29}$, $\mu_{\rm LMC}^{\rm NGC~5904}=18.503_{-0.19}^{+0.24}$ while for the $PW_{(G,G_{\rm BP},G_{\rm RP})}(Z)$ relation $\mu_{\rm LMC}^{\rm FS}$  is $18.489_{-0.10}^{+0.11}$ mag. The largest difference of $0.045$ mag between the $\mu_{\rm LMC}$ estimates associated to the $PWZ$ and $PW$ models translates into a difference of $1.04$ kpc in the estimation of the distance to the LMC.

 
 Our 
 eight credible intervals for the LMC distance modulus are 
plotted in Figure \ref{fig:LMC-mu-comp}.  
These 
eight estimates are in good
 agreement with the value of 
 $\mu_{\rm LMC}$=$18.477\pm0.026$ mag, reported by \cite{Pietrzynski-et-al-2019}  using a sample of 20 eclipsing binary stars in the LMC and with the two recent estimates by \cite{Riess-et-al-2021} of $18.461\pm 0.016$ and $18.511\pm 0.024$ mag derived from their one- and two-parameter solutions using a $PW$ 
  calibration model 
 trained with 75 MW Cepheids and Gaia EDR3 parallaxes. 
  This comparison has the sole purpose of testing the reliability of our relations and how Gaia parallax improvements are reflected in the reassessment of the RRL relations.
  The errors related to the distance moduli of the LMC are rather large, in particular those obtained with the relations based on the GC RRL sample, compared with the more accurate values present in the literature but also, with those obtained with the relations based on the field RRL sample. The 
  LMC distance inferred from  our relations based on the EDR3 parallaxes of the GC RRLs is simply a sanity check, based on data pushed to their limit and it will be superseded by future {\it Gaia} releases and more accurate data.

\section{Summary and conclusions} 
\label{sec:summary} 
  Combining a Bayesian hierachical approach with trigonometric parallaxes  from  the recent {\it Gaia} Early Data Release 3, we have derived new 
 RR Lyrae luminosity-metallicity ($M_{V}-{\rm [Fe/H]}$ and $M_{G}-{\rm [Fe/H]}$) relations and the first empirical  RRL period-Wesenheit-metallicity   relation in the \textit{Gaia} bands ($PW_{\left(G,G_{\rm{BP}},G_{\rm{RP}}\right)}Z$).
 
 Our analysis is based on two different samples: 
 291 MW field RRLs and 
385 RRLs in 15 Galactic globular clusters spanning  metallicity ([Fe/H]) ranges from $-$2.84 to $+$0.07 dex and from $-$2.39 to $-$0.36 dex, respectively. We have used for the RRLs $V$ mean magnitudes, pulsation periods and metal abundances from the literature as well as $G$, $G_{\rm BP}$ and $G_{\rm RP}$ intensity-averaged magnitudes  from the {\it Gaia} DR2 \texttt{vari\_rrlyrae} table 
\citep{Clementini-et-al-2019}.
Our $M_{V}-{\rm [Fe/H]}$ relation 
shows a strong  dependence of the luminosity on metallicity that is higher than previously found in literature. This relation is however in good agreement with the $M_{V}-{\rm [Fe/H]}$ relation derived by \citet{Muraveva-et-al-2018a} which was  calibrated on the {\it Gaia} DR2 parallaxes for the same sample of field RRLs. 

The {\it Gaia} EDR3 parallaxes are still affected by  a zero-point offset which, however, is smaller than found for DR2.  
\citet{Lindegren-et-al-2020} using quasars (QSOs) 
estimate that the EDR3 parallaxes are systematically smaller, with a global median offset of $\sim$ $-$0.017 mas and systematic variations at a level of $\sim$ 0.010 mas.  Recent analysis of the EDR3 parallaxes, based on independent methods, report different values for the mean parallax zero-point offset: \citet{Groenewegen2021}, on a sample of sources with independent trigonometric parallaxes, found 
a mean offset (EDR3 -  independent trigonometric parallaxes) of about $-$0.039 mas, while \citet{Bhardwaj-et-al-2020} using theoretical PLZ relations for RRLs obtained a small parallax zero-point offset of $-$0.007 $\pm$ 0.003 mas in EDR3 which becomes significantly larger, $-$0.032 $\pm$ 0.004 mas, when the authors apply the corrections proposed by  \citet{Lindegren-et-al-2020}.
Using our HMs 
we have obtained independent estimations of the EDR3 parallax offset distribution using the MW field and GC RRLs. We find offset values in the range from $-$0.033 to $-$0.024 mas. These values are in agreement with the results provided both by \citet{Lindegren-et-al-2020} and \citet{Groenewegen2021}.\\
To test the reliability of our relations we applied them to estimate the distance modulus of the LMC finding values well consistent with the very accurate estimate based on  eclipsing binaries by \citet{Pietrzynski-et-al-2019}.


The 
updated $M_{V}-{\rm [Fe/H]}$  and $M_{G}-{\rm [Fe/H]}$ relations based on field RRLs derived in this paper improve the results reported in table 4 of \cite{Muraveva-et-al-2018a} 
, being now the intrinsic dispersion of both relations reduced by an order of magnitude. \\
 The relations based on EDR3 parallaxes presented in this paper are expected to further improve with 
the upcoming full {\it Gaia} data release 3 (DR3) in 2022.  
DR3 will provide, both more accurate time-series photometry of RRLs in the $G$, $G_{\rm BP}$ and $G_{\rm RP}$ bands along with photometric ${\rm [Fe/H]}$ metallicities derived from the  light curves as well as  astrophysical parameters and, in particular, individual chemical abundances and metal content (${\rm [Fe/H]}$) of the brightest sources (RRLs included) inferred from the $G_{\rm BP}$ and $G_{\rm RP}$ spectra.


\section*{Acknowledgements} 
 This work made use of data from the European Space Agency
(ESA) mission Gaia (\url{https://www.cosmos.esa.int/gaia}), processed
by the Gaia Data Processing and Analysis Consortium (DPAC; \url{https://www.cosmos.esa.int/web/gaia/dpac/consortium}). Funding for
the DPAC has been provided by national institutions, in particular
the institutions participating in the Gaia Multilateral Agreement.
 This research also made use of TOPCAT \citep{Taylor-2005} an interactive graphical viewer and editor for tabular data.We thank the
anonymous referee for the useful comments and suggestions that helped to improve the paper.
 
\section*{Data availability}

The data underlying this article are available in the article and in its online supplementary material or will be shared on reasonable request to the corresponding author.

\appendix 
\section{RR Lyrae data samples} 
\label{sec:appendix} 
 \begin{landscape}
 \begin{table}
  \caption{Information on the 291 Galactic field RRLs used in this study.   The table is published in its entirety as Supporting Information with the electronic version of the article. A portion is shown here for guidance regarding its form and content.}
  \tiny
  \label{tab:data-F}
  \sisetup{round-mode=places}
    \begin{tabular}{ll S[round-precision=4] S[round-precision=4]S[round-precision=3] S[round-precision=3] cc S[round-precision=3] S[round-precision=3] S[round-precision=3] S[round-precision=3] S[round-precision=4] S[round-precision=4]S[round-precision=4] S[round-precision=4]S[round-precision=4] S[round-precision=4]c}
 \hline
  Name &  sourceid$_{EDR3}$ & RA$_{EDR3}$ & {Dec$_{EDR3}$}& $\varpi_{EDR3}$& $\sigma_{\varpi_{EDR3}}$&Type &P& $V$& $\sigma_{V}$&A$_{V}$ &$\sigma_{A_{V}}$& $G$& $\sigma_{G}$& $G_{BP}$& $\sigma_{G_{BP}}$& $G_{RP}$& $\sigma_{G_{RP}}$&$[Fe/H]$\\ 
 &   & {(degree)} & {(degree)}& {(mas)}& {(mas)}& &{(days)}& {(mag)}& {(mag)} & {(mag)}& {(mag)}& {(mag)} & {(mag)}&{(mag)}&{(mag)}&{(mag)}&{(mag)}&{(dex)}\\ 
  \hline\\
   AA~Aql & 4224859720193721856  & 309.5627935710882 &    -2.8903444172439583 &0.7205640306818607 &0.0178503    &    AB    &0.3619 &11.822&   0.012&      0.27&    0.04320000171661377& 11.815822236294224   &0.0002 & 11.867027333027078  & 0.0031086642845804795&   11.467014975268985&   0.0003  &  $-$0.58  \\
     AA~CMi & 3111925220109675136  & 109.3298613269804     &1.7277926287510361   &   0.871767673994095   &    0.017745633    &  AB  &  0.4764  & 11.552  & 0.018   &   0.257   &0.04111999988555908    & 11.491178405549473   &0.0002  & 11.780208292438505   &0.0004  & 11.025686468473275  & 0.0003  & $-$0.55   \\
       AA~Leo & 3915944064285661312 &  174.80926143283443    &10.327207251564337     & 0.40875002006098654  &  0.022513429     & AB   & 0.5987 &  12.332  & 0.012      &0.115  & 0.018400000333786012    &12.21930730049371    &0.0004  & ""         &          ""  &          ""          &         ""                      &$-$1.47  \\
       AA~Mic  &  6773950920933073536   &  315.17318979298756  &    -39.18108882405454   &     0.21742387754588027   &    0.02568103    &     AB   &   0.4859  &   13.943   &  0.019   &     0.114  &   0.018240000009536742   &    13.842517117731603   &  0.0001    & 14.129363332236778  &   0.0015388037056199774   &  13.559620202035557   &  0.001560433704297763    &  $-$1.17   \\
       AB~Aps& 5779372319228046720&229.9093647419467&-78.6774708410973&0.2185867203923223&0.017674234& AB&0.4819&14.163&0.037&0.362&0.057919998168945316&14.10265419669739&0.00014300198339726346&  14.337235140550257&0.0006915008413300452   & 13.705214927632012 &0.000752067248104543 &    $-$1.30 \\ 
       AB~UMa& 1546016672688675200   &182.81067834805881  &  47.82875800196694     &  0.9940273454614617     & 0.018994704     & AB  &  0.5996 &  10.899  & 0.018      &0.068  & 0.010880000591278076 &   10.787574892292092  & 0.00008372118885610475    &11.052949861665663  & 0.0000736905710768544    & 10.39516640492385  &  0.00010877707420911401 &   $-$0.72\\
         AE~Leo& 3976785295394882688  & 171.5508656576936  &   17.66090237716889 &      0.370573260590598    &   0.018770542    &  AB  &  0.6268  & 12.43 &   0.019 &     0.071  & 0.011360000371932983    &12.242947868282048  & 0.0001525139701076803& 12.460834805534146  & 0.0008082092028680209 &  11.954146669967274  & 0.001106553752653826 &    $-$1.71   \\
         AF~Vel& 5360400630327427072  & 163.26066565306127  &  -49.9063846039836 &       0.8308065737995423  &    0.015529353    &  AB   & 0.5275  & 11.389 &  0.008 &     0.407 &  0.0651200008392334  &    11.25959522012131  &  0.00015017506327039265 &  11.487839019095796 &  0.0003797340065412976    & 10.858545787094254  & 0.00023786832958362964  & $-$1.64  \\ 
         AG~Tuc & 4705269305654137728 &  13.722443824474775  &  -66.70802016414046     & 0.28862854290091483   &  0.014340932    &  AB    &0.6026 &  12.861  & 0.036     & 0.046 &  0.007360000014305115    &12.778369312644912   & 0.00020409700094213033 &   13.088480709104484  & 0.00071498311474017 & 12.46220127130735   & 0.0004993905868927675   & $-$1.95  \\
           AN~Leo& 3817504856970457984  & 170.8436733138869  &  6.6346286387559275    &  0.4482523640295588  &    0.01665443     &  AB  &  0.5721 &  12.528  & 0.012   &   0.186  & 0.029760000705718995 &12.415892419225687  & 0.0001458236667627452 &    12.639526994546983 &  0.0004122116832388668 &   11.993166538367042  & 0.0006531206614462322  & $-$1.14 \\
           \\
           \hline  \\
\end{tabular}
\footnotesize
\\
\end{table}
\end{landscape}

\begin{landscape}
 \begin{table}
 \caption{Sample of 385 RRLs in the Galactic GCs considered  in this study. The table is published in its entirety as Supporting Information with the electronic version of the article.  
A portion is shown here for guidance regarding its form and content.}
 \label{tab:data-C}
 \tiny
 \begin{tabular}{lllcccccccccc}
  \hline
 Cluster& Name &  sourceid$_{EDR3}$ & RA$_{EDR3}$ & Dec$_{EDR3}$& $\varpi_{EDR3}$& Type &P& $V$& $G$& $G_{BP}$& $G_{RP}$\\
       &&  (degree)&(degree)& (mas) & &(days) & (mag) &(mag)&(mag)& (mag)\\
  \hline\\
 NGC~3201&V1&5413575310452568064&154.4284&$-$46.4438&0.1647$\pm$0.0192& RRAB&0.6048&14.810$\pm$0.006&14.6033$\pm$0.0001&14.9781$\pm$0.0008&14.0203$\pm$0.0005\\
 NGC~3201&V2&5413575520915335936&154.4166&$-$46.4439&0.1840$\pm$0.0224&RRAB&0.5326&14.800$\pm$0.006& 14.6296$\pm$0.0002&15.0855$\pm$0.0025&14.1143$\pm$0.0008\\ 
 NGC~3201&V3&5413575417830877696&154.4769&$-$46.4237&0.2009$\pm$0.0197& RRAB&0.5994&14.910$\pm$0.006&14.6589$\pm$0.0001&15.0761$\pm$0.0009&14.0678$\pm$0.0007\\
 NGC~3201&V4&5413575795793380864&154.4667&$-$46.4112&0.1412$\pm$0.0248&RRAB&0.6300&14.800$\pm$0.006&14.5671$\pm$0.0002&15.0079$\pm$0.0011&14.0107$\pm$0.0007\\
 NGC~3201&V5&5413575555275148672&154.4220&$-$46.4184&0.1942$\pm$0.0251&RRAB&0.5013&14.730$\pm$0.006&14.6018$\pm$0.0002&14.8527$\pm$0.0043&14.0792$\pm$0.0012\\
 NGC~3201&V6&5413574730641610112&154.3587&$-$46.4506&0.1726$\pm$0.0256&RRAB&0.5253&14.740$\pm$0.006&14.5643$\pm$0.0001&14.8966$\pm$0.0013&14.0671$\pm$0.0060\\
 NGC~3201&V7&5413573974727032704&154.3687&$-$46.4635&0.1909$\pm$0.0195&RRAB&0.6303&14.690$\pm$0.006&14.4969$\pm$0.0001&14.8249$\pm$0.0022&13.9259$\pm$0.0028\\
 NGC~3201&V8&5413574077806291456&154.3781&$-$46.4387&0.1272$\pm$0.0253&RRAB&0.6286&14.740$\pm$0.006&14.5398$\pm$0.0001&14.9168$\pm$0.0008&13.9482$\pm$0.0006\\
 NGC~3201&V9&5413574112159737472&154.3845&$-$46.4368& 0.1731$\pm$0.0220 & RRAB &0.5254& 14.820$\pm$0.006 & 14.6330$\pm$0.0001 & 15.0181$\pm$0.0010 & 14.0982$\pm$0.0008\\
 NGC~3201&V10&5413576826585163264&154.3339&$-$46.3472& 0.2115$\pm$0.0200 & RRAB &0.5352 & 14.840$\pm$0.006 & 14.6532$\pm$0.0002 & 15.0723$\pm$0.0125 & 14.1087$\pm$0.0006\\
 \\
  \hline
  \end{tabular}
  \normalsize
\end{table}
\end{landscape}




\bibliographystyle{mnras}
\bibliography{references-PLZ-GC-EDR3} 

\bsp	
\label{lastpage}
\end{document}